\documentclass[]{aastex701}
\usepackage{soul}
\usepackage{color}
\usepackage{subcaption}  

\begin{document}

\title{Primary and Secondary Star Formation in the Spiral Waves of NGC 628}

\author[0000-0002-1723-6330]{Bruce G. Elmegreen}\affiliation{Katonah, NY 10536, USA}
\email[show]{belmegreen@gmail.com}  

\author[0000-0002-1000-6081]{Sean T. Linden} \affiliation{School of Physics and Astronomy, Rochester Institute of Technology, 84 Lomb Memorial Drive, Rochester, NY 14623, USA} \affiliation{Steward Observatory, University of Arizona, 933 N Cherry Avenue, Tucson, AZ 85721, USA}\email{stlsps@rit.edu>}

\author[0000-0001-8348-2671]{Kelsey E. Johnson}\affiliation{Department of Astronomy, University of Virginia, Charlottesville, VA, USA}\email{kej7a@virginia.edu}

\author[0000-0002-8192-8091]{Angela Adamo}\affiliation{Department of Astronomy, The Oskar Klein Centre, Stockholm University, AlbaNova, SE-10691 Stockholm, Sweden}\email{adamo@astro.su.se}

\author[0000-0002-5189-8004]{Daniela Calzetti}
\affiliation{Department of Astronomy, University of Massachusetts Amherst, 710 North Pleasant Street, Amherst, MA 01003, USA}\email{calzetti@astro.umass.edu>}

\author[0009-0008-4009-3391]{Varun Bajaj}
\affiliation{Space Telescope Science Institute, 3700 San Martin Drive, Baltimore, MD 21218, USA}\email{vbajaj@stsci.edu>}

\author[0009-0003-6182-8928]{Giacomo Bortolini}
\affiliation{Department of Astronomy, The Oskar Klein Centre, Stockholm University, AlbaNova, SE-10691 Stockholm, Sweden}\email{giacomo.bortolini@astro.su.se}

\author[0000-0001-6464-3257]{Matteo Correnti}
\affiliation{INAF Osservatorio Astronomico di Roma, Via Frascati 33, 00078 Monteporzio Catone, Rome, Italy}
\affiliation{ASI-Space Science Data Center, Via del Politecnico, I-00133 Rome, Italy}\email{matteo.correnti@inaf.it}

\author[0000-0002-5259-4774]{Ana Duarte-Cabral}
\affiliation{Department of Physics and Astronomy, School of Physical, Chemical and Environmental Sciences, Cardiff University, The Parade, CF24 3AA Cardiff, UK}\email{adc@astro.cf.ac.uk}

\author[0000-0002-3247-5321]{Kathryn Grasha}
\affiliation{Research School of Astronomy and Astrophysics, Australian National University, Canberra, ACT 2611, Australia}
\affiliation{ARC Centre of Excellence for All Sky Astrophysics in 3 Dimensions (ASTRO 3D), Australia}\email{kathryn.grasha@anu.edu.au}

\author[0000-0002-3005-1349]{G\"oran \"Ostlin}
\affiliation{Department of Astronomy, The Oskar Klein Centre, Stockholm University, AlbaNova, SE-10691 Stockholm, Sweden}\email{ostlin@astro.su.se}

\author[0000-0002-8222-8986]{Alex Pedrini}
\affiliation{Department of Astronomy, The Oskar Klein Centre, Stockholm University, AlbaNova, SE-10691 Stockholm, Sweden}\email{alex.pedrini@astro.su.se}

\author[0000-0002-2918-7417]{Jenna Ryon}
\affiliation{Space Telescope Science Institute, 3700 San Martin Drive, Baltimore, MD 21218, USA}\email{ryon@stsci.edu}

\author[0000-0002-0806-168X]{Linda J. Smith}
\affiliation{Space Telescope Science Institute, 3700 San Martin Drive Baltimore, MD 21218, USA}\email{lsmith@stsci.edu}

\begin{abstract}
To determine the role of spiral arms and stellar feedback on the formation of stars, we examine the positions of optically identified and emerging star clusters in NGC 628. We find that the youngest clusters extend further into the spiral arms at small radii than at large radii, as expected when the angular flow rate through a wave-like arm increases with greater distance from corotation.  The embedded or emerging star clusters (eYSCs) are identified relative to four distinct morphological features in the $7.7\mu$m image: spiral arm filaments, spiral arm spurs, spiral interarm feathers, and bubbles.  Spiral arm-related eYSCs are presumably first-generation stars made from interstellar gas that was directly compressed by the density wave, while clusters along the rims of bubbles are secondary generations produced by feedback compression.  The mass functions of the first and second generations differ: eYSCs in spiral arm filaments extend to significantly higher masses than eYSCs in bubbles. Secondary star formation  continues downstream from primary star formation,  removing the expected age gradient through the arms. Bubble eYSCs account for approximately one-third of all the emerging cluster mass.  The outer arms have a more mixed distribution of spiral- and bubble-related eYSCs than the inner arms, potentially providing a way to distinguish wave and material arms and to find corotation in wave-like arms. Among the 17 most massive clusters with $\log M(M_\odot)>4$, eight of them are in four tight pairs; in at least one case, the dynamical time for the pair is comparable to the cluster ages. 

\end{abstract}

\section{Introduction}
The connection between star formation and spiral arms has been discussed since the time of \cite{baade51} and \cite{morgan53} when OB associations were first observed along the arms of M31 and the Milky Way \citep[see reviews in][]{vallee17,hou21}. Blue arms as a characteristic feature of spiral galaxies suggested generally enhanced star formation rates, and this led to the triggering model in compressed density wave gas \citep{roberts69}. The concentration of molecular clouds in spiral arms \citep{heyer15} and an increase in the luminosity of spiral-arm star-forming regions support this idea \citep[e.g.,][]{rumstay83,rand92,thilker00,scoville01,lee11,azimlu11,higdon13,santoro22,linden26}. 

NGC 628 is a good example of a face-on grand-design spiral galaxy. At a distance of 9.84 Mpc (\citet{anand21}, see also \citet{ck25}), it has been surveyed extensively for molecules, star formation, and young stellar clusters \citep[e.g.,][]{larsen04,grasha15,leroy23}. Here we examine the locations of these optically identified clusters, as well as the locations of highly-extincted clusters, relative to the density wave seen in old stars at $1.5\mu$m and relative to the dust emission at $7.7\mu$m.  Our goal is to determine the types of regions where star clusters form and the influence of spiral waves and star-formation feedback on cluster mass. 

We begin with a detailed summary of star formation, spiral arms, feedback bubbles and young clusters in NGC 628 (Sect. \ref{sect:overview}). We then summarize our data sources in Section \ref{sect:data}. The first step in our analysis is to determine the location of the crest in the old stellar spiral (Sect. \ref{sect:spirals}). The positions of young optical clusters are examined with respect to this crest to look for evidence of a time sequence or steady flow through the spiral pattern (Sect. \ref{sect:optical}).  We consider the emerging clusters next (Sect. \ref{sect:embedded}) and determine their locations with respect to three main spiral wave features, i.e., the main arms, bright spurs that jut off from the main arms, and interarm feathers; these features are explained in Sect. \ref{sect:spurs}. We also identify emerging clusters that are at the edges of bubbles (cf. Sects. \ref{sect:bubbles}, \ref{sect:eysc_bubbles}).  This characterization of the youngest star-forming regions is used to determine the relative numbers and masses of clusters that recently formed by gas collapse associated with specific dynamical processes. In Section \ref{sect:cmf}, we compare the cluster mass functions for each environment to examine the claim that spiral arms make more massive clusters (Sect. \ref{sect:armSF}). From the relative cluster masses we also suggest approximate proportions of interstellar turbulence that is driven by spiral arms and feedback (Sect. \ref{sect:local}).  A surprising discovery is the presence of three tight binary clusters and one moderately tight binary among the 17 most massive clusters (Sect. \ref{sect:pairs}). The results are summarized in Section \ref{sect:sum}.

\section{Spiral Arms, Clusters and Star Formation in NGC 628}
\label{sect:overview}

\subsection{Arm and Interarm Star Formation in NGC 628}
\label{sect:armSF}

There is a long history comparing star formation properties in the arm and interarm regions of spiral galaxies \citep[e.g.,][]{elmegreen95,querejeta21,suncalzetti24}. For NGC 628, \cite{santoro22} found from ground-based observations that the luminosity function for spiral-arm HII regions is slightly shallower than it is for interarm HII regions, suggesting a higher proportion of the most luminous regions in the arms. This trend was significant for about half of their 13 galaxies with clear spirals. \cite{vieira25} similarly observed higher molecular cloud surface densities in the arms of NGC 628 than in the interarms, but only inside a galactocentric radius of 2.5 kpc.  This radius corresponds to the radial range where there appears to be a steady density wave.  \cite{williams21} derived a corotation radius of $4.5\pm2.0$ kpc ($94^{\prime\prime}$) for the inner spiral in NGC 628 where the pattern speed is constant.  \cite{speights25} derived a similar corotation radius, 4.8 kpc, if the same distance is used (9.84 Mpc).  \cite{sakhibov21} found a slightly larger corotation radius from kinematics and the displacement of HII regions from clusters. In comparison, the largest full circle in the JWST F770W image has a radius of $\sim3$ kpc.

\cite{foyle10} measured the star formation rate per unit molecular mass, defined as the star formation efficiency, SFE, in the arms and interarms of three galaxies, including NGC 628, and found less than a 10\% increase in the arms. \cite{querejeta24} observed SFEs for 28 spiral galaxies and saw values only slightly higher in the arms unless the stellar arm/interarm contrast was large, $\gtrsim2$. NGC 628 was one of their weak-arm galaxies where the arm/interarm SFE contrast was close to unity while the molecular surface density contrast was a factor of $\sim2$. \cite{suncalzetti24} compared arm and interarm SFE in NGC 628 and NGC 4321, finding less than a factor of 2 difference.  \cite{meidt21} found a larger arm/interarm contrast for CO than stars in NGC 628. \cite{romanelli25} measured these properties for 22 spiral galaxies and obtained about the same result; for NGC 628, the arm/interarm contrast for the efficiency of star formation was $\sim0.9$, which comes from the product of the contrasts of the star formation rate surface density, $\sim1.7$, and the CO cloud lifetime, $\sim1.2$, divided by the contrast of the CO dense-cloud surface density, $\sim2.4$. A similarly low arm/interarm contrast for SFE in the Milky Way was found by \cite{moore12}.

Variations in star formation properties with spiral phase have also been noted for NGC 628.
\cite{speights25} showed that H$\alpha$ tends to lie $\sim40^\circ$ further in angle downstream along the arms than CO emission at about half the corotation radius, corresponding to a distance of 1.7 kpc. \cite{querejeta25} found the same relative positions for 24 galaxies, and for NGC 628, the H$\alpha-$CO offset was $343\pm1223$ pc on average for the inner regions (the dispersion represents the scatter in offset for different radii).  \cite{gusev14} observed different star formation rate densities in the two arms of NGC 628. Regarding cluster ages, \cite{shabani18} found no offset in the azimuthal distribution of star clusters with different ages, in agreement with simulations of star formation in spiral galaxies by \cite{grand12}. 

\subsection{Bubbles in NGC 628}
\label{sect:bubbles}

Perhaps the most interesting feature of gas and star formation in NGC 628 is the ubiquity of bubble structures in warm dust emission in both the arm and interarm regions \citep{bagetakos11,watkins23a,watkins23b,nishimoto25}. These bubbles are sometimes classified as a filamentary network \citep{thilker23,meidt23}. Simulations of such a filamentary network for a galaxy like NGC 628 are in \cite{koletic26}.  For the bubbles associated with star formation, there are often aging clusters and OB associations inside the bubbles with younger clusters and star-forming regions along the periphery \citep{barnes23,egorov23,mayya23,watkins23a,watkins23b}.  Many bubbles form by feedback pressure from previous generations of stars \cite[e.g.,][]{mayya23,jimenez24} and trigger new star formation in the accumulated gas \citep[][but see Porter et al. 2024]{watkins23b,barnes23}. \cite{thilker23} found that 60\% of the clusters younger than 5 Myr are associated with filaments, while loose OB associations are anti-associated with the filaments, suggesting an age or dispersal sequence from the high to the low density gas. \cite{watkins23a} observed that bubble radii increase downstream in the wave, suggesting continued feedback into the cavities as they age, and that 30\% of bubbles have at least one smaller bubble on the rim. \cite{egorov23} similarly found that in NGC 628, 40\% of the young stellar associations are located on superbubble rims. With secondary star formation and continued triggering throughout an arm, it is clear why the youngest clusters have no downstream age gradient, i.e., new clusters are constantly forming along the swept-up rims of feedback-driven cavities. 

\subsection{Star Clusters in NGC 628}
\label{sect:clusters}

The clusters themselves can be traced from the embedded phase \citep{lada03} to the emerging and optical phases with ages and masses determined from photometry. In NGC 628, \cite{turner19} measured a smaller separation between dense dust clouds seen with mm-wave continuum emission and optical HST clusters younger than 10 Myr compared to older clusters.  JWST revealed even younger cluster evolution. \cite{pedrini24} showed that polycyclic aromatic hydrocarbon (PAH) emission in NGC 628 decreases in the vicinity of embedded clusters after 3 Myr, and that gas and dust generally disperse faster around more massive clusters, although the trend with mass in NGC 628 was slightly less than for three other galaxies \citep{pedrini26}. \cite{rodriguez25} also derived 3 Myr as an average time for PAH fading around embedded clusters in 19 galaxies.  This time interval is before the supernovae phase begins, and is consistent with the observation that photoionization dominates feedback in NGC 628 embedded clusters \citep{vieira26}. Gas clearing on 10's of pc scales occurs at high cluster masses \citep{pedrini26} in $\sim5$ Myr for a variety of galaxies \citep{chevance22,kim25,Ramambason26}. On these longer timescales, supernovae can be important for clearing gas around the most massive stars. In all cases, these clearing times are short compared to the flow-through time in a spiral arm, which can be 20 Myr or more in the inner regions of a galaxy, and the expansion time of a superbubble, which can be 20 Myr \citep{barnes23} to 50 Myr \citep{mayya23} depending on size and location. This comparison implies that embedded clusters and the youngest optical clusters can be used to indicate where stars form relative to spiral arm features and gas bubbles. 

The optical clusters observed in NGC 628 with HST and the emerging clusters observed with JWST are also correlated with each other in position, with a two-point correlation function having a power-law decline  up to several hundred pc for the youngest clusters \citep{grasha15,grasha17a,menon21,lapeer26}. The emerging clusters tend to be near each other and not randomly placed \citep{buckner26} because star formation is hierarchical, following structural patterns of turbulence with smaller scales evolving faster than larger scales \citep{elmegreen00}. There is also a correlation between cluster separation and age difference that extends up to several hundred pc and $\sim60$ Myr in NGC 628 \citep{grasha17b}. Broadband intensities sampling not just clusters but also unbound young regions are correlated from $\sim3$ pc to $\sim100$ pc \citep{elmegreen06}.  The maximum scale of these correlations seems to be important as an indicator of the transition between feedback-driven structure and gravity-driven structure related to spiral waves and gas instabilities, as suggested for NGC 628 by \cite{chanu26}. A similar transition from ``Swiss cheese'' (hole) to ``meatball'' (clump) structures was noted in NGC 628 by \cite{sandstrom23} at a gas surface density of $10\;M_\odot$ pc$^{-2}$, suggesting again a physical transition from feedback to self-gravity. The mean spacing between filaments in NGC 628 was also attributed to gas self-gravity by \cite{meidt23}. Shear is likely to be important at the transition scale \citep{grasha17b}.

\subsection{Spiral arms, Spurs, and Interarm Feathers}
\label{sect:spurs}

The filaments discussed by \cite{thilker23} and \cite{meidt23} in the context of $7.7\mu$m emission include dust structures that are much longer and more continuous than bubble walls and which trace the spiral arm and interarm regions for a kpc or more in length. These structures were recognized long ago from ground based optical images and HI maps. \cite{lynds70} cataloged interarm dust ``feathers'' in 7 galaxies, including NGC 628, where she measured a spiral pitch angle of $50^\circ$ (we derive the same value in Eq. \ref{eq:feather} below). That work could only trace these feathers over their darkest extent, which was 0.5 kpc inside the bright main arms. At the same time, \cite{weaver70a} noted bright structures outlined by young stars and HII regions that he called ``spurs'' or ``interarm features''. \cite{weaver70b} observed in HI that the Sun is located in an ``offshoot or spur of the Sagittarius arm'' in the Milky Way with a pitch angle for both young stars and gas of $20^\circ$ to $25^\circ$ \citep[see also][]{xu16,hou21}, compared to the main spirals which have a pitch angle of $12.5^\circ$.  \cite{delmegreen80b} studied spurs in 7 galaxies using B- and I-band photographic plates. The 5 spurs  she identified in NGC 628 were bright in both bands and extended for an average length of 3.5 kpc at an average pitch angle of $63^\circ$. They spanned close to the full extent of the interarm regions from one arm to the next. She concluded that because the extents and pitch angles of Lynd's dark feathers and her bright spurs were about the same, these object are probably the same phenomenon. She added that the presence of spurs in I-band images, their near-universal pitch angles, and their relatively short shear times if they were material objects, suggested they were long-lived features connected with the main spiral arms.

A dynamical origin for spurs was first offered by \cite{balbus88}, who suggested they are the result of gaseous gravitational processes in the expanding, reverse-shear environment that exists inside and downstream from a stellar density wave. His analysis also suggested there should be a regular spacing of massive spiral arm star-forming regions at the bases of these spurs, similar to those observed at the time in other galaxies \citep{elmegreen83}. In this interpretation, spurs are the bright star-forming parts of the instability where the gas is densest, while feathers are dusty gaseous extensions into the interarm regions. Both are pinned to the stellar spiral and consistent with D.M. Elmegreen's interpretation of spurs as relatively long-lived features.  Computer simulations of self-gravitating gas flow through a spiral arm confirmed these ideas \citep{kim02a,kim02b,kim06a}. In more global simulations, \cite{shetty06} found that spurs jut outward from the convex sides of spiral arms inside corotation and inward from the concave sides of the arms outside corotation.  An analytical model of self-gravitating spiral arm gas flow by \cite{lee12} found seven dimensionless quantities that characterized their structure, suggesting their use as gas flow diagnostics.  The key ingredient of these models is the kinematic motion of gas in a spiral arm, as self-gravity is not needed to make a spur if the spiral shock is strong enough \citep{dobbs06}. The walls of downstream feedback bubbles can also look like spurs \citep{kim20}. 

Interarm feathers measuring several kpc in length were first observed in M51 at $15\mu$m with 300 pc resolution by the ISOCAM Instrument on the Infrared Space Observatory \citep{block97}. They were first seen at high resolution optically in M51 with HST \citep{scoville01} and in M81 using the Infrared Array Camera on Spitzer \citep{willner04}.  In an archival survey, \cite{lavigne06} found interarm ``feathers'' in 45 Sb and Sc type galaxies imaged with HST, including NGC 628; they concluded on the basis of feather spacing and surface density dependence that gaseous self-gravity was important in their formation. Interarm feathers were also observed in the earliest high-resolution CO galaxy maps \citep{corder08,koda09,schinnerer13,schinnerer17}, where they were seen to form stars. Their presence implied the associated molecular clouds can live for a long time as they pass through the interarm region \citep{koda09,koda23}. \cite{chandar11} measured cluster ages of 100 to 200 Myr in the interarm feathers of M51.

Long interstellar filaments are also observed in the Milky Way, the longest being the ``Radcliff wave,'' which is a 2.7 kpc long distribution of molecular clouds near the solar neighborhood  \citep{alves20,li22b}. The ``Gangotri wave'' is another, at least 2 kpc long at a distance of $\sim5$ kpc in the direction of the Galactic center; it could be a molecular interarm feather \citep{veena21}. ``Nessie'' is at least $\sim160$ pc long \citep{jackson10,goodman14} at a distance of 3.1 kpc in the midplane of the Scutum–Centaurus Arm.  \cite{li13} traced a 500 pc long star-forming filament in CO and $24\mu$m emission at a distance of $\sim10$ kpc in the Perseus arm, where \cite{du17} also identified four new molecular filaments $\sim100$ pc long.  The ``Cassiopeia Filament'' is $\sim390$ pc long \citep{chen23}. \cite{he26} found a 80 pc filament at 5.2 kpc distance. \cite{ge23} cataloged 88 molecular filaments, and found 8 in spiral arms and 3 in the local spur. \cite{ssun24} determined the molecular fractions in several long filaments from HI self-absorption and \cite{feher24} compared the dense gas fractions in an arm and interarm filament.   Some of these molecular filaments are large enough to be related to galactic-scale processes, but the smaller filaments are not likely to have the same origin \citep{arzoumanian11,battersby14,zucker15,zucker18}. As suggested by \cite{thilker23} for NGC 628, many of the smaller filaments could be swept up by young stellar feedback \citep[c.f.,][]{schneider79}.  In the F770W images of NGC 628 shown here, the longest dust filaments extend for a kpc or more.

An important consequence of the presence of spurs with young stars and interarm feathers of gas is that they imply the interstellar medium is flowing through the spiral arm from one side to the other. This can only happen if the arm is a density wave and not purely material. Material arms accumulate gas and disk stars from both sides, forming stars in the middle, with little flow through the arm and no spurs or feathers \citep{dobbs10,wada11}.

\subsection{Massive Clouds in Spiral Arms}
\label{sect:massiveclouds}

The cloud masses at the main-arm anchor points for spurs and feathers were theoretically predicted to be the interstellar turbulent Jeans mass, which is $\sim10^7\;M_\odot$ \citep[e.g.,][]{kim02a}. This is much larger than a typical giant molecular cloud, but such large masses were known, as they had been observed in HI emission along the spiral arms of the Milky Way \citep{mcgee64,henderson82,elmegreen87a} and other galaxies \citep[e.g.,][]{allen73}, as well as in CO and dust cloud observations of other galaxies \citep{rand90,delmegreen80}. The HI structures were called superclouds by \cite{belmegreen80} to distinguish them from giant molecular clouds, which tend to be condensations or core regions inside the larger atomic structures \citep{grabelsky87,elmegreen87a,elmegreen87b,lada88,park23}.  CO clouds of this mass were called giant molecular associations \citep{rand90}.  The largest molecular clouds in a galaxy tend to be larger in the arms than in the interarm regions \cite[e.g.,][]{bazzi26}.

\section{Data Sources}
\label{sect:data}

\subsection{Near- and Mid-Infrared Imaging}

This work makes use of JWST NIRcam and MIRI imaging obtained as part of the Feedback in Extragalactic Star Clusters (FEAST) survey of NGC~628, which combines $0.3 - 7.7\mu$m observations to study the earliest stages of stellar cluster evolution. The fully reduced mosaics, data reduction procedures, and photometric calibrations are described in detail by \cite{pedrini24} and \cite{gregg24} (see also A. Adamo et al, in prep.). Throughout this paper we adopt a distance of 9.84~Mpc for NGC~628 \citep{anand21}, corresponding to a physical scale of 47.7~pc~arcsec$^{-1}$.

The stellar mass distribution used to define the spiral density wave is traced with the JWST/NIRCam F150W image. At $1.5\mu$m the emission is dominated by evolved stellar populations and is minimally affected by dust extinction, providing an excellent tracer of the stellar mass distribution \cite[e.g.,][]{meidt14,lee23}. The final F150W mosaic has a pixel scale of 0.04~arcsec~px$^{-1}$ and is registered to a common astrometric reference frame for all FEAST targets, Gaia DR3.

The morphology of the dense interstellar medium is traced using the JWST/MIRI F770W image, which has a pixel scale of 0.08~arcsec~px$^{-1}$. Emission in this filter is dominated by the strong $7.7\mu$m PAH feature together with warm dust continuum, producing a high-resolution view of the dusty structures associated with recent star formation \citep[e.g.,][]{peeters04,gregg26}. We use the F770W image to identify spiral-arm filaments, spiral-arm spurs, inter-arm feathers, and feedback-driven bubbles. Because the youngest stellar clusters remain spatially coincident with their natal dust emission for only a few Myr, the F770W morphology provides a direct view of the environments in which these clusters form \citep[e.g.,][]{chevance22,pedrini25}. No correction has been made for stellar continuum in the F770W images because that is a minor contribution.

\subsection{Star Cluster Catalogs and Derived Physical Properties}

The analysis presented here combines the optical and emerging star cluster catalogs presented by \cite{linden26}. Together, these catalogs provide a nearly continuous census of stellar cluster evolution, extending from deeply embedded and emerging young stellar clusters (eYSCs) that remain associated with their natal dust clouds to optically visible clusters that have largely emerged from their birth material. Full details of the cluster identification, photometric analysis, and catalog construction are presented by \cite{linden26}; here we provide a brief summary.

The optical catalog was constructed from archival \textit{HST} imaging observations. Candidate star clusters were initially identified using the Source Extractor Package (SEP) on the ACS F555W image. The measured concentration-index (CI) was then used to separate compact stellar clusters from individual stars, chance super-positions, and background galaxies. The remaining cluster candidates were visually inspected following the same procedure as in the Legacy Extragalactic UV Survey \citep[LEGUS -][]{calzetti15,adamo17}. The combined HST and JWST NIRCam photometry up to $2\mu$m was then used to construct spectral energy distributions (SEDs) for all 1329 candidates.

The eYSC catalog was constructed from JWST/ NIRCam observations using continuum-subtracted narrow-band imaging of hydrogen recombination lines (F187N - Pa$\alpha$, and F405N - Br$\alpha$) and the $3.3\mu$m PAH feature (traced by the F335M filter). Details of the continuum subtraction for these emission features are presented in \cite{gregg24}. Candidate eYSCs were first identified as sources with bright hydrogen recombination-line emission, with eYSC-I sources exhibiting compact and spatially coincident $3.3\mu$m PAH emission, and eYSC-II sources exhibiting diffuse or extended $3.3\mu$m PAH emission. These classifications are interpreted as an evolutionary sequence in which the surrounding photo-dissociation region is disrupted and dispersed as stellar feedback clears the natal material, producing a transition from eYSC-I to eYSC-II as the cluster becomes more optically-visible \citep{knutas25,pedrini26,linden26}. The combined emerging star cluster catalog contains 835 eYSC-I and 741 eYSC-II sources. In the final catalog, 197 sources that fall outside the field of view of the JWST/MIRI F770W image are removed from further analysis. In combination with optically-selected sources, our complete catalogs provide a nearly continuous view of cluster evolution from the embedded phase through the exposed optical stage across the FEAST footprint of NGC~628.

Physical properties for both catalogs were derived using the \textsc{CIGALE} SED fitting framework. The optical and emerging star cluster catalogs were fit using independently optimized model grids that account for differences in the evolutionary stages of the two populations. Both of the adopted model grids include an instantaneous single stellar population, nebular line and continuum emission, and dust attenuation. The models used to fit emerging stellar clusters also include infrared dust emission, allowing the ultraviolet through near-infrared SEDs of eYSCs to be modeled self-consistently. The complete model grids and fitting parameters are presented in Tables~2 and~3 of \cite{linden26}. \textsc{CIGALE} constructs probability distribution functions for every fitted parameter, and the ages, stellar masses, extinctions, and associated uncertainties adopted throughout this work correspond to the marginalized Bayesian estimates. Because both the optical and embedded catalogs were analyzed within the same framework, their derived physical properties can be compared directly across the full evolutionary sequence investigated in this paper. Both catalogs are complete for masses down to $10^3\;M_\odot$ 
\citep{knutas25,pedrini25}.

\section{The Stellar Spiral Wave}
\label{sect:spirals}

The position of the stellar spiral was determined from the JWST F150W image, which shows old stars in the disk. For all radii between 700 and 1500 pixels in one pixel steps, corresponding to 1335.8 pc and 2864.2 pc from the center, we fit the azimuthal intensity profile as a function of angle, $\theta$, by a double sine function, $A\sin(2\theta)+B\cos(2\theta)$, and then used the fitting constants $A$ and $B$ to determine the two positions of peak intensity, $\theta_{\rm peak}$ and $\theta_{\rm peak}+\pi$. These azimuthal angles were then fit linearly to the log of the radius to determine the best-fit logarithmic spiral. The spiral deviates from this logarithmic form inside and outside our limiting radii; the edge of the F770W image is at a radius of $\sim1700$ pixels. In making these fits, we replaced all the pixels that exceeded the mean plus $3\times$ times the rms in each azimuthal scan with that threshold value to avoid excessively bright pixels from small sources.  The resulting stellar spiral differs slightly from a spiral defined by star formation and warm dust \citep[e.g.,][]{querejeta21} because young stars tend to lag the old stellar wave more at smaller radii inside corotation, giving the star formation ridge a slightly smaller pitch angle than the wave (see below).

\begin{figure*}
\begin{center}
\includegraphics[width=5.5cm]{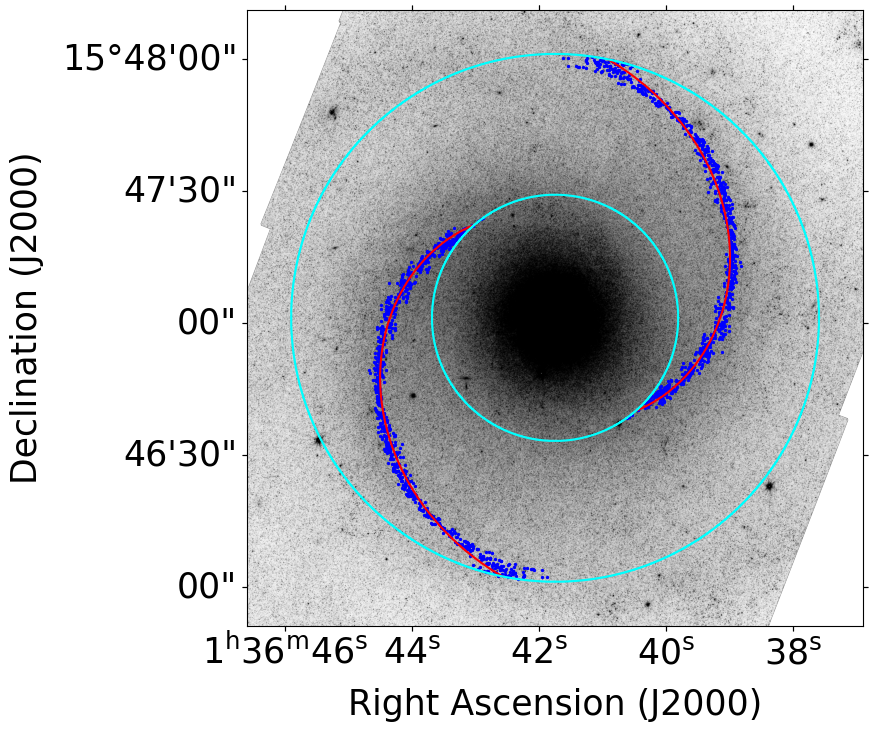} 
\includegraphics[width=5.1cm]{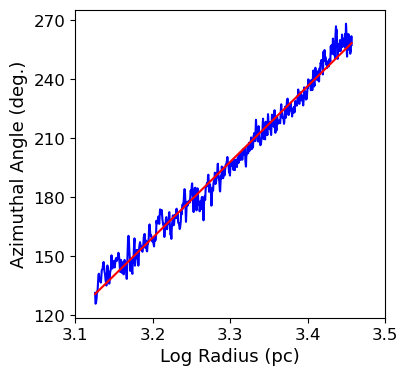}  
\includegraphics[width=6.8cm]{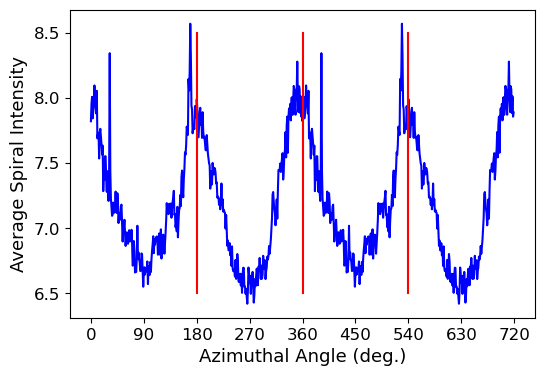}  
\caption{(Left:) F150W image of NGC 628 with blue points showing the positions of the peaks in the spiral wave and red curves showing the logarithmic spiral fit to these peaks. The inner and outer radii of the fit are indicated by cyan circles (at $28^{\prime\prime}$ and $60^{\prime\prime}$). The spiral enclosed by these circles is supposed to be a density wave with a steady pattern and corotation at $94^{\prime\prime}$, which is slightly outside this field of view.  (Middle:) Points and fitted straight line for the density wave peak in angle-log Radius coordinates.  Azimuthal angle increases counter clockwise from the west. The starting angle for the spiral, $130.9^\circ$, corresponds to the inner radius of the eastern arm. (Right:) Spiral wave profile at $1.5\mu$m displayed with two full cycles, made by averaging all the azimuthal intensity profiles between the radial limits, shifted in azimuth to align all the peaks. The red lines are at $180^\circ$ intervals for illustration of the two sides of the galaxy; these are the angles of the peaks after fitting each azimuth to an $m=2$ sine function. 
}
\label{fig:crest}
\end{center}
\end{figure*}

\begin{figure*}
\begin{center}
\includegraphics[width=4.4cm]{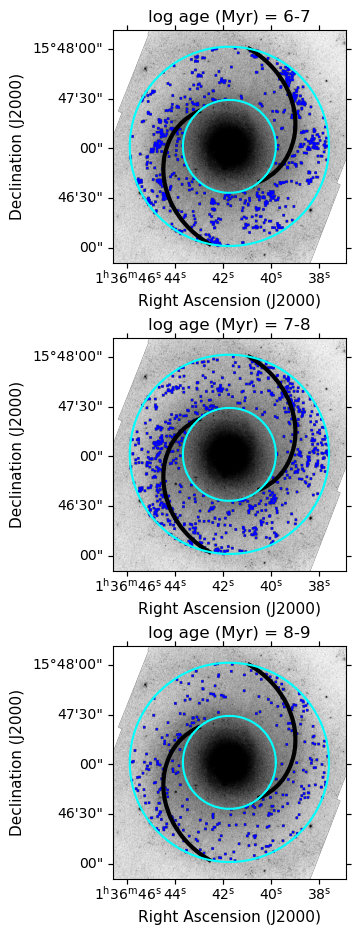}  
\includegraphics[width=8.cm]{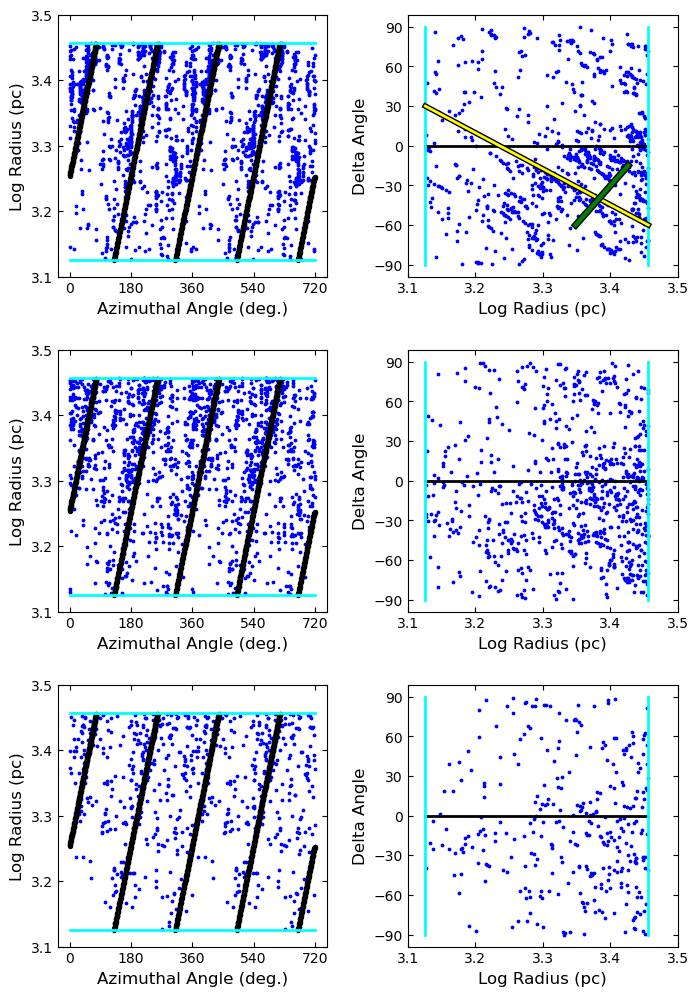}
\caption{Optical clusters in three age groups (top to bottom) shown (left) in position on the F150W image, along with the fitted spiral arms and radial limits; (middle) plotted in a coordinate plane of $\log_{10} {\rm Radius}$ versus Azimuthal Angle, with the spiral arms as straight black lines and the radial limits as cyan lines, and (right) plotted in a coordinate plane of angular difference from the spiral arm peak at each radius versus $\log_{10} {\rm Radius}$, again with the spiral arms as a black line and the radial limits as cyan lines. The Azimuthal Angle and Delta Angle decrease in the clockwise direction, which means trailing the arm peak as the gas flows through.  The galaxy and $\log_{10}{\rm Radius}$-angle view clearly show large groupings of young clusters downstream from the peak of the spiral wave, and more so for younger clusters. The Delta Angle versus $\log_{10}{\rm Radius}$ view shows these groupings too, as concentrations of points that start at around $-60^\circ$ at mid-radius and move up toward a Delta Angle of $\sim0$ at the largest radius (in the direction of the green line). The thin diagonal streaks of points that go from left to right and top to bottom on the right-hand panels (parallel to the yellow line, which is drawn to guide the eye) are spiral interarm feathers with a pitch angle of $\sim50^\circ$. 
}
\label{fig:opticalclusters}
\end{center}
\end{figure*}

Figure \ref{fig:crest}-left shows the F150W image with inner and outer radial limits as cyan circles, positions of the spiral arm peaks as blue squares, and the best fit logarithmic spiral as red curves. The grayscale darkness for the galaxy image is proportional to the logarithm of the intensity. If $\theta=0$ is pointed to the west and $\theta$ increases counter-clockwise, then the fit is
\begin{equation}
\theta_{\rm peak}=``{\rm Azimuthal\;Angle}"=C + 385.4 \log_{10}({\rm Radius})
\label{eq:pitch1}
\end{equation}
for angle in degrees. For a radius in pc, $C=-1073.9$  assuming a distance of 9.84 Mpc, and for a radius in arcmin, $C=258.5$. This equation corresponds to the spiral arc in the lower left part of the galaxy, which goes from $130.9^\circ$ to $258.6^\circ$; for the northwest arc, add $180^\circ$ to $\theta_{\rm peak}$.  Figure \ref{fig:crest}-middle shows the spiral peaks and fit on $\theta-\log_{10}({\rm Radius})$ coordinates. The corresponding pitch angle is $\arctan(d \ln({\rm Radius})/d\theta_{\rm radians})=18.9^\circ$. 

\cite{honig15} determined the pitch angle of the NGC 628 spiral arm as outlined by HII regions. Their result was $27.6\pm0.6^\circ$ for the southern inner spiral at 2.56 kpc mean radius and $15.2^\circ$ for the southern outer spiral at 5.75 kpc mean radius. The pitch angle for the inner spiral is significantly larger than ours, which is based on the stellar distribution in the spiral wave itself, suggesting that the inner-disk HII regions get further from the stellar arm crest with increasing radius. 

The right-hand panel of Figure \ref{fig:crest} shows the average intensity of all the azimuthal scans (again cropped at mean$+3\times$rms), made by shifting and stacking the 800 azimuthal scans so each has its wave crest at the same position, which is where the plotted azimuthal angle is zero. The large spikes are small bright sources that remain even with the cropping.  The double-sine curve shows the spiral density wave profile over two full cycles for clarity. The difference between the peak and valley wave intensities, which for this $1.5\mu$m image corresponds to stellar surface densities, is 18\%. This is a weak spiral wave, which accounts for the difficulty in observing it directly on the left-hand image. 

\section{Optical Clusters: Evidence for Density Wave Flow}
\label{sect:optical}

Figure \ref{fig:opticalclusters} shows the distribution of optical clusters in three age bins measured in years: $\log_{10} {\rm age (yr)}=6$ to 7, 7 to 8, and 8 to 9. The left-hand panels show the distributions of these clusters on the F150W image, with the log-spiral shown for comparison and the inner and outer circle limits as cyan circles.  The middle panels show their distributions on a $\log_{10}({\rm Radius})-\theta$ plot, duplicated over two $360^\circ$ intervals for visualization purposes. The younger clusters are more confined to the trailing half of spiral wave crest, which corresponds to smaller azimuthal angles.  The older clusters are more randomly distributed.  The cluster density is higher at the top of each $\log_{10}({\rm Radius})-\theta$ plot because the physical lengths of the arcs are longer at larger radius.  The right-hand panels show the distributions of the cluster position angles relative to the peak of the stellar wave. This is the angular lag in degrees from the wave crest, with negative values corresponding to negative azimuthal angles, which are downstream in the spiral arm flow (clockwise in the images). Both arms are included in these right-hand panels, without distinction, so the ordinate values from $-90^\circ$ to $+90^\circ$ are relative to either arm.

\begin{figure*}
\begin{center}
\includegraphics[width=16.cm]{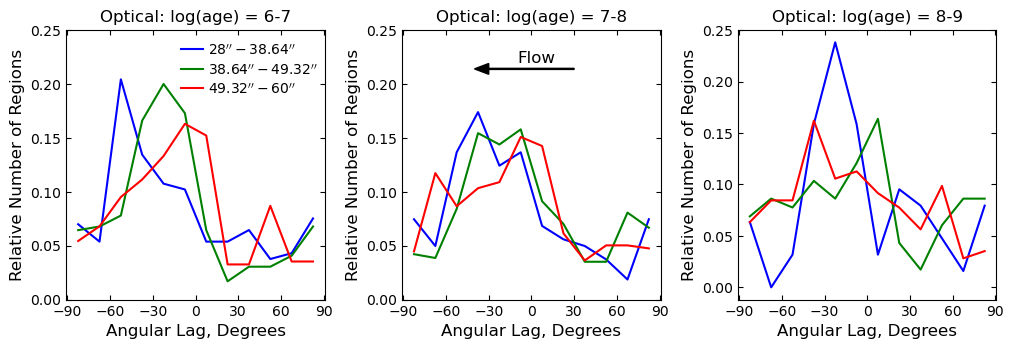} 
\caption{The distributions of azimuthal angle to the nearest spiral wave peak are plotted for optical clusters in three age groups (left to right) and three radial bins, which are the inner (blue curves), middle (green curves) and outer thirds (red curves) of the interval between the radial limits.  For the youngest clusters, the trend of less negative Angular Lag with increasing radius verifies that the clusters are moving through a somewhat fixed, rotating spiral pattern with clusters further from corotation (blue lines) moving further in angle than clusters near corotation (red curves) during their young lives. This pattern disappears for the older clusters. The direction of downstream flow is indicated by the arrow.
}
\label{fig:opticalangles}
\end{center}
\end{figure*}

Normalized distribution functions for the angular lags of the optical clusters are shown in Figure \ref{fig:opticalangles} with young clusters on the left ($\log_{10}{\rm age(yr)}=6$ to 7), middle-age clusters in the center ($\log_{10}{\rm age (yr)}=7$ to 8) and old clusters on the right ($\log_{10}{\rm age (yr)}=8$ to 9). As in Figure \ref{fig:opticalclusters}, negative lag on the abscissa is downstream in the spiral-arm flow. The color corresponds to radius in the galaxy.  Blue is for pixel radius from 700 to 966, green is from 966 to 1233 and red is from 1233 to 1500 (these four distances correspond to $28^{\prime\prime}$, $38.64^{\prime\prime}$, $49.32^{\prime\prime}$ and $60^{\prime\prime}$ or 1335 pc, 1843 pc, 2353 pc and 2862 pc). Figure \ref{fig:opticalangles} shows a positive shift in angular lag with radius for the youngest clusters, and a similar but smaller shift for the middle-age clusters. These shifts result from the faster angular rate of all objects relative to the crest of the stellar wave at smaller radii. The shifts in distance from the crest are more constant with radius for the middle ages, as shown in Figure \ref{fig:opticalclusters}-middle. The older clusters in Figure \ref{fig:opticalangles} have no obvious shift in angular lag with radius. They are also more uniformly distributed in azimuthal angle, as shown by the bottom center panel in Figure \ref{fig:opticalclusters}.

There are two patterns in the right-hand panels of Figure \ref{fig:opticalclusters}.  The largest-scale pattern is the same one shown by Figure \ref{fig:opticalangles} in which large clumps of points systematically shift from negative angular differences such as $-60^\circ$ to near-zero angular differences at larger radii (green line). This is the result of the faster flow through the arm at smaller radii, as just mentioned. The second pattern consists of thin streaks of points that trail about half-way vertically from the upper left to the lower right in the figure, particularly at the youngest ages (yellow line), but also to a lesser extent at middle ages. The direction of these streaks corresponds to increasing spiral arm flow-through position (more negative Delta Angle) for larger radii, which means the associated features have higher pitch angles than the main spirals, corresponding to young clusters in interarm feathers. We will see these feathers more clearly in the F770W image shown in the next section. The feathers form star clusters like the main arms, but the cluster-concentrations at each formation site are smaller. 

\begin{figure*}
\begin{center}
\includegraphics[width=18.cm]{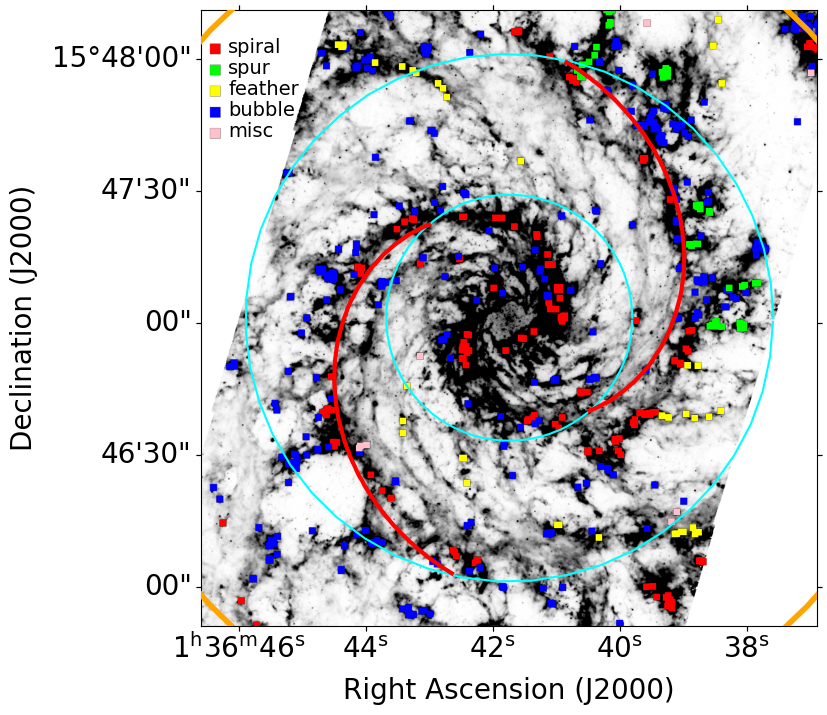}  
\caption{eYSC superposed as squares on the F770W image of NGC 628. The two red curves denote the spiral arm fit and the cyan circles are the inner and outer radii of the fit, as in Figure \ref{fig:crest}. The colors correspond to morphology of the associated $7.7\mu$m emission: blue for clusters in bubble walls, red for clusters in long dust filaments that follow the main arms, green for clusters in prominent spurs and yellow for clusters in faint interarm feathers. Some $7.7\mu$m spurs and feathers have no eYSCs. The F770W image is shown with a linear intensity stretch. }
\label{fig:8mubig}
\end{center}
\end{figure*}

The pitch angle for the feathers may be obtained from Figure \ref{fig:opticalclusters}. They typically cover $\sim90^\circ$ between the two limiting radii in the figure. This is shown by the yellow line. If we use the starting azimuthal angle of the main spiral of $130.9^\circ$ at $\log_{\rm 10} {\rm Radius}(pc)=3.126$ for the low-radius point of a sample interarm feather, and we use the ending azimuthal angle of the main spiral arm of $258.6^\circ$ at $\log_{\rm 10}{\rm Radius}(pc)=3.457$ minus the excursion of an interarm feather in Figure \ref{fig:opticalclusters} over the same radial interval, $\sim90^\circ$, then we can determine the total angular and radial differences between the endpoints of a typical feather, and therefore the pitch angle $i$ from the equation $\tan (i) = \Delta \ln {\rm Radius}/\Delta \theta$.  The result is
\begin{equation}
i = \arctan {{\ln(10)(3.457-3.126)}\over{(\pi/180)\times(258.5-90-130.9)}}=49.3^\circ
\label{eq:feather}
\end{equation}
which is the same as that determined for NGC 628 feathers ($50^\circ$) by \cite{lynds70} based on optical dust features.

\section{Emerging Clusters}
\label{sect:embedded}

Embedded and emerging clusters are in a younger stage than optical clusters, so their positions relative to the spiral arms should be even better evidence of differential gas flow in a stellar density wave. Emerging clusters also highlight the clouds where they are forming, so the associated cloud morphologies and locations might indicate the star formation processes.  Here we consider the 1576 emerging clusters of types eYSC-I or eYSC-II in the NIRCam survey of NGC 628 by \cite{linden26}. As mentioned in Section \ref{sect:data}, emerging clusters of type eYSC-I have compact and spatially coincident $3.3\mu$m PAH emission, while emerging clusters of type  eYSC-II have diffuse or extended $3.3\mu$m PAH emission, which is interpreted as a later stage when the surrounding photo-dissociation region has been partially dispersed. In this section the combined eYSC catalog is superimposed on the F770W JWST image to examine the cloud structures. We consider all 1379 of the eYSC that occur in that image without distinction between the two types, including those outside the $60^{\prime\prime}$ limit used for the optical clusters discussed above.  Any age differences between the two types that result from SED fits \citep{lapeer26} are small compared to the age range of optical clusters, and these eYSC age differences are also not as important as the fact that they are still with their primordial clouds while the optical clusters are generally not.

Figure \ref{fig:8mubig} shows the inner part of the galaxy at $7.7\mu$m with a red curve for the spiral arm fit and the two cyan circles at $28^{\prime\prime}$ and $60^{\prime\prime}$, as above. Figures \ref{fig:8mu_2x1} and \ref{fig:8mu_2x1_feathers} in the appendix show enlargements of the inner part and Figures \ref{fig:8mu_top} and \ref{fig:8mu_bottom} in the appendix show the northern and southern regions beyond corotation, where there is an orange circle at $94^{\prime\prime}$ (4.5 kpc).  The radii in pixels for these circles are half what they were for the F150W image, as the pixels for F770W are $0.08^{\prime\prime}$. In the figures, the eYSC are indicated by colored symbols corresponding to their suggested locations in spiral arm filaments, spiral arm spurs, bubbles, and interarm feathers. 
The definitions of these locations are based on the gas features, as described next.

\subsection{Spiral Arm Filaments, Spurs, Bubbles and Interarm Feathers}
\label{sect:eysc_bubbles}

The general features of NGC 628 were summarized in Section \ref{sect:overview}. Here we highlight four main types of features and attempt to associate each eYSC with a particular  type.

We distinguish as {\it spiral arm filaments} the $7.7\mu$m emission that follows the main 2-arm spiral over the entire F770W image. {\it Spurs} extend nearly perpendicular to these arms and are sometimes massive and without curvature. {\it Bubbles} have curvature and may be closed loops.  The largest bubbles can have edges that look like spurs and the smaller spurs can have adjacent structures that look like bubble walls, so these two types are sometimes ambiguous. We tend to identify the most massive and straightest of the features as spurs.  {\it Interarm feathers} are faint thin streaks of $7.7\mu$m emission with pitch angles over twice that of the main arms ($\sim50^\circ$ compared to $18.9^\circ$; Sect. \ref{sect:optical}, \ref{sect:spirals}). Most feathers do not contain cataloged clusters. They differ from spurs in being fainter, longer, and with lower pitch angles, but sometimes this distinction is ambiguous, especially where the feathers meet the convex sides of the arms. Interarm feathers sometimes intersect to make bubble-like structures, and interarm bubbles can become stretched so much that their long sides begin to look like feathers. 

With a general understanding of the origin of all these features, and an understanding of what spiral waves, shear, feedback and star formation can do to them, we can usually disentangle the various processes and make a best guess for the type of region that contains an eYSC. Because of the inevitable ambiguities, we limit our discussion of eYSCs to the most general comparisons, such as the relative masses in each type of $7.7\mu$m feature and the mass distribution functions.  We would like to understand the dynamical origin of particular star-forming clouds but cannot do that yet in all cases. With higher resolution HI and molecular line emission, enabling the detection of gas structures and motions on scales of $\sim10$ pc and few-km s$^{-1}$, there might be a better way to get this dynamics in the future.

\subsection{eYSC Separated into the Four Types of $7.7\mu$m Morphologies}

Figure \ref{fig:8mubig} and Figures \ref{fig:8mu_2x1} to \ref{fig:8mu_bottom} show the F770W images, with coloured symbols to represent eYSC positions relative to morphological features.  There are 703 small blue squares that show eYSC inside the rims of what we identify as bubbles. In most cases, there is a distinct curvature to the embedding gas that gives it the bubble-like appearance. These bubbles are often the same objects as those previously studied for this galaxy (Sect. \ref{sect:bubbles}), but here they were identified independently by one of the authors (B.G.E) using direct visualization.  Additional observations such as H$\alpha$ or young stars were not used to distinguish bubbles with pressure sources because we found that many large bubbles with peripheral star formation in dwarf irregular galaxies are too old to show their pressure sources \citep{elmegreen26}. Also, large cavities can be made entirely by gas self-gravity \citep[e.g.,][]{grisdale17,zhao24}. Many of the smaller bubbles are inside the rims of larger bubbles, as also noted by \cite{watkins23a}, \cite{egorov23} and others.  

The 466 red and 47 yellow squares are eYSC that are either in the main spiral arms (red) or interarm feathers (yellow). These eYSC are all inside $7.7\mu$m peaks that represent dense gas condensations. They are sometimes positioned with a regular spacing. In the main spiral arms, the condensations are so large that each one typically contains several clusters. The interarm feathers have much smaller $7.7\mu$m condensations and typically only one eYSC in each.

The 145 green squares are eYSC inside relatively short and thick spurs that jut out from the main spirals at large pitch angles. These spurs also have large $7.7\mu$m clouds and several clusters in each, like the main-arm condensations at the red squares. Pink squares are for 18 eYSC that could not be classified into one of these types. 

\begin{figure*}
\begin{center}
\includegraphics[width=8.cm]{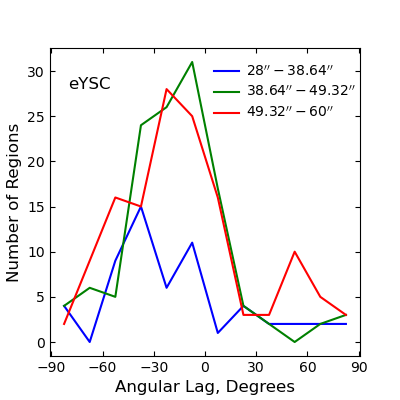} 
\caption{The distribution of Angular Lag between eYSC and the peak of the spiral wave at the same radius, to be compared with the optical clusters in Figure \ref{fig:opticalangles}. The colors correspond to radial ranges inside the two limits: blue to green to red with increasing radius, divided into three equal parts.  eYSC show the same trend as the young optical clusters where those at the smallest radius extend further in angle into the downstream flow (leftward on this plot). 
}
\label{fig:8muangles}
\end{center}
\end{figure*}

Most spurs and interarm feathers stand out from bubbles and main spiral arms.  The spur in Figure \ref{fig:8mu_top} at $1^{\rm h}36^{\rm m}41.1^{\rm s}$, $15^\circ 48^{\prime} 44.9^{\prime\prime}$ is larger and brighter at $7.7\mu$m than most bubble rims, and it also contains a massive collection of 39 eYSC. On the other hand, the smaller spurs on the right of Figure \ref{fig:8mu_2x1} have disturbed gas morphologies all around them, some of which could be from bubbles, but we call these spurs anyway because there is no indication that the spur features themselves have curvature or form part of a loop. Figure \ref{fig:8mu_top} has a clear interarm feather connecting the two spiral arms, and this differs from the spurs in the same figure because of the longer length and thinner structure for the feather. 

The difference in cloud morphology for the eYSC in bubbles versus the spiral filaments and spurs, combined with their different positions relative to the spiral density wave, illustrates two distinct star formation processes \citep[e.g., see simulations in][]{pillsworth25}: stellar spiral waves inside corotation slow down and compress the entering gas into dense filaments that collapse into primary star clusters (red squares), and feedback from slightly older primary clusters and their associated young stars makes bubbles several Myr later, which also collapse to make clusters (blue squares), perhaps after colliding with other bubbles \citep[e.g.,][]{chernin95,zhao24}. Spiral arm spurs and feathers are like main-arm filaments in being primary sites of star formation, i.e., directly related to gas compression in the complex flow of a density wave.  Similarly, the spurs can have their own secondary star formation, as shown on the top of Figure \ref{fig:8mu_top} where blue squares representing eYSC in bubbles surround green squares representing eYSC in the spur. Some bubbles could also be cavities produced by escaping pressures from eYSCs that are embedded in the bubble walls, as in the blister model \citep{israel78}.

Because all of these environments produce stars, they all contain extremely young sources, so if we consider only the eYSC, there are no age gradients. The older generations of stars that made the bubbles are dispersed associations and optical clusters inside and around the bubbles.

The spiral arms delineated by the eYSC they contain, as denoted by red squares in Figure \ref{fig:8mubig}, have slightly smaller pitch angles than the underlying old stellar waves, as delineated by the red curves in that figure. At the bottom of Figure \ref{fig:8mubig} where the red spiral curve ends, there are several red-square eYSC at a distance of $54.1^{\prime\prime}$ from the center of the galaxy, just inside the cyan circle which is at $60^{\prime\prime}$. This smaller distance corresponds to $\log_{10} {\rm Radius (pc)}=3.412$. The azimuthal angle there is essentially the same as at the end of the red spiral, which is $258.5^\circ$.  At the top of the eastern red spiral in Figure \ref{fig:8mubig}, close to where it meets the inner cyan circle at 
$28^{\prime\prime}$, there is another grouping of red-square eYSC at a slightly larger distance of $32.3^{\prime\prime}$, corresponding to $\log_{10} {\rm Radius (pc)}=3.187$. The azimuthal angle there is $130.9^\circ$. These limits correspond to a pitch angle, $i$, for a logarithmic spiral defined by the associated eYSC, 
\begin{equation}
i=\arctan{{\ln(10)(3.412-3.187)}\over{(\pi/180)\times(258.5-130.9)}}=13.1^\circ,
\label{eq:pitch2}
\end{equation}
which is less than the pitch angle of $18.9^\circ$ for the density wave in old stars from equation (\ref{eq:pitch1}).

\subsection{Distribution of Relative Azimuthal Angles}
\label{sect:angle}

\begin{figure*}
\begin{center}
\includegraphics[width=14.cm]{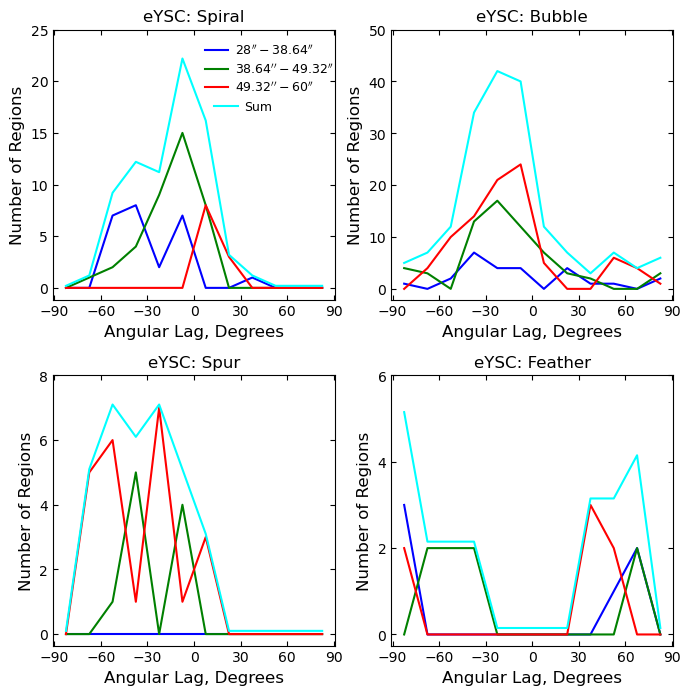} 
\caption{The distribution of Angular Lag between eYSC and the peak of the spiral wave at the same radius for the four categories of clusters depending on the morphology of their associated clouds.  The colors correspond to radial ranges inside the two limits: blue to green to red with increasing radius, divided into three equal parts.  Cyan curves are the sum of the others. The spiral eYSC tend to lag downstream of the density wave peak by $\sim30^\circ$ for the two smaller radial intervals but they are slightly ahead of the peak at large radius (red curve), which means their pitch angle is smaller than that of the stellar spiral. The bubbles are typically downstream of the wave peak, the spurs are consistently downstream, and the feathers are ahead of or behind the spiral peak by $30^\circ$ to $60^\circ$. The downstream direction is to the left on these plots.
}
\label{fig:8m_2x2}
\end{center}
\end{figure*}

The distribution of azimuthal angles relative to the peak of the stellar spiral wave is shown in Figure \ref{fig:8muangles} for all the eYSCs inside the $60^{\prime\prime}$ radius.  The same three radial intervals are used as in Figure \ref{fig:opticalangles} for the optical clusters,  i.e., $28^{\prime\prime}-38.64^{\prime\prime}$ for the blue curve, $38.64^{\prime\prime}-49.32^{\prime\prime}$ for the green curve, and $49.32^{\prime\prime}-60^{\prime\prime}$ for the red curve. In this case, the total numbers of clusters are plotted, rather than the relative numbers. There is a slight shift toward more negative angular lags at smaller radii, as before, but not as much as for the youngest optical clusters. The red peak at positive lag is from the feathers.

Figure \ref{fig:8m_2x2} shows the lag distributions in the same radial intervals, now divided into the four F770W morphology classes. Cyan curves give the total counts; note the vertical scales differ for the four types, with bubbles having the largest counts. All of these counts are much smaller than the total numbers of eYSC because these are only the clusters inside the $60^{\prime\prime}$ radius. Most of the spurs and feathers are outside of that.  The spiral eYSC span a $\sim90^\circ$ range around the peak of the stellar spiral, with a trend to lag downstream at the two smaller radial intervals and lead ahead of the peak at large radius. This pattern corresponds to a smaller pitch angle for the spiral arm eYSC than stellar arms. The bubbles are typically downstream of the wave by $\sim10^\circ$ to $\sim40^\circ$, the spurs are consistently downstream, and the feathers are ahead of or behind the spiral peak by $30^\circ$ to $60^\circ$. 

The spiral phase distributions for eYSC are similar to those for optical clusters in Figure \ref{fig:opticalangles}, where there was also a phase difference with galactocentric radius. For eYSC the phase difference results from a difference in morphology of the primordial clouds, since eYSC are still associated with those clouds; the bubble regions are mostly downstream from the spiral arm filaments, and the interarm feathers lead the arms in azimuth. With optical clusters, the phase difference illustrates their relative position in the arms without revealing the reason. Now, the eYSC in Figures \ref{fig:8muangles} and \ref{fig:8m_2x2} show there is continuous downstream triggering in bubbles and spurs, as well as a far-downstream, or effectively upstream, triggering in feathers. 

The eYSC start to show a different distribution outside the $60^{\prime\prime}$ radius, which is the second cyan circle from the top in Figure \ref{fig:8mu_top} and the top cyan circle in Figure \ref{fig:8mu_bottom}.  Recall that the orange circle is supposed to be the corotation radius. The stellar spiral cannot be traced with the F150W image outside $\sim60^{\prime\prime}$, so between the outer two cyan circles we do not know as precisely where the density wave is, although the optical and $7.7\mu$m spirals are clear. In Figure \ref{fig:8mu_top} and the lower part of Figure \ref{fig:8mu_bottom} there are prominent spurs shown by clusterings of green squares. There are also bright $7.7\mu$m feathers with and without eYSC, shown by yellow squares. Large bubbles, including half of the phantom bubble \citep{mayya23}, are common outside $60^{\prime\prime}$. 

An interesting feature of these outer regions is that the outer arms, i.e., at the top of Figure \ref{fig:8mu_top} and the bottom of Figure \ref{fig:8mu_bottom}, have their bubble-related clusters surrounding the spiral arm clusters without any clear downstream trend, unlike the inner arms. This is consistent with these outer arms being close to corotation, or being material arms. The bright spur at the top of Figure \ref{fig:8mu_top} has its bubbles clearly to the right of it, which suggests a clockwise flow of gas through the spur as in the inner galaxy. So perhaps this spur is still inside corotation or corotation is not precisely defined. The spur at the bottom of Figure \ref{fig:8mu_bottom} is smaller than the two spurs in the middle and top of Figure \ref{fig:8mu_top}.

\subsection{Mass Distribution Functions}
\label{sect:cmf}

\begin{figure*}
\begin{center}
\includegraphics[width=8.cm]{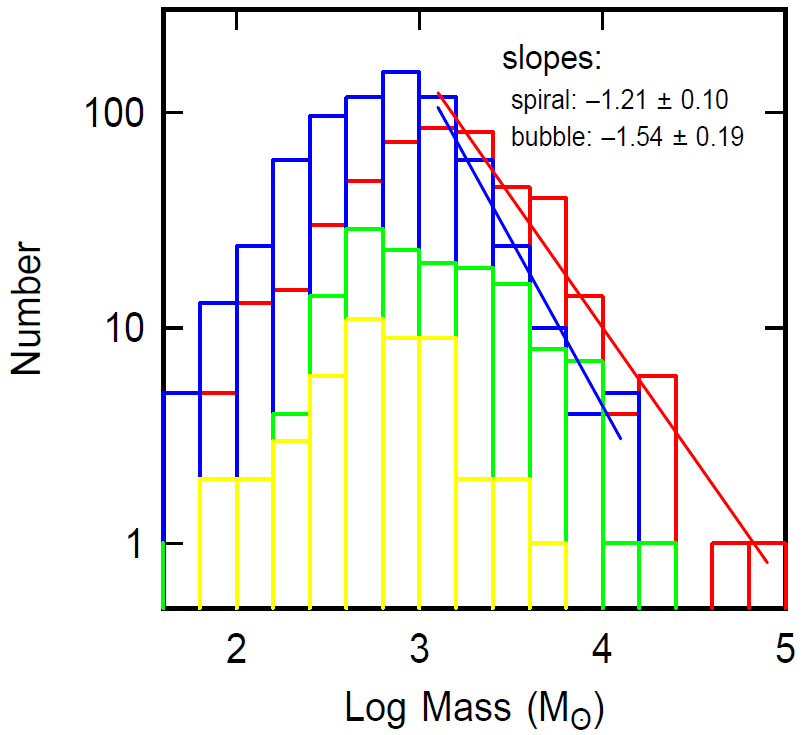} 
\includegraphics[width=7.5cm]{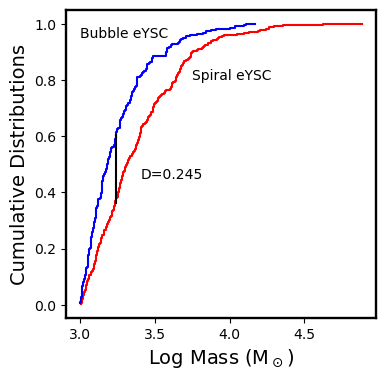} 
\caption{Mass distribution for eYSCs separated into the four types: red for eYSCs in spiral arm filaments, blue in bubbles, green in spurs and yellow in interarm feathers. The detection limit is about $10^3\;M_\odot$, below which there are many missing clusters. The slopes fit to the histogram midpoints are shown. The maximum mass of spiral arm eYSCs is higher than all the others by at least a factor of 3. }
\label{fig:8m_mass}
\end{center}
\end{figure*}

Figure \ref{fig:8m_mass}(left) presents the mass distributions for all of the eYSCs throughout the F770W image, with color code as in the galaxy figures: red, green and yellow for eYSCs in spiral filaments, spurs, and feathers, respectively, and blue for eYSCs in bubble walls.  We do not confine ourselves to be inside the $60^{\prime\prime}$ circle here, as we do not need azimuthal information to associate the eYSC with their $7.7\mu$m cloud morphologies. 

The figure shows that the spiral arm eYSCs extend to higher masses by a factor of $\sim5$ than the bubble or spur eYSCs, while the most massive eYSCs in the feathers are lower in mass than the most massive spiral arm eYSCs by a factor of $\sim17$. The slopes for the spiral arm and bubble populations on these $\log-\log$ plots are $-1.21\pm0.10$ and $-1.54\pm0.19$ with R-squared values of 0.953 and 0.944, respectively.  Clearly the survey contains more massive eYSC in the spiral sample than in the bubble sample. A previous study by \cite{linden26} with the same cluster list also found that eYSC inside the spiral arms delineated by \cite{querejeta21} have more massive members than the eYSC outside these arms. The mass function of the optical clusters in NGC 628 has a similar slope above $10^3\;M_\odot$, $-1.32^{+0.13}_{-0.16}$, with no evident truncation at high mass \citep{tang26}.

A Kolmogorov-Smirnov (KS) test was used to compare the spiral arm and bubble eYSC mass distributions using both the binned data and the actual mass values for $\log_{10}M(M_\odot)>3.0$ where they are best sampled. The cluster numbers are 277 and 221, respectively.  For the binned distributions shown on the left in Figure \ref{fig:8m_mass}, the supremum of the two normalized cumulative distribution functions is 0.33. The resulting high value of $p=0.73$ means that the null hypothesis (that the spiral arm and bubble cluster populations come from the same intrinsic mass distribution function) cannot be rejected with certainty, i.e., the bubbles might not be able to produce the same proportion of massive clusters as the spirals, but there are not enough mass bins to be sure about this. On the other hand, for the unbinned masses with cumulative distribution functions on the right of Figure \ref{fig:8m_mass}, the $p$ value is $5.5\times10^{-7}$, which rejects the null hypothesis: these mass functions are intrinsically different. However, the figure shows that the vertical difference is a maximum at $\log_{10} M(M_\odot)\sim3.24$, which misses the most interesting difference between these two functions at $\log{10} M(M_\odot)>4.2$. 

To consider these cluster mass functions in another way, we directly evaluated the probability that a spiral-like mass function with a logarithmic slope $d\log N/d\log M=-1.21$ and the observed number of bubble eYSC with $\log_{10} M(M_\odot)>3$, namely 221, has no clusters above $\log_{10} M(M_\odot)=4.2$ as in Figure \ref{fig:8m_mass}. We randomly generated distribution functions $N(M)dM\propto M^{-2.21}dM$ for 221 clusters with $\log_{10}M(M_\odot)$ between 3 and 5 and discarded any result with a cluster mass greater than $\log_{10} M(M_\odot)=4.2$. Typically this took several thousand trials to get one example with no $\log_{10} M(M_\odot)>4.2$. We did this process $1000$ times to get an average number of trials before the first one appeared with no $\log_{10} M(M_\odot)>4.2$. This average is 1245, so the probability of this happening is the inverse, 0.08\%.  For a standard cluster mass function with a logarithmic slope of $-1$ \citep[e.g.,][]{krumholz19}, the average number before the first one with no $\log_{10} M(M_\odot)>4.2$ clusters is $2.18\times10^5$ and the inverse is the probability $4.6\times10^{-6}$.  These small probabilities suggest that the limiting mass of $\log_{10} M(M_\odot)=4.2$ for the bubbles is not a size-of-sample effect for this many clusters (i.e., from the statistics of sampling a function out to infinite possible mass), but a physical effect where the bubble walls do not provide the right environment for the most massive clusters in NGC 628 to form, but the spiral arms do.  The two KS tests were not as conclusive as this random sampling test because either the binned cluster numbers are too small or the main recognized differences occurred at too low a mass. A typical deviation between two cumulative functions that are the same is $\sim1/\sqrt(N_{red})$ where $N_{red}=(N_s\times N_b)/(N_s+N_b)$ is the reduced number for spiral and bubble numbers $N_s=277$ and $N_b=221$. For us, $N_{red}=123$ and the typical deviation is 9\%, which is too large to give a near-certain result on the function similarities or dissimilarities, as stated in the previous paragraph. 

Another possibility for the different mass functions for eYSC in bubbles and spiral arm filaments is that clusters with $\log M(M_\odot) > 4.2$  emerge from primordial bubble clouds faster than they emerge from primordial spiral arm clouds. Then young clusters would be more likely to be put in the optical category if they formed in bubbles, removing them from the embedded category. This possibility needs further testing with more precise ages for all of the youngest clusters.

\begin{center}
\begin{deluxetable}{lccccc}
\tablenum{1} \tablecolumns{6} \tablewidth{315pt} \tablecaption{Cluster Formation Sites \label{tab:sites}} 
\tablehead{
\colhead{} &
\colhead{Number} &
\colhead{Number Fraction} &
\colhead{$\log_{10}$ Mass} & 
\colhead{Mass Fraction}&
\colhead{$\log_{10}$ Max. Mass} \\
\colhead{} &
\colhead{} & 
\colhead{} & 
\colhead{($M_\odot$)} & 
\colhead{} &
\colhead{($M_\odot$)} 
}
\startdata
Spiral Arms & 466  & 0.34 & 6.04 & 0.51 & 4.89 \\
Bubbles     & 703  & 0.51 & 5.85 & 0.33 & 4.17 \\
Spurs       & 145  & 0.11 & 5.44 & 0.13 & 4.24 \\
Feathers    & 47   & 0.03 & 4.62 & 0.02 & 3.66 \\
Misc        & 18   & 0.01 & 4.31 & 0.01 & 3.96 \\
\enddata
\end{deluxetable}
\end{center}

Table \ref{tab:sites} lists the summed masses of the eYSC in the four categories: $\log_{10} M (M_\odot)=6.04$ for spiral filaments, 5.85 for bubbles, 5.44 for spurs, and 4.62 for interarm feathers. Of the total number of eYSCs, 1576, there are 466, 703, 145 and 47 in these categories, respectively. The remaining 215 were outside the border of the F770W image (197) or had no obvious connection to spiral arms or bubbles (miscellaneous category, with 18 eYSC). With these mass totals, the fractional masses in the 4 main categories are 0.51, 0.33, 0.13 and 0.02, respectively. Cluster formation sites are dominated in mass by the main spiral arm filaments and they are dominated in number by the bubbles, which represent 51\%.  The maximum eYSC masses in these categories are $\log_{10} M(M_\odot)=4.89$, $4.17$, $4.24$, and $3.66$. 

\subsection{Cluster Separations: Four Anomalous Massive Cluster Pairs}
\label{sect:pairs}

\begin{figure*}
\begin{center}
\includegraphics[width=10.cm]{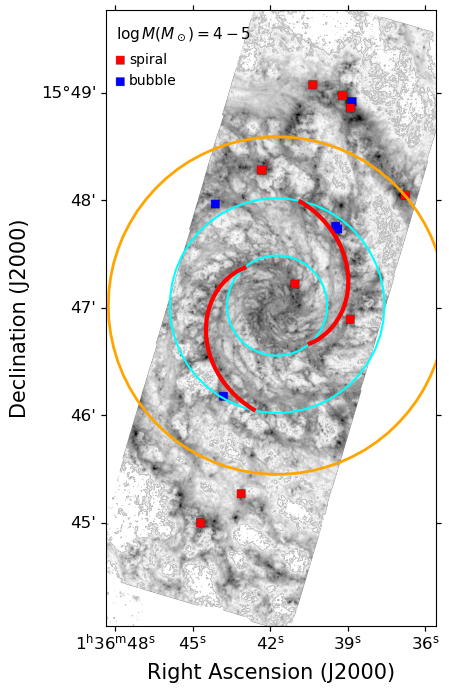}  
\caption{Positions of the 17 eYSCs having $\log_{10} M(M_\odot)=4$ to 5. There are 12 eYSC in the spiral arms (red squares) and 5 in bubbles (blue squares). This image shows only 9 red and 4 blue squares, however, because 6 of the spiral arm clusters and 2 of the bubble clusters are in very close pairs that cannot be distinguished here.  The F770W image has a logarithmic intensity stretch.} 
\label{fig:massiveclusters}
\end{center}
\end{figure*}

\cite{lapeer26} determined the two-point correlation function for eYSC in NGC 628 using the same catalog as we use here. They found a power-law correlation for scales between 10 pc and $\sim1000$ pc, which indicates that eYSC space densities are higher near a typical eYSC than the average space density on larger scales. The number of massive eYSC with $\log_{10} M (M_\odot) >4$ is so low, however, that a two-point correlation function for just these clusters is not statistically meaningful. To see where these clusters formed, we show their positions in Figure \ref{fig:massiveclusters}.  As for the other figures, the spiral arm eYSC are red squares and the bubble eYSC are blue squares. 

\begin{figure*}
\begin{center}
\includegraphics[width=18.cm]{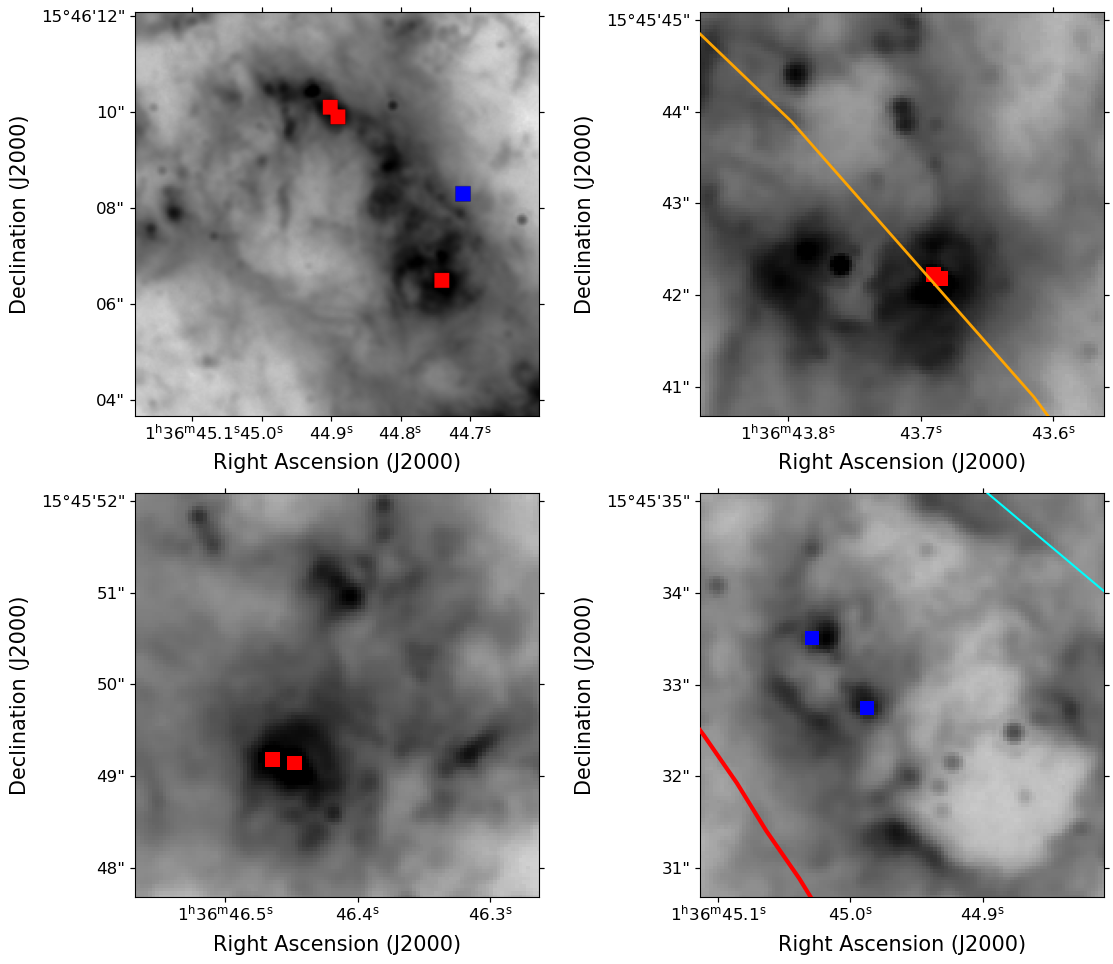}  
\caption{Enlargements of the regions around the 4 cluster pairs with red squares for spiral arm locations and blue squares for bubbles. These panels are 801 pc (top left) and 401 pc on a side.  The eYSC pairs in Table \ref{tab:massive} are in order from top left to top right to lower left to lower right.  The F770W image has a logarithmic intensity stretch.
}
\label{fig:massiveblowup}
\end{center}
\end{figure*}

There are 12 spiral arm eYSC and 5 bubble eYSC with $\log_{10} M(M_\odot)>4$, but Figure \ref{fig:massiveclusters} shows only 9 red squares and 4 blue squares. This is because 3 of the massive spiral arm eYSC and 1 of the massive bubble eYSC are so close to another massive eYSC that they cannot be distinguished individually in this way. Clearly this is not part of a smooth hierarchy among the few massive eYSC, which are generally spread out over the whole disk. Rather, it is something bimodal in that some massive eYSC are single, and probably part of the two-point correlation hierarchy with all the other eYSC, and other massive eYSC are binary with unusually small separations. Binary clusters have also been observed in the Milky Way \citep[e.g.,][]{subramaniam95,vereshchagin22,chen26} and Large Magellanic Cloud \citep[e.g.,][]{bhatia90,priyat19}. Models of their formation are in \cite{darma21}. 

Figure \ref{fig:massiveblowup} shows the binary eYSC on larger scales in maps that are 210 pixels (top-left) and 110 pixels wide. For $0.08^{\prime\prime}$ pixels on the F770W image, these sides measure 801 pc and 401 pc. Going clockwise from the top-left in the figure, the projected cluster separations are $0.5216^{\prime\prime}$, $0.1579^{\prime\prime}$, $1.93700^{\prime\prime}$, and $0.48586^{\prime\prime}$, which correspond to 24.9 pc, 7.53 pc, 92.4 pc, and 23.2 pc at the distance of 9.84 Mpc. These and other numbers related to the eYSC pairs are in Table \ref{tab:massive}. 

The two closest massive eYSC, in the top right panel of Figure \ref{fig:massiveblowup}, have masses of $M_{\rm 1}=1.36\times10^4\;M_\odot$ and $M_{\rm 2}=1.55\times10^4\;M_\odot$. If their true separation is the projected value, 7.53 pc, and they are on circular orbits around each other, then the nominal orbital rate is $(GM_{\rm tot}/a^3)^{0.5}=1.57 $ rad Myr$^{-1}$ for total mass $M_{\rm tot}=M_{\rm 1}+M_{\rm 2}$ and half-separation $a=3.77$ pc.  The actual separation of these clusters is probably larger than their projected separation if they have different positions on the line of sight. If one cluster is in the midplane and the other is somewhere with equal probability along the line of sight between the midplane and the disk half thickness, $H$, then the expected ratio of the true half-separation $a_{\rm true}$ to the observed half-separation $a$ is 
\begin{equation}
{{a_{\rm true}}\over{a}}={1\over\theta_{\rm max}}\int_0^{\theta_{\rm max}}{{d\theta}\over{\cos\theta}}={{0.5}\over{\arctan(H/a)}}\ln{{(H^2+a^2)^{0.5}+H}\over{{(H^2+a^2)^{0.5}-H}}},
\end{equation}
which equals 1.45 for $H/a=3$ and 1.68 for $H/a=5$. This change would decrease the orbital rate by factors of 0.57 and 0.46. The SED-fitted ages of these two clusters are 4 and 5 Myr, so the average age multiplied by the nominal orbital rate is 7.1 radians, with a value that is perhaps half of that after correcting for projection effects. In any case, the orbital rate is comparable to various other dynamical interaction rates, and it is high enough that the eYSC could have interacted significantly in their $\sim 4.5$ Myr lifetimes. They could even be in the process of coalescing. 

\begin{figure*}
\begin{center}
\includegraphics[width=10.cm]{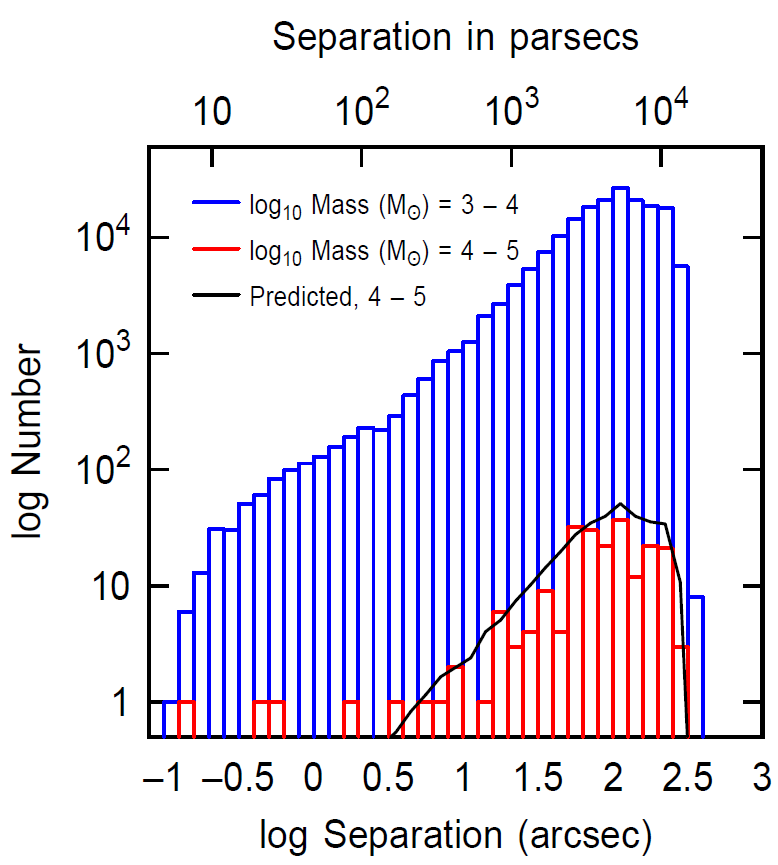}  
\caption{Separation distributions for clusters in the mass range $\log_{10} M(M_\odot)=3$ to 4 (blue) and $\log_{10} M(M_\odot)=4$ to 5 (red). The black curve is 1/525 times the blue histogram, showing the expected distribution for the higher mass clusters if they follow the same pattern as the low mass clusters. The four red peaks on the left are the close cluster pairs. The expected distribution, extrapolated to below 1 count, is used to estimate the likelihood of the close cluster pairs. }
\label{fig:separations}
\end{center}
\end{figure*}

The fiducial orbital rate for the eYSC pair in the top left of Figure \ref{fig:massiveblowup} is 0.31 rad Myr$^{-1}$, which, when multiplied by their average age becomes 1.6 radians, significantly smaller than for the pair in the top right. Similarly, for the pair in the bottom left the rate is 0.51 rad Myr$^{-1}$ and the rate times the average age is 1.4 radians. The $\sim1.5$ radian excursions are small enough for these pairs to avoid significant physical encounters before they break out from their cloud cores. For the bubble eYSC pair in the lower right of the figure, the orbital rate is 0.035 rad Myr$^{-1}$ and their estimated excursion around each other is only 0.1 radian.

The appearance of these four tight eYSC pairs is statistically unlikely considering how the other massive clusters form all over the galaxy disk; three of the four pairs are much too closely spaced to have happened by chance. This is true even though the two-point correlation function for all eYSC masses \citep{lapeer26} shows significant dense eYSC groupings on the $\sim10$ pc scale with a gradually decreasing density on larger scales.   We demonstrate how unusual the eYSC pairs are by plotting in Figure \ref{fig:separations} the distributions of the separations between all pairs of eYSC (regardless of formation environment) in the mass ranges of $\log_{10} M(M_\odot)=3$ to 4 (blue histogram) and $\log_{10} M(M_\odot)=4$ to 5 (red histogram). Analogous distribution functions for cluster pair separations in the LMC, without consideration of cluster mass as in Figure \ref{fig:separations}, are in \cite{dieball02} and \cite{priyat19}. The binary eYSC in the high mass range stand out as peaks at small separation in the red histogram.  There are no similar low-separation peaks among the lower-mass group. Aside from these low-separation peaks, the red histogram looks like the blue histogram, but shifted down in counts because of the smaller number of high mass eYSC. The predicted shift downward is shown by the black curve. This shift was determined from the square of the ratio of the number of high mass eYSC ($10^4-10^5\;M_\odot$) to the number of low mass eYSC ($10^3-10^4\;M_\odot$), and that ratio comes from the eYSC mass function, which has fairly good statistical accuracy. Weighting the two mass function slopes for bubbles and spiral arms by the number of eYSC in each bin gives an average slope in the mass range from $\log_{10}M(M_\odot)=$ 3 to 5 of $(1.21\times277+1.54\times221)/(277+221)=1.36$. Thus the relative number at $\log_{10}M(M_\odot)=$ 4 to 5 eYSC compared $\log_{10} M(M_\odot)=$ 3 to 4 eYSC is $10^{-1.36}=22.9$, and the square of this is 525. The black curve in Figure \ref{fig:separations} is $1/525$ times the blue curve. This scaled mass function agrees well with the observed function for the high-mass eYSC, showing that they have the same type of distribution on scales larger than $\sim1$ arcsec (47 pc) and differ only because there are fewer high mass eYSC. These distributions increase with separation because the circumference of the annulus within each separation interval increases at first, and then the distributions decrease at large separations because the edge of the image starts to matter.

At small separations, the red peaks in Figure \ref{fig:separations} stand out as peculiar. We consider $1/525$ times the number of low mass eYSC to be the average number of high mass eYSC expected in the same separation interval. This would give the continuation of the black curve to values less than 1 cluster pair.  For the massive eYSC pair in the red histogram at log-separation range $-0.9$ to $-0.8$, the number of eYSC in the lower mass interval (blue histogram) is 6, in which case the expected number of massive pairs in this range is $6/525=0.0114$.  We consider this to be the probability of one massive pair in that separation range, simply by Poisson statistics (i.e., the probably of one object in a bin, $\lambda^1\exp(-\lambda)$, is the average number in all of the bins, $\lambda$, when this average is much less than 1, because then $\exp(-\lambda)\sim1$).  Similarly, for the separation ranges $-0.4$ to $-0.3$, $-0.3$ to $-0.2$, and $0.2$ to $0.3$, where there are also massive pairs, the expected probabilities are 0.114, 0.16, and 0.36.  The first two are fairly low probabilities, suggesting special circumstances to make these eYSC pairs. 

The probability that there are four eYSC pairs like we see among the most massive cluster group ($\log_{10} M(M_\odot)=4$ to 5) all closer than $0.5^{\prime\prime}$, which is where the red histogram dips below 1 count, is not just the product of these 4 particular probabilities because we would claim as peculiar any four pair separations in this separation range, not just the particular ones we see. To calculate the probability there are four such pairs, we took the average value of the products of the individual probabilities of all possible combinations of 4 eYSC pairs within the 15 considered separation intervals below $0.5^{\prime\prime}$. This average is 0.00104, which we consider to be the probability of seeing these 4 eYSC pairs or other similar 4 eYSC pairs closer than $0.5^{\prime\prime}$ among the most massive group. 

There is evidently a specific mechanism that forms close cluster pairs in regions where massive clusters form \citep[e.g.,][]{darma21}.  Considering the discussion above, some of these eYSC pairs may be so close that they coalesce or significantly distort each other with tidal forces. This seems to be a property of the massive eYSC and not their environments, as both spiral arm and bubble eYSC have examples of close pairs. Massive cluster coalescence has been illustrated in some dynamical simulations \citep[e.g.,][]{lahen19}.

Figure \ref{fig:separations} shows a much larger number of low-mass projected eYSC pairs than high-mass eYSC pairs, down to the resolution limit of $\sim0.1^{\prime\prime}$. These low-mass eYSC may interact with each other and with even lower-mass clusters. Considering the orbital rate equation given above, their dynamical times will be larger by the inverse square root of mass, which is only a factor of $\sim3$. Interactions between low- and high-mass clusters should happen too, with essentially the same dynamical time as for high-mass pairs. However, only $\sim0.27$\% of the $\log M(M_\odot)=3$ to 4 eYSC cluster pairs have projected separations less than $1^{\prime\prime}$ (48 pc) while $\sim5$ times as many, 1.4\%, of the $\log M(M_\odot)=4$ to 5 cluster pairs are within this separation. This comparison suggests again that the massive eYSC pairs are unusually close compared to the low-mass eYSC pairs. 

The eYSC also have nearby optical clusters that are young and massive. Some could have formed in the same cloud cores. \cite{buckner26} compared optically identified clusters with eYSC and found that 136 objects listed in both catalogs are probably the same cluster, based on their positions being closer than $0.16^{\prime\prime}$. If we discount these, then the average number of optically identified clusters with ages less than 5 Myr and masses larger than $10^3\;M_\odot$ that are within $0.63^{\prime\prime}$ of an eYSC (but further away than $0.16^{\prime\prime}$) in the same mass range is 0.18, with a maximum number of such neighbors equal to 5. This $0.63^{\prime\prime}$ is the maximum separation of the three closest massive clusters in figure \ref{fig:separations}, i.e., $-0.2$ on the histogram. The average number of optical clusters with $\log M(M_\odot)>4$ that are within this distance of an eYSC with $\log M(M_\odot)>4$ is 0.0032 (i.e., 5 eYSC), with no massive eYSC having more than 1 massive optical cluster.  Only one of the massive pairs in Table \ref{tab:massive}, the second one, has equally massive optical clusters at an equally close distance. The log masses of these optical clusters in $M_\odot$ are 4.11 and 4.15, their ages are 3.0 Myr and 2.7 Myr, and their minimum separations from one of the stars in the eYSC pair are 0.52 and 0.31 arcsec. 

\begin{center}
\begin{deluxetable}{llcccccc}
\tablenum{2} \tablecolumns{8} \tablewidth{315pt} \tablecaption{Massive Cluster Pairs
\label{tab:massive}} 
\tablehead{
\colhead{ID} &
\colhead{RA} & 
\colhead{DEC} & 
\colhead{$\log_{10}$Mass} & 
\colhead{Age\tablenotemark{a} } & 
\colhead{Separation} &
\colhead{Orbital rate} & 
\colhead{Age$\times$Orbital Rate} \\
\colhead{} &
\colhead{(deg)} & 
\colhead{(deg)} & 
\colhead{($M_\odot$)} & 
\colhead{(Myr)} &
\colhead{($^{\prime\prime}$/pc)}&
\colhead{(rad Myr$^{-1}$)} & 
\colhead{}
}
\startdata
1. & 24.163359909271907 & 15.816244445522209 & $4.36\pm0.02$ & $5.0\pm1.0$  & $0.52^{\prime\prime}$ & 0.31 & 1.59 \\
   & 24.163464459956451 & 15.816348729518175 & $4.26\pm0.10$ & $5.2\pm1.0$ & 24.9 pc &      &      \\
2. & 24.153373728815318 & 15.800859314509015 & $4.13\pm0.02$ & $4.0\pm1.0$ & $0.16^{\prime\prime}$ & 1.57 & 7.05 \\ 
   & 24.153333796442620 & 15.800838161037220 & $4.19\pm0.02$ & $5.0\pm1.0$ & 7.53 pc &      &      \\
3. & 24.176478640611364 & 15.804734920655388 & $4.89\pm0.03$ & $4.2\pm1.0$ & $0.49^{\prime\prime}$ & 0.51 & 1.43 \\
   & 24.176341238852032 & 15.804707792445788 & $4.05\pm0.05$ & $1.5\pm1.0$ & 23.2 pc &      &      \\
4. & 24.164169242422613 & 15.795581264374906 & $4.10\pm0.02$ & $6.0\pm1.0$ & $1.94^{\prime\prime}$ & 0.035 & 0.21 \\
   & 24.164513726356500 & 15.796005091850327 & $4.17\pm0.02$ & $6.0\pm1.0$ & 92.4 pc &       &      \\ 
\enddata
\tablenotetext{a}{Error values for age from the bayes.sfh.age\_err keyword in the results output of CIGALE are all less than 1 Myr for these clusters, but we take instead an ``error floor'' of 1 Myr that is the age resolution in our model grid.}
\end{deluxetable}
\end{center}

\section{Discussion}
\label{sect:discussion}

Observations at $1.5\mu$m of the spiral density wave in NGC 628, and observations at $7.7\mu$m of the warm interstellar dust, combined with catalogs of the optical and emerging young stellar clusters, reveal the kinematics of gas flow through the arms and the triggering mechanisms for star formation. This section summarizes these observations in a larger context.

\subsection{Spiral Waves and Star Formation}

Several observations discussed in the previous sections support the classical picture where gas entering a spiral density wave arm gets compressed and collapses gravitationally into massive and dense cloud complexes \citep[e.g.,][]{roberts69,elmegreen79}. These observations include the greater angular displacement downstream further from corotation for both optical and emerging clusters (Figs. \ref{fig:opticalangles} and \ref{fig:8muangles}). Another confirmation is the presence of massive spurs and low-density interarm feathers (Figs \ref{fig:opticalclusters} and \ref{fig:8mubig}), as predicted for spiral arm gas collapse with expanding and shearing flows downstream \citep{balbus88}. All of these features contain star formation, as observed here by the presence of young clusters. There would be non-clustered star formation too, but the fractional mass of young stars that form outside of clusters may not change much in different environments \citep[e.g.,][]{chandar17,cook23}.   

Further evidence for spiral-arm gas flow and arm-enhanced star formation is shown by the difference between the azimuthal distributions of young and old clusters, comparing the top and bottom panels in Figure \ref{fig:opticalclusters}.  The young and intermediate-age clusters are more concentrated in the arms than the old clusters, which are more uniformly distributed. Old clusters, like old stars, follow the gravitational attraction of the density wave and synch up their epicyclic orbits with the wave pattern. After enough time, old objects with similar in-plane velocity dispersions should have similar arm/interarm contrasts. Since the wave itself has only an 18\% intensity variation (Fig. \ref{fig:crest}) and the $\sim10^{8.5}$ year old clusters should be approaching this small variation as they join the wave potential, their alignment with the spiral would hardly be noticed in Figure \ref{fig:opticalclusters} because of their small numbers. Young clusters have the same kinematics as the gas, which is compressed more by the spiral arm because of its lower velocity dispersion.

The difference between quasi-stationary density waves, where the pattern speed is about constant for several rotations \citep{mark76,saha16}, and transient spiral arms that wrap up either with the inward group velocity \citep{toomre69} or because they are material arms, and the distinction between inside and outside of corotation in the quasi-steady case, could be evident using the cluster-flow method illustrated by Figures \ref{fig:opticalangles} and \ref{fig:8muangles}. Transient spiral arms should not have the differential downstream flows that these figures illustrate, and the flow directions should be opposite outside corotation if the quasi-steady spiral is still present there (and not overwhelmed by material arms, for example). We suggested in Section \ref{sect:angle}, based on Figures \ref{fig:8mu_top} and \ref{fig:8mu_bottom}, that the outer arms in NGC 628 are in fact close to the predicted corotation radius \citep{speights25} on the basis of the more scattered distribution of bubble clusters around spiral arm clusters. 

\subsection{No Age Gradients Because of Secondary Star Formation}

The observations also explain the common lack of age gradients through spiral arms: there is secondary star formation in the rims of bubbles all through the arms, and there is star formation in spurs and interarm feathers, which occur downstream as well. This does not mean there is no evidence for gas flow through the arms, as this flow shows up in a different way, i.e., the greater downstream displacement for equal age clusters that are further from corotation.  \cite{watkins23a} also observed that bubbles got bigger further downstream. Evidence for the conventional age sequence should still be sought in the recent star formation histories of the downstream stellar populations, especially inside feedback-driven bubbles. Stellar populations that are further downstream from the wave crest at a given distance from corotation should have older bursts of star formation associated with the earlier time of wave passage. At corotation, there should be a long history of star formation going back to the time when the associated wave pattern began, which could be several orbit times. These differences in star formation histories might not be evident in the average spiral arm color, which could be dominated by the youngest populations.

\subsection{Energy Input and Turbulence}
\label{sect:turbulence}

What was not present in the classical picture was the proper source for re-energizing the interstellar medium (ISM) after rapid dissipation in the density wave shock. In \cite{roberts69} and other early analytical work, the gas was assumed to be isothermal or adiabatic, which means that decompression in a wave-enforced flow is automatically accompanied by unspecified energy input, regardless of star formation. This type of equation of state leads to a minimum scale for gravitational collapse, i.e., the Jeans length, and including rotation, a maximum Toomre Q value for radial collapse in a disk. The present observations show that energy lost in forming the main spiral arm filament is returned to the ISM in the form of giant bubbles from star formation feedback. Star-formation feedback breaks apart gravitating clouds and starts the gas moving again inside the downstream flow. Because this type of re-energizing comes after star formation, it cannot enter the equation of state before star formation, or be involved with a stability condition before star formation \citep{elmegreen11}.  

We can also see from the juxtaposition of interarm feathers, spiral arm filaments, and bubbles that the feathers, along with debris from old disrupted bubbles, deliver gas to the spiral arm filaments, which then collapse, form stars and break apart into new bubbles as this gas flows downstream. The bubble rims also form stars and break apart, making scattered debris for the next spiral arm.  The compression of interarm gas into spiral arm filaments generates turbulence \citep[e.g.,][]{kim06b}, and the expansion of bubbles generates turbulence \citep[e.g.][]{lau25}. If the fractional gas mass that forms clusters is the same everywhere, then the fractional cluster masses in our main environmental regions correspond to the fractional gas masses in these regions, i.e., 51\% for the main spiral arm filaments and 33\% for the bubbles (Sect. \ref{sect:cmf}). For similar pressures in these two regions, these same proportions would roughly correspond to the proportions of turbulence generated by spiral arms versus bubble expansions. If the pressures in the spiral arm filaments are larger than in the bubbles, as might be the case to explain the larger eYSC masses in the arms (see below), then the fraction of ISM turbulence generated by the density wave could be larger than 51\%.  Considering also spurs as a source of post-shock turbulence, the density wave could contribute more than 51\%+13\%=64\% of the total turbulence having a kinematic origin (as opposed to turbulence from gravitational collapse), with bubble feedback contributing less than 33\%, which is the bubble cluster mass fraction. 

\subsection{Local physical conditions and the products of star formation}
\label{sect:local}

The difference between the spiral arm and bubble cluster mass functions (Sect. \ref{sect:cmf}) suggests that young clusters do not have a universal mass distribution function. Either the CMF slope varies with environment or the upper mass limit does. Also, the upper mass limit for the bubble clusters is apparently not just a size-of-sample limit because there are 5 clusters in the most-massive distribution interval instead of the expected 1 cluster if the power-law were to smoothly run out of clusters \citep{whitmore00,larsen02,billett02}.  This suggests that the high-mass end of the eYSC population is sensitive to local physical conditions.

Pressure could be important for cluster mass \citep{elmegreen97,pillsworth26}.  For example, if cluster mass scales weakly with radius, then more massive clusters have higher internal pressures and may require higher pressures to form.  The internal pressure for cluster mass $M$ and radius $R$ is $P\sim GM^2/R^4$.  In NGC 628, $R\propto M^\alpha$ for optical clusters with $\alpha$ measured between $\sim0.25$ by one method and $\sim0.5$ by another \citep{ryon17}. \cite{brown21} measured $\alpha=0.24$ as an average for 31 galaxies. With $\alpha=0.24$, $P\propto M^{1.04}$ and the factor of $10^{4.89-4.17}=5.2$ between the most massive cluster masses in the spirals and bubbles (Sect. \ref{sect:cmf}) corresponds to a factor of $\sim5.6$ in pressure, with the spiral arm clusters having the greater value.  This pressure factor might be enough to affect the formation of massive clusters, but these clusters are deeply embedded inside giant cloud complexes, and considering all the dynamical processes that are involved with cloud formation, cloud collapse, individual star formation and the mutual interactions between stars in a dense cluster, the outcome of cluster formation based on pressure alone is difficult to predict.

Nevertheless, we observe that feedback-driven bubbles produce secondary star formation in swept-up shells, where the available gas reservoirs, pressures, confinement times, and shell column densities may be insufficient to form the most massive clusters in NGC 628. Giant shells also slow down as they age, decreasing the ram pressure along the rim with time as the clusters form. By contrast, spiral density-wave filaments assemble large coherent gas structures from continuously incoming gas, producing high gas surface densities, large filament line densities and cloud masses, and sustained high ram pressures. The combination of these differences, along with shear differences in the cloud environment \citep{reina17} could raise the maximum mass of an eYSC even if pressure alone does not contribute much.  For example, simulations \citep{koletic26} and observations of other galaxies \citep{zhou25} suggest that more massive clouds make more massive clusters. 

There are other examples where maximum cluster mass depends on environment.  Radial variations in the upper mass limit in M83 have been interpreted as evidence that gas pressure and surface density regulate the cluster mass scale \citep{adamo15}. Also in highly perturbed regions on galactic scales there can be unusually massive clusters.  Radio and infrared observations of embedded clusters in Henize 2-10 showed that the earliest stages of massive cluster formation occur in extremely dense, highly pressurized environments \citep{johnson03,johnson18,costa21}. In the Antennae, ALMA observations of the ``Firecracker'' cloud and related molecular complexes showed that clouds capable of forming massive bound clusters require high masses within small radii and external pressures orders of magnitude above those of typical Galactic molecular clouds \citep{johnson15,finn19,krahm24}. More recently, ALMA-LEGUS comparisons have shown that galaxies and subgalactic regions with similar star formation rates can differ in their molecular cloud properties and their production of massive clusters \citep{finn24a,finn24b}.  Other examples are in the dwarf galaxies NGC 5253 \citep{lopez12,calzetti15} and NGC 1569 \citep{johnson12} where super star clusters have apparently formed at the collision sites of extragalactic inflowing gas streams. These results support a picture in which the formation of the most massive clusters is not always a consequence of forming more clusters overall, but sometimes depends on whether the local ISM can assemble sufficiently massive, dense, pressurized, and short-free-fall-time structures before feedback disperses them.

Another aspect of the eYSC distributions shown by Figures \ref{fig:8mu_2x1}, \ref{fig:8mu_2x1_feathers}, \ref{fig:8mu_top}, and \ref{fig:8mu_bottom} is that both the spiral arm and bubble eYSC are often in aggregates of 5 or more other eYSC in the same cloud complex. Figure \ref{fig:separations} shows this for the $\log_{10} M(M_\odot)=3$ to 4 mass range by the steadily rising distribution of separations closer than $\sim100$ pc and beyond. This 100 pc is the typical size of a massive cloud complex.  For the ten separation ($S$) intervals between $\log_{10} S =-0.5$ and $\log_{10} S=0.5$ in Figure \ref{fig:separations}, the total number of pairs is 1334. If these small-separation counts are dominated by $N_{\rm GMC}$ giant clouds with $C$ clusters in each, then $N_{\rm GMC}C^2/2\sim1334$. Sample solutions are $N_{\rm GMC}\sim106$ for $C=5$ and $N_{\rm GMC}\sim42$ for $C=8$.  When there are several clusters in this mass range for each giant cloud complex (and many smaller clusters too), then most clusters in the whole galaxy are forming inside more-or-less isolated environments that are influenced by turbulence and gravitational fragmentation.  This is the situation usually envisioned for the origin of the $dN/dM\propto M^{-2}$ cluster mass function, i.e., that it comes from what is effectively random sampling in a hierarchically fractal gas \citep{fleck96,elmegreen96,elmegreen97}.  Because the clouds have their own mass function, which has a similar slope and possibly a similar explanation \citep{elmegreen96}, the combined cluster mass function from all of the $N_{\rm GMC}$ clouds with their independent $C$ clusters has the same total power law effectively made by randomly sampling the whole ensemble. In this case the mass function slope would not depend on whether eYSCs are from bubbles or spiral arm filaments, but the limiting values of this function could depend on this difference, e.g., through pressure, accumulated mass, etc. 

eYSC in the interarm feathers have a maximum mass of only $10^{3.66}=4600\;M_\odot$, which is 3.2 times lower than the maximum for the bubble eYSC. The feather eYSC are all denoted by yellow squares in our figures, and the corresponding $7.7\mu$m emission is usually faint with only small clouds around the clusters. This is an extrapolation of the cluster formation process in the other direction, toward lower pressures and slower dynamics, presumably leading to lower maximum cluster masses. The spurs have eYSC masses up to $10^{4.24}\;M_\odot$, which is between the bubble and spiral filament maximum masses, reflecting the large cloud sizes seen in the F770W images around the spurs and possibly higher dynamical pressures in the spurs than the bubbles. 

Maximum cluster masses might also be related to the formation of close binary clusters. In the current sample, only the most massive eYSC ($\log_{10} M(M_\odot)=4$ to 5) have binaries that stand out in a statistically significantly way for their closeness (Fig. \ref{fig:separations}). The closest pair is separated by only 7.53 pc (Table \ref{tab:massive}), which is comparable to the average radii of equally massive clusters elsewhere \citep{brown21} and suggestive of a near-future coalescence to make an even more massive cluster.  Evidently the conditions in dense massive (spiral-arm) molecular cloud complexes are conducive to excessively strong and fast collapse, which leads to massive young stellar concentrations in universally hierarchical patterns, including transient binary systems.

\section{Summary}
\label{sect:sum}
The positions of optical star clusters of various ages have been determined relative to the stellar density wave in NGC 628, and the positions of emerging star clusters (eYSC), which are generally younger than the optical clusters and still mostly obscured, have been determined relative to the density wave and four types of gas dynamical features, as revealed by the F770W JWST image. These dynamical features are spiral arm filaments, spiral arm spurs, interarm feathers and bubbles. 

Evidence for gas flow through the density wave has been found in the presence of a greater angular displacement downstream of the arm crest for optically detected clusters of a given age when they are further from the corotation radius, i.e., at smaller radius. The same trend occurs for eYSC.  This observation differs from the age or color gradients that have long been sought but rarely found as evidence for this flow.  As a result, the optical and emerging clusters have a slightly lower pitch angle ($13.1^\circ$) than the underlying stellar wave ($18.9^\circ$). The oldest optical clusters ($\log_{10} {\rm age}({\rm yrs}) \sim8$ to 9) do not show this differential azimuthal shift, indicating that they have traversed the spiral wave at least once already. 

The positions of eYSC relative to certain gas dynamical features are usually clear, but some bubble pieces that are adjacent to spurs or are stretched so much by shear that they look like interarm feathers make these associations imperfect. Nevertheless, we find agreement with previous studies that show more massive clusters in the spiral arms than elsewhere. With our identification of eYSC in bubbles versus spiral arms, we find a probability of only 0.08\% that the bubbles randomly lack eYSC that are as massive as the most massive eYSC in the spiral arms.  The bubbles seem physically unable to make clusters more massive than $\log M(M_\odot)\sim4.2$, while spiral eYSC have masses up to $\log M(M_\odot)\sim5$. This probability assumes that the high-mass slope of the mass function for bubble eYSC is the same as the high-mass slope for spiral arm eYSC. If they both have the conventional slope of $-1$ on a log-log plot, then the lack of such massive clusters in the bubble population has a probability of only 0.00046\%.

Table \ref{tab:sites} summarized the locations of the eYSC: 51\% of the eYSC mass is in the main spiral arms, 13\% and 2\% is in spurs and feathers indirectly connected with these arms, and 33\% is along the rims of bubbles.  If we consider that the mass fraction of a cloud in the form of eYSC is about constant, and the turbulent energy in the cloud, before self-gravity contributes to this energy, comes from the dynamical processes that formed the cloud, then these mass fractions are also the fractions of pre-collapse turbulence in the dense component of the interstellar medium. If the greater eYSC masses in the spiral arms correspond to greater pressures and turbulent energy densities, then the fraction of turbulent energy from the spiral density wave exceeds these mass fractions. For the three types of features associated with the density wave, this total turbulent energy would then exceed 66\%, with less than 33\% from star-formation feedback. These numbers should be considered representative only for NGC 628, since the strength and nature of spiral arms varies from galaxy to galaxy. For example, \cite{tress20} compared simulations of an interacting galaxy like M51 with an analogous non-interacting galaxy that was flocculent and found that the total star formation rates  and the star formation rates per molecule were similar, but the M51 case had a high fraction of star formation associated with strong spiral arms and the flocculent case had most of the star formation in feedback-driven bubbles. 

The 17 eYSCs more massive than $\log_{10} M(M_\odot)=4.0$ contain 4 close pairs that have no analogy in the $\log_{10} M(M_\odot)=3$ to 4 sample.  The closest pair has a projected separation of 7.53 pc and a true separation likely to be about twice that. Considering their masses and using this separation as an orbital semi-major axis, the corresponding dynamical rate for their interaction is only 1.57 rad Myr$^{-1}$ (half that with de-projection) and the product of this rate and their average age is 7.1 radians (half that with de-projection).  This large product implies the two clusters are strongly interacting, even considering the line-of-sight positional uncertainty. They could coalesce, for example.   Observations of tidal interactions or a common halo of young stars could help clarify this possibility.
 
{\it Acknowledgements:} 
We are grateful to the referee for constructive comments and the suggestion that massive clusters may emerge from bubble walls faster than from spiral arm filaments.
This work is based in part on observations made with the NASA/ESA/CSA James Webb Space Telescope, which is operated by the Association of Universities for Research in Astronomy, Inc., under NASA contract NAS 5-03127. 
These observations are associated with program \#1783. 
Support for program \#1783 was provided by NASA through a grant from the Space Telescope Science Institute, which is operated by the Association of Universities for Research in Astronomy, Inc., under NASA contract NAS 5-03127. 
Archival data presented in this paper were obtained from the Mikulski Archive for Space Telescopes (MAST) at the Space Telescope Science Institute. 
K.G. is supported by the Australian Research Council through the Discovery Early Career Researcher Award (DECRA) Fellowship (project number DE220100766) funded by the Australian Government and by the Australian Research Council Centre of Excellence for All Sky Astrophysics in 3 Dimensions (ASTRO 3D), through project number CE170100013. 
This project is supported by the Australian Research Council through the Discovery Program (project number DP240101472) funded by the Australian Government.
A.D.C. acknowledges support from a Royal Society University Research Fellowship (URF/R1/19160 and URF/R/241028).
A.A. and A.P. acknowledge support from the Swedish National Space Agency (SNSA) through grant 2021-00108. 
A.A. acknowledges support by the Swedish research council Vetenskapsr{\aa}det (2021-05559). 
The specific observations analyzed can be accessed via \dataset[doi:10.17909/6dc1-9h53]{https://doi.org/10.17909/6dc1-9h53}

{\it Facility:} JWST (NIRCam, MIRI). Software: JWST Calibration Pipeline \citep{bushouse22, greenfield16}; SAOImage DS9 \citep{joye03}; FORTRAN, and VossPlot \citep{voss95}.

\bibliography{refs}{}

@ARTICLE{block97,
       author = {{Block}, David L. and {Elmegreen}, Bruce G. and {Stockton}, Alan and 
       {Sauvage}, Marc},
        title = "{Macromolecules, Very Small Grains, and Large Dust Particles in the 
        Whirlpool Galaxy M51 and Its Companions: A Unified View}",
      journal = {\apjl},
         year = 1997,
        month = sep,
       volume = {486},
       number = {2},
        pages = {L95-L98},
          doi = {10.1086/310855},
       adsurl = {https://ui.adsabs.harvard.edu/abs/1997ApJ...486L..95B}
}

@ARTICLE{morgan53,
       author = {{Morgan}, W.~W. and {Whitford}, A.~E. and {Code}, A.~D.},
        title = "{Studies in Galactic Structure. I. a Preliminary Determination of the Space Distribution of the Blue Giants.}",
      journal = {\apj},
         year = 1953,
        month = sep,
       volume = {118},
        pages = {318},
          doi = {10.1086/145754},
       adsurl = {https://ui.adsabs.harvard.edu/abs/1953ApJ...118..318M}
}

@ARTICLE{vallee17,
       author = {{Vall{\'e}e}, Jacques P.},
        title = "{A guided map to the spiral arms in the galactic disk of the Milky Way}",
      journal = {The Astronomical Review},
         year = 2017,
        month = oct,
       volume = {13},
       number = {3-4},
        pages = {113-146},
          doi = {10.1080/21672857.2017.1379459},
archivePrefix = {arXiv},
       eprint = {1711.05228},
 primaryClass = {astro-ph.GA},
       adsurl = {https://ui.adsabs.harvard.edu/abs/2017AstRv..13..113V}
}

@ARTICLE{roberts69,
       author = {{Roberts}, W.~W.},
        title = "{Large-Scale Shock Formation in Spiral Galaxies and its Implications on Star Formation}",
      journal = {\apj},
         year = 1969,
        month = oct,
       volume = {158},
        pages = {123},
          doi = {10.1086/150177},
       adsurl = {https://ui.adsabs.harvard.edu/abs/1969ApJ...158..123R}
}

@ARTICLE{heyer15,
       author = {{Heyer}, Mark and {Dame}, T.~M.},
        title = "{Molecular Clouds in the Milky Way}",
      journal = {\araa},
         year = 2015,
        month = aug,
       volume = {53},
        pages = {583-629},
          doi = {10.1146/annurev-astro-082214-122324},
       adsurl = {https://ui.adsabs.harvard.edu/abs/2015ARA&A..53..583H}
}

@ARTICLE{rand92,
       author = {{Rand}, Richard J.},
        title = "{The H II Region Luminosity Function of M51}",
      journal = {\aj},
         year = 1992,
        month = mar,
       volume = {103},
        pages = {815},
          doi = {10.1086/116103},
       adsurl = {https://ui.adsabs.harvard.edu/abs/1992AJ....103..815R}
}

@ARTICLE{rumstay83,
       author = {{Rumstay}, K.~S. and {Kaufman}, M.},
        title = "{H II regions and star formation in M83 and M 33.}",
      journal = {\apj},
         year = 1983,
        month = nov,
       volume = {274},
        pages = {611-631},
          doi = {10.1086/161474},
       adsurl = {https://ui.adsabs.harvard.edu/abs/1983ApJ...274..611R}
}

@INPROCEEDINGS{baade51,
       author = {{Baade}, W. and {Mayall}, N.~U.},
        title = "{Distribution and Motions of Gaseous Masses in Spirals}",
    booktitle = {Problems of Cosmical Aerodynamics},
         year = 1951,
        month = jan,
        pages = {165},
       adsurl = {https://ui.adsabs.harvard.edu/abs/1951pca..conf..165B}
}

@ARTICLE{higdon13,
       author = {{Higdon}, J.~C. and {Lingenfelter}, R.~E.},
        title = "{The Galactic Spatial Distribution of OB Associations and their Surrounding Supernova-generated Superbubbles}",
      journal = {\apj},
         year = 2013,
        month = oct,
       volume = {775},
       number = {2},
          eid = {110},
        pages = {110},
          doi = {10.1088/0004-637X/775/2/110},
archivePrefix = {arXiv},
       eprint = {1302.1223},
 primaryClass = {astro-ph.HE},
       adsurl = {https://ui.adsabs.harvard.edu/abs/2013ApJ...775..110H}
}

@ARTICLE{lee11,
       author = {{Lee}, Jong Hwan and {Hwang}, Narae and {Lee}, Myung Gyoon},
        title = "{H II Region Luminosity Function of the Interacting Galaxy M51}",
      journal = {\apj},
         year = 2011,
        month = jul,
       volume = {735},
       number = {2},
          eid = {75},
        pages = {75},
          doi = {10.1088/0004-637X/735/2/75},
archivePrefix = {arXiv},
       eprint = {1104.3669},
 primaryClass = {astro-ph.CO},
       adsurl = {https://ui.adsabs.harvard.edu/abs/2011ApJ...735...75L}
}

@ARTICLE{azimlu11,
       author = {{Azimlu}, M. and {Marciniak}, R. and {Barmby}, P.},
        title = "{A New Catalog of H II Regions in M31}",
      journal = {\aj},
         year = 2011,
        month = oct,
       volume = {142},
       number = {4},
          eid = {139},
        pages = {139},
          doi = {10.1088/0004-6256/142/4/139},
archivePrefix = {arXiv},
       eprint = {1108.4044},
 primaryClass = {astro-ph.CO},
       adsurl = {https://ui.adsabs.harvard.edu/abs/2011AJ....142..139A}
}

@ARTICLE{santoro22,
       author = {{Santoro}, Francesco and {Kreckel}, Kathryn and {Belfiore}, Francesco and {Groves}, Brent and {Congiu}, Enrico and {Thilker}, David A. and {Blanc}, Guillermo A. and {Schinnerer}, Eva and {Ho}, I.-Ting and {Kruijssen}, J.~M. Diederik and {Meidt}, Sharon and {Klessen}, Ralf S. and {Schruba}, Andreas and {Querejeta}, Miguel and {Pessa}, Ismael and {Chevance}, M{\'e}lanie and {Kim}, Jaeyeon and {Emsellem}, Eric and {McElroy}, Rebecca and {Barnes}, Ashley T. and {Bigiel}, Frank and {Boquien}, M{\'e}d{\'e}ric and {Dale}, Daniel A. and {Glover}, Simon C.~O. and {Grasha}, Kathryn and {Lee}, Janice and {Leroy}, Adam K. and {Pan}, Hsi-An and {Rosolowsky}, Erik and {Saito}, Toshiki and {Sanchez-Blazquez}, Patricia and {Watkins}, Elizabeth J. and {Williams}, Thomas G.},
        title = "{PHANGS-MUSE: The H II region luminosity function of local star-forming galaxies}",
      journal = {\aap},
         year = 2022,
        month = feb,
       volume = {658},
          eid = {A188},
        pages = {A188},
          doi = {10.1051/0004-6361/202141907},
archivePrefix = {arXiv},
       eprint = {2111.09362},
 primaryClass = {astro-ph.GA},
       adsurl = {https://ui.adsabs.harvard.edu/abs/2022A&A...658A.188S}
}

@ARTICLE{linden26,
       author = {{Linden}, Sean T. and {Adamo}, Angela and {Pedrini}, Alex and {Calzetti}, Daniela and {Faustino Vieira}, Helena and {Elmegreen}, Bruce E. and {Gallagher}, John S. and {Johnson}, Kelsey and {Messa}, Matteo and {Grasha}, Kathryn and {Duarte-Cabral}, Ana and {Bik}, Arjan and {Bortolini}, Giacomo and {Correnti}, Matteo and {Lapeer}, Drew and {Lai}, Thomas S.-Y. and {Ostlin}, Goran and {Smith}, Linda J. and {Tosi}, Monica},
        title = "{Feedback in Extragalactic Star Clusters (FEAST): Spectral Energy Distributions and the Physical Properties of Star Clusters in NGC 628 with CIGALE}",
      journal = {arXiv e-prints},
         year = 2026,
        month = jun,
          eid = {arXiv:2606.13391},
        pages = {arXiv:2606.13391},
          doi = {10.48550/arXiv.2606.13391},
archivePrefix = {arXiv},
       eprint = {2606.13391},
 primaryClass = {astro-ph.GA},
       adsurl = {https://ui.adsabs.harvard.edu/abs/2026arXiv260613391L}
}

@ARTICLE{watkins23a,
       author = {{Watkins}, Elizabeth J. and {Barnes}, Ashley T. and {Henny}, Kiana and {Kim}, Hwihyun and {Kreckel}, Kathryn and {Meidt}, Sharon E. and {Klessen}, Ralf S. and {Glover}, Simon C.~O. and {Williams}, Thomas G. and {Keller}, Benjamin W. and {Leroy}, Adam K. and {Rosolowsky}, Erik and {Lee}, Janice C. and {Anand}, Gagandeep S. and {Belfiore}, Francesco and {Bigiel}, Frank and {Blanc}, Guillermo A. and {Boquien}, M{\'e}d{\'e}ric and {Cao}, Yixian and {Chandar}, Rupali and {Chen}, Ness Mayker and {Chevance}, M{\'e}lanie and {Congiu}, Enrico and {Dale}, Daniel A. and {Deger}, Sinan and {Egorov}, Oleg V. and {Emsellem}, Eric and {Faesi}, Christopher M. and {Grasha}, Kathryn and {Groves}, Brent and {Hassani}, Hamid and {Henshaw}, Jonathan D. and {Herrera}, Cinthya and {Hughes}, Annie and {Jeffreson}, Sarah and {Jim{\'e}nez-Donaire}, Mar{\'\i}a J. and {Koch}, Eric W. and {Kruijssen}, J.~M. Diederik and {Larson}, Kirsten L. and {Liu}, Daizhong and {Lopez}, Laura A. and {Pessa}, Ismael and {Pety}, J{\'e}r{\^o}me and {Querejeta}, Miguel and {Saito}, Toshiki and {Sandstrom}, Karin and {Scheuermann}, Fabian and {Schinnerer}, Eva and {Sormani}, Mattia C. and {Stuber}, Sophia K. and {Thilker}, David A. and {Usero}, Antonio and {Whitmore}, Bradley C.},
        title = "{PHANGS-JWST First Results: A Statistical View on Bubble Evolution in NGC 628}",
      journal = {\apjl},
         year = 2023,
        month = feb,
       volume = {944},
       number = {2},
          eid = {L24},
        pages = {L24},
          doi = {10.3847/2041-8213/aca6e4},
archivePrefix = {arXiv},
       eprint = {2212.00811},
 primaryClass = {astro-ph.GA},
       adsurl = {https://ui.adsabs.harvard.edu/abs/2023ApJ...944L..24W}
}

@ARTICLE{watkins23b,
       author = {{Watkins}, E.~J. and {Kreckel}, K. and {Groves}, B. and {Glover}, S.~C.~O. and {Whitmore}, B.~C. and {Leroy}, A.~K. and {Schinnerer}, E. and {Meidt}, S.~E. and {Egorov}, O.~V. and {Barnes}, A.~T. and {Lee}, J.~C. and {Bigiel}, F. and {Boquien}, M. and {Chandar}, R. and {Chevance}, M. and {Dale}, D.~A. and {Grasha}, K. and {Klessen}, R.~S. and {Kruijssen}, J.~M.~D. and {Larson}, K.~L. and {Li}, J. and {M{\'e}ndez-Delgado}, J.~E. and {Pessa}, I. and {Saito}, T. and {Sanchez-Blazquez}, P. and {Sarbadhicary}, S.~K. and {Scheuermann}, F. and {Thilker}, D.~A. and {Williams}, T.~G.},
        title = "{Quantifying the energetics of molecular superbubbles in PHANGS galaxies}",
      journal = {\aap},
         year = 2023,
        month = aug,
       volume = {676},
          eid = {A67},
        pages = {A67},
          doi = {10.1051/0004-6361/202346075},
archivePrefix = {arXiv},
       eprint = {2302.03699},
 primaryClass = {astro-ph.GA},
       adsurl = {https://ui.adsabs.harvard.edu/abs/2023A&A...676A..67W}
}

@ARTICLE{thilker23,
       author = {{Thilker}, David A. and {Lee}, Janice C. and {Deger}, Sinan and {Barnes}, Ashley T. and {Bigiel}, Frank and {Boquien}, M{\'e}d{\'e}ric and {Cao}, Yixian and {Chevance}, M{\'e}lanie and {Dale}, Daniel A. and {Egorov}, Oleg V. and {Glover}, Simon C.~O. and {Grasha}, Kathryn and {Henshaw}, Jonathan D. and {Klessen}, Ralf S. and {Koch}, Eric and {Kruijssen}, J.~M. Diederik and {Leroy}, Adam K. and {Lessing}, Ryan A. and {Meidt}, Sharon E. and {Pinna}, Francesca and {Querejeta}, Miguel and {Rosolowsky}, Erik and {Sandstrom}, Karin M. and {Schinnerer}, Eva and {Smith}, Rowan J. and {Watkins}, Elizabeth J. and {Williams}, Thomas G. and {Anand}, Gagandeep S. and {Belfiore}, Francesco and {Blanc}, Guillermo A. and {Chandar}, Rupali and {Congiu}, Enrico and {Emsellem}, Eric and {Groves}, Brent and {Kreckel}, Kathryn and {Larson}, Kirsten L. and {Liu}, Daizhong and {Pessa}, Ismael and {Whitmore}, Bradley C.},
        title = "{PHANGS-JWST First Results: The Dust Filament Network of NGC 628 and Its Relation to Star Formation Activity}",
      journal = {\apjl},
         year = 2023,
        month = feb,
       volume = {944},
       number = {2},
          eid = {L13},
        pages = {L13},
          doi = {10.3847/2041-8213/acaeac},
archivePrefix = {arXiv},
       eprint = {2301.00881},
 primaryClass = {astro-ph.GA},
       adsurl = {https://ui.adsabs.harvard.edu/abs/2023ApJ...944L..13T}
}

@ARTICLE{meidt23,
       author = {{Meidt}, Sharon E. and {Rosolowsky}, Erik and {Sun}, Jiayi and {Koch}, Eric W. and {Klessen}, Ralf S. and {Leroy}, Adam K. and {Schinnerer}, Eva and {Barnes}, Ashley. T. and {Glover}, Simon C.~O. and {Lee}, Janice C. and {van der Wel}, Arjen and {Watkins}, Elizabeth J. and {Williams}, Thomas G. and {Bigiel}, F. and {Boquien}, M{\'e}d{\'e}ric and {Blanc}, Guillermo A. and {Cao}, Yixian and {Chevance}, M{\'e}lanie and {Dale}, Daniel A. and {Egorov}, Oleg V. and {Emsellem}, Eric and {Grasha}, Kathryn and {Henshaw}, Jonathan D. and {Kruijssen}, J.~M. Diederik and {Larson}, Kirsten L. and {Liu}, Daizhong and {Murphy}, Eric J. and {Pety}, J{\'e}r{\^o}me and {Querejeta}, Miguel and {Saito}, Toshiki and {Sandstrom}, Karin M. and {Smith}, Rowan J. and {Sormani}, Mattia C. and {Thilker}, David A.},
        title = "{PHANGS-JWST First Results: Interstellar Medium Structure on the Turbulent Jeans Scale in Four Disk Galaxies Observed by JWST and the Atacama Large Millimeter/submillimeter Array}",
      journal = {\apjl},
         year = 2023,
        month = feb,
       volume = {944},
       number = {2},
          eid = {L18},
        pages = {L18},
          doi = {10.3847/2041-8213/acaaa8},
archivePrefix = {arXiv},
       eprint = {2212.06434},
 primaryClass = {astro-ph.GA},
       adsurl = {https://ui.adsabs.harvard.edu/abs/2023ApJ...944L..18M}
}

@ARTICLE{leroy23,
       author = {{Leroy}, Adam K. and {Sandstrom}, Karin and {Rosolowsky}, Erik and {Belfiore}, Francesco and {Bolatto}, Alberto D. and {Cao}, Yixian and {Koch}, Eric W. and {Schinnerer}, Eva and {Barnes}, Ashley. T. and {Be{\v{s}}li{\'c}}, Ivana and {Bigiel}, F. and {Blanc}, Guillermo A. and {Chastenet}, J{\'e}r{\'e}my and {Chen}, Ness Mayker and {Chevance}, M{\'e}lanie and {Chown}, Ryan and {Congiu}, Enrico and {Dale}, Daniel A. and {Egorov}, Oleg V. and {Emsellem}, Eric and {Eibensteiner}, Cosima and {Faesi}, Christopher M. and {Glover}, Simon C.~O. and {Grasha}, Kathryn and {Groves}, Brent and {Hassani}, Hamid and {Henshaw}, Jonathan D. and {Hughes}, Annie and {Jim{\'e}nez-Donaire}, Mar{\'\i}a J. and {Kim}, Jaeyeon and {Klessen}, Ralf S. and {Kreckel}, Kathryn and {Kruijssen}, J.~M. Diederik and {Larson}, Kirsten L. and {Lee}, Janice C. and {Levy}, Rebecca C. and {Liu}, Daizhong and {Lopez}, Laura A. and {Meidt}, Sharon E. and {Murphy}, Eric J. and {Neumann}, Justus and {Pessa}, Ismael and {Pety}, J{\'e}r{\^o}me and {Saito}, Toshiki and {Sardone}, Amy and {Sun}, Jiayi and {Thilker}, David A. and {Usero}, Antonio and {Watkins}, Elizabeth J. and {Whitcomb}, Cory M. and {Williams}, Thomas G.},
        title = "{PHANGS-JWST First Results: Mid-infrared Emission Traces Both Gas Column Density and Heating at 100 pc Scales}",
      journal = {\apjl},
         year = 2023,
        month = feb,
       volume = {944},
       number = {2},
          eid = {L9},
        pages = {L9},
          doi = {10.3847/2041-8213/acaf85},
archivePrefix = {arXiv},
       eprint = {2212.10574},
 primaryClass = {astro-ph.GA},
       adsurl = {https://ui.adsabs.harvard.edu/abs/2023ApJ...944L...9L}
}

@ARTICLE{sandstrom23,
       author = {{Sandstrom}, Karin M. and {Koch}, Eric W. and {Leroy}, Adam K. and {Rosolowsky}, Erik and {Emsellem}, Eric and {Smith}, Rowan J. and {Egorov}, Oleg V. and {Williams}, Thomas G. and {Larson}, Kirsten L. and {Lee}, Janice C. and {Schinnerer}, Eva and {Thilker}, David A. and {Barnes}, Ashley T. and {Belfiore}, Francesco and {Bigiel}, F. and {Blanc}, Guillermo A. and {Bolatto}, Alberto D. and {Boquien}, M{\'e}d{\'e}ric and {Cao}, Yixian and {Chastenet}, J{\'e}r{\'e}my and {Chevance}, M{\'e}lanie and {Chiang}, I-Da and {Dale}, Daniel A. and {Faesi}, Christopher M. and {Glover}, Simon C.~O. and {Grasha}, Kathryn and {Groves}, Brent and {Hassani}, Hamid and {Henshaw}, Jonathan D. and {Hughes}, Annie and {Kim}, Jaeyeon and {Klessen}, Ralf S. and {Kreckel}, Kathryn and {Kruijssen}, J.~M. Diederik and {Lopez}, Laura A. and {Liu}, Daizhong and {Meidt}, Sharon E. and {Murphy}, Eric J. and {Pan}, Hsi-An and {Querejeta}, Miguel and {Saito}, Toshiki and {Sardone}, Amy and {Sormani}, Mattia C. and {Sutter}, Jessica and {Usero}, Antonio and {Watkins}, Elizabeth J.},
        title = "{PHANGS-JWST First Results: Tracing the Diffuse Interstellar Medium with JWST Imaging of Polycyclic Aromatic Hydrocarbon Emission in Nearby Galaxies}",
      journal = {\apjl},
         year = 2023,
        month = feb,
       volume = {944},
       number = {2},
          eid = {L8},
        pages = {L8},
          doi = {10.3847/2041-8213/aca972},
archivePrefix = {arXiv},
       eprint = {2212.11177},
 primaryClass = {astro-ph.GA},
       adsurl = {https://ui.adsabs.harvard.edu/abs/2023ApJ...944L...8S}
}

@ARTICLE{barnes23,
       author = {{Barnes}, Ashley. T. and {Watkins}, Elizabeth J. and {Meidt}, Sharon E. and {Kreckel}, Kathryn and {Sormani}, Mattia C. and {Tre{\ss}}, Robin G. and {Glover}, Simon C.~O. and {Bigiel}, Frank and {Chandar}, Rupali and {Emsellem}, Eric and {Lee}, Janice C. and {Leroy}, Adam K. and {Sandstrom}, Karin M. and {Schinnerer}, Eva and {Rosolowsky}, Erik and {Belfiore}, Francesco and {Blanc}, Guillermo A. and {Boquien}, M{\'e}d{\'e}ric and {Brok}, Jakob den and {Cao}, Yixian and {Chevance}, M{\'e}lanie and {Dale}, Daniel A. and {Egorov}, Oleg V. and {Eibensteiner}, Cosima and {Grasha}, Kathryn and {Groves}, Brent and {Hassani}, Hamid and {Henshaw}, Jonathan D. and {Jeffreson}, Sarah and {Jim{\'e}nez-Donaire}, Mar{\'\i}a J. and {Keller}, Benjamin W. and {Klessen}, Ralf S. and {Koch}, Eric W. and {Kruijssen}, J.~M. Diederik and {Larson}, Kirsten L. and {Li}, Jing and {Liu}, Daizhong and {Lopez}, Laura A. and {Murphy}, Eric J. and {Neumann}, Lukas and {Pety}, J{\'e}r{\^o}me and {Pinna}, Francesca and {Querejeta}, Miguel and {Renaud}, Florent and {Saito}, Toshiki and {Sarbadhicary}, Sumit K. and {Sardone}, Amy and {Smith}, Rowan J. and {Stuber}, Sophia K. and {Sun}, Jiayi and {Thilker}, David A. and {Usero}, Antonio and {Whitmore}, Bradley C. and {Williams}, Thomas G.},
        title = "{PHANGS-JWST First Results: Multiwavelength View of Feedback-driven Bubbles (the Phantom Voids) across NGC 628}",
      journal = {\apjl},
         year = 2023,
        month = feb,
       volume = {944},
       number = {2},
          eid = {L22},
        pages = {L22},
          doi = {10.3847/2041-8213/aca7b9},
archivePrefix = {arXiv},
       eprint = {2212.00812},
 primaryClass = {astro-ph.GA},
       adsurl = {https://ui.adsabs.harvard.edu/abs/2023ApJ...944L..22B}
}

@ARTICLE{williams22,
       author = {{Williams}, Thomas G. and {Sun}, Jiayi and {Barnes}, Ashley T. and {Schinnerer}, Eva and {Henshaw}, Jonathan D. and {Meidt}, Sharon E. and {Querejeta}, Miguel and {Watkins}, Elizabeth J. and {Bigiel}, Frank and {Blanc}, Guillermo A. and {Boquien}, M{\'e}d{\'e}ric and {Cao}, Yixian and {Chevance}, M{\'e}lanie and {Egorov}, Oleg V. and {Emsellem}, Eric and {Glover}, Simon C.~O. and {Grasha}, Kathryn and {Hassani}, Hamid and {Jeffreson}, Sarah and {Jim{\'e}nez-Donaire}, Mar{\'\i}a J. and {Kim}, Jaeyeon and {Klessen}, Ralf S. and {Kreckel}, Kathryn and {Kruijssen}, J.~M. Diederik and {Larson}, Kirsten L. and {Leroy}, Adam K. and {Liu}, Daizhong and {Pessa}, Ismael and {Pety}, J{\'e}r{\^o}me and {Pinna}, Francesca and {Rosolowsky}, Erik and {Sandstrom}, Karin M. and {Smith}, Rowan and {Sormani}, Mattia C. and {Stuber}, Sophia and {Thilker}, David A. and {Whitmore}, Bradley C.},
        title = "{PHANGS-JWST First Results: Spurring on Star Formation: JWST Reveals Localized Star Formation in a Spiral Arm Spur of NGC 628}",
      journal = {\apjl},
         year = 2022,
        month = dec,
       volume = {941},
       number = {2},
          eid = {L27},
        pages = {L27},
          doi = {10.3847/2041-8213/aca674},
archivePrefix = {arXiv},
       eprint = {2212.00032},
 primaryClass = {astro-ph.GA},
       adsurl = {https://ui.adsabs.harvard.edu/abs/2022ApJ...941L..27W}
}

@ARTICLE{egorov23,
       author = {{Egorov}, Oleg V. and {Kreckel}, Kathryn and {Glover}, Simon C.~O. and {Groves}, Brent and {Belfiore}, Francesco and {Emsellem}, Eric and {Klessen}, Ralf S. and {Leroy}, Adam K. and {Meidt}, Sharon E. and {Sarbadhicary}, Sumit K. and {Schinnerer}, Eva and {Watkins}, Elizabeth J. and {Whitmore}, Brad C. and {Barnes}, Ashley T. and {Congiu}, Enrico and {Dale}, Daniel A. and {Grasha}, Kathryn and {Larson}, Kirsten L. and {Lee}, Janice C. and {M{\'e}ndez-Delgado}, J. Eduardo and {Thilker}, David A. and {Williams}, Thomas G.},
        title = "{Quantifying the energy balance between the turbulent ionised gas and young stars}",
      journal = {\aap},
         year = 2023,
        month = oct,
       volume = {678},
          eid = {A153},
        pages = {A153},
          doi = {10.1051/0004-6361/202346919},
archivePrefix = {arXiv},
       eprint = {2307.10277},
 primaryClass = {astro-ph.GA},
       adsurl = {https://ui.adsabs.harvard.edu/abs/2023A&A...678A.153E}
}

@ARTICLE{jimenez24,
       author = {{Jim{\'e}nez}, S. and {Silich}, S. and {Mayya}, Y.~D. and {Zaragoza-Cardiel}, J.},
        title = "{What Holes in the Gas Distribution of Nearly Face-on Galaxies Can Tell Us about the Host Disk Parameters: The Case of the NGC 628 Southeast Superbubble}",
      journal = {\apj},
         year = 2024,
        month = jan,
       volume = {960},
       number = {1},
          eid = {81},
        pages = {81},
          doi = {10.3847/1538-4357/ad0cb8},
archivePrefix = {arXiv},
       eprint = {2311.08178},
 primaryClass = {astro-ph.GA},
       adsurl = {https://ui.adsabs.harvard.edu/abs/2024ApJ...960...81J}
}

@ARTICLE{mayya23,
       author = {{Mayya}, Y.~D. and {Alzate}, J.~A. and {Lomel{\'\i}-N{\'u}ez}, L. and {Zaragoza-Cardiel}, J. and {G{\'o}mez-Gonz{\'a}lez}, V.~M.~A. and {Silich}, S. and {Fern{\'a}ndez-Arenas}, D. and {Vega}, O. and {Ovando}, P.~A. and {Rodr{\'\i}guez}, L.~H. and {Rosa-Gonz{\'a}lez}, D. and {Luna}, A. and {Zamora-Avil{\'e}s}, M. and {Rosales-Ortega}, F.},
        title = "{The stellar population responsible for a kiloparsec-size superbubble seen in the JWST 'phantom' images of NGC 628}",
      journal = {\mnras},
         year = 2023,
        month = jun,
       volume = {521},
       number = {4},
        pages = {5492-5507},
          doi = {10.1093/mnras/stad636},
archivePrefix = {arXiv},
       eprint = {2302.12704},
 primaryClass = {astro-ph.GA},
       adsurl = {https://ui.adsabs.harvard.edu/abs/2023MNRAS.521.5492M}
}

@ARTICLE{querejeta21,
       author = {{Querejeta}, M. and {Schinnerer}, E. and {Meidt}, S. and {Sun}, J. and {Leroy}, A.~K. and {Emsellem}, E. and {Klessen}, R.~S. and {Mu{\~n}oz-Mateos}, J.~C. and {Salo}, H. and {Laurikainen}, E. and {Be{\v{s}}li{\'c}}, I. and {Blanc}, G.~A. and {Chevance}, M. and {Dale}, D.~A. and {Eibensteiner}, C. and {Faesi}, C. and {Garc{\'\i}a-Rodr{\'\i}guez}, A. and {Glover}, S.~C.~O. and {Grasha}, K. and {Henshaw}, J. and {Herrera}, C. and {Hughes}, A. and {Kreckel}, K. and {Kruijssen}, J.~M.~D. and {Liu}, D. and {Murphy}, E.~J. and {Pan}, H.-A. and {Pety}, J. and {Razza}, A. and {Rosolowsky}, E. and {Saito}, T. and {Schruba}, A. and {Usero}, A. and {Watkins}, E.~J. and {Williams}, T.~G.},
        title = "{Stellar structures, molecular gas, and star formation across the PHANGS sample of nearby galaxies}",
      journal = {\aap},
         year = 2021,
        month = dec,
       volume = {656},
          eid = {A133},
        pages = {A133},
          doi = {10.1051/0004-6361/202140695},
archivePrefix = {arXiv},
       eprint = {2109.04491},
 primaryClass = {astro-ph.GA},
       adsurl = {https://ui.adsabs.harvard.edu/abs/2021A&A...656A.133Q}
}

@ARTICLE{querejeta24,
       author = {{Querejeta}, Miguel and {Leroy}, Adam K. and {Meidt}, Sharon E. and {Schinnerer}, Eva and {Belfiore}, Francesco and {Emsellem}, Eric and {Klessen}, Ralf S. and {Sun}, Jiayi and {Sormani}, Mattia and {Be{\v{s}}li{\'c}}, Ivana and {Cao}, Yixian and {Chevance}, M{\'e}lanie and {Colombo}, Dario and {Dale}, Daniel A. and {Garc{\'\i}a-Burillo}, Santiago and {Glover}, Simon C.~O. and {Grasha}, Kathryn and {Groves}, Brent and {Koch}, Eric. W. and {Neumann}, Lukas and {Pan}, Hsi-An and {Pessa}, Ismael and {Pety}, J{\'e}r{\^o}me and {Pinna}, Francesca and {Ramambason}, Lise and {Razza}, Alessandro and {Romanelli}, Andrea and {Rosolowsky}, Erik and {Ruiz-Garc{\'\i}a}, Marina and {S{\'a}nchez-Bl{\'a}zquez}, Patricia and {Smith}, Rowan and {Stuber}, Sophia and {Ubeda}, Leonardo and {Usero}, Antonio and {Williams}, Thomas G.},
        title = "{Do spiral arms enhance star formation efficiency?}",
      journal = {\aap},
         year = 2024,
        month = jul,
       volume = {687},
          eid = {A293},
        pages = {A293},
          doi = {10.1051/0004-6361/202449733},
archivePrefix = {arXiv},
       eprint = {2405.05364},
 primaryClass = {astro-ph.GA},
       adsurl = {https://ui.adsabs.harvard.edu/abs/2024A&A...687A.293Q}
}

@ARTICLE{querejeta25,
       author = {{Querejeta}, Miguel and {Meidt}, Sharon E. and {Cao}, Yixian and {Colombo}, Dario and {Emsellem}, Eric and {Garc{\'\i}a-Burillo}, Santiago and {Klessen}, Ralf S. and {Koch}, Eric W. and {Leroy}, Adam K. and {Ruiz-Garc{\'\i}a}, Marina and {Schinnerer}, Eva and {Smith}, Rowan and {Stuber}, Sophia and {Thorp}, Mallory and {Williams}, Thomas G. and {Boquien}, M{\'e}d{\'e}ric and {Dale}, Daniel A. and {Faesi}, Chris and {Gleis}, Damian R. and {Grasha}, Kathryn and {Hughes}, Annie and {Jim{\'e}nez-Donaire}, Mar{\'\i}a J. and {Kreckel}, Kathryn and {Liu}, Daizhong and {Neumann}, Justus and {Pan}, Hsi-An and {Pinna}, Francesca and {Razza}, Alessandro and {Saito}, Toshiki and {Sun}, Jiayi and {Usero}, Antonio},
        title = "{Azimuthal offsets in spiral arms of nearby galaxies}",
      journal = {\aap},
         year = 2025,
        month = sep,
       volume = {701},
          eid = {A183},
        pages = {A183},
          doi = {10.1051/0004-6361/202556175},
archivePrefix = {arXiv},
       eprint = {2509.01668},
 primaryClass = {astro-ph.GA},
       adsurl = {https://ui.adsabs.harvard.edu/abs/2025A&A...701A.183Q}
}

@ARTICLE{vieira26,
       author = {{Faustino Vieira}, Helena and {Adamo}, Angela and {Shane}, Neville and {Smith}, Linda J. and {Bik}, Arjan and {Lai}, Thomas S.-Y. and {Pedrini}, Alex and {Hunt}, Leslie K. and {Linden}, Sean T. and {Bortolini}, Giacomo and {Buckner}, Anne S.~M. and {Calzetti}, Daniela and {Correnti}, Matteo and {Duarte-Cabral}, Ana and {Grasha}, Kathryn and {Johnson}, Kelsey and {Lapeer}, Drew and {Messa}, Matteo and {{\"O}stlin}, G{\"o}ran and {Roos}, Linn and {Sabbi}, Elena},
        title = "{FEAST: A NIRSpec/MOS Survey of Emerging Young Star Clusters in NGC 628}",
      journal = {\apj},
         year = 2026,
        month = apr,
       volume = {1001},
       number = {2},
          eid = {128},
        pages = {128},
          doi = {10.3847/1538-4357/ae523f},
archivePrefix = {arXiv},
       eprint = {2603.09866},
 primaryClass = {astro-ph.GA},
       adsurl = {https://ui.adsabs.harvard.edu/abs/2026ApJ..1001..128F}
}

@ARTICLE{koletic26,
       author = {{Koletic}, Tamara R. and {Pillsworth}, Rachel and {Pudritz}, Ralph E.},
        title = "{Revealing the Connection between the Filamentary Hierarchy and Star Cluster Formation in a Simulated NGC 628 Galaxy}",
      journal = {\apj},
         year = 2026,
        month = feb,
       volume = {997},
       number = {2},
          eid = {337},
        pages = {337},
          doi = {10.3847/1538-4357/ae285a},
archivePrefix = {arXiv},
       eprint = {2511.10486},
 primaryClass = {astro-ph.GA},
       adsurl = {https://ui.adsabs.harvard.edu/abs/2026ApJ...997..337K}
}

@ARTICLE{pedrini24,
       author = {{Pedrini}, Alex and {Adamo}, Angela and {Calzetti}, Daniela and {Bik}, Arjan and {Gregg}, Benjamin and {Linden}, Sean T. and {Bajaj}, Varun and {Ryon}, Jenna E. and {Ali}, Ahmad A. and {Bortolini}, Giacomo and {Correnti}, Matteo and {Elmegreen}, Bruce G. and {Elmegreen}, Debra Meloy and {Gallagher}, John S. and {Grasha}, Kathryn and {Gutermuth}, Robert A. and {Johnson}, Kelsey E. and {Melinder}, Jens and {Messa}, Matteo and {{\"O}stlin}, G{\"o}ran and {Sabbi}, Elena and {Smith}, Linda J. and {Tosi}, Monica and {Faustino Vieira}, Helena},
        title = "{FEAST: Feedback in Emerging extragAlactic Star ClusTers: JWST Spots Polycyclic Aromatic Hydrocarbon Destruction in NGC 628 during the Emerging Phase of Star Formation}",
      journal = {\apj},
         year = 2024,
        month = aug,
       volume = {971},
       number = {1},
          eid = {32},
        pages = {32},
          doi = {10.3847/1538-4357/ad534d},
archivePrefix = {arXiv},
       eprint = {2406.01666},
 primaryClass = {astro-ph.GA},
       adsurl = {https://ui.adsabs.harvard.edu/abs/2024ApJ...971...32P}
}

@ARTICLE{chanu26,
       author = {{Chanu}, Athokpam Langlen and {Amrutha}, S. and {Chingangbam}, Pravabati and {Park}, Changbom},
        title = "{Morphological Complexity of NGC 628{\textemdash}A Multiwavelength Multiscale Analysis Using the Ordinal Pattern Framework}",
      journal = {\apj},
         year = 2026,
        month = may,
       volume = {1003},
       number = {1},
          eid = {50},
        pages = {50},
          doi = {10.3847/1538-4357/ae5c98},
archivePrefix = {arXiv},
       eprint = {2604.08409},
 primaryClass = {astro-ph.GA},
       adsurl = {https://ui.adsabs.harvard.edu/abs/2026ApJ..1003...50C}
}

@ARTICLE{buckner26,
       author = {{Buckner}, Anne S.~M. and {Duarte-Cabral}, Ana and {Adamo}, Angela and {Linden}, Sean T. and {Cignoni}, Michele and {Bajaj}, Varun and {Bik}, Arjan and {Bortolini}, Giacomo and {Calzetti}, Daniela and {Correnti}, Matteo and {Elmegreen}, Bruce G. and {Elmegreen}, Debra M. and {Faustino Vieira}, Helena and {Gallagher}, III, John S. and {Grasha}, Kathryn and {Gregg}, Benjamin and {Gutermuth}, Rob A. and {Johnson}, Kelsey and {Krumholz}, Mark and {Lapeer}, Drew and {Messa}, Matteo and {{\"O}stlin}, G{\"o}ran and {Pedrini}, Alex and {Ryon}, Jenna E. and {Smith}, Linda J. and {Tosi}, Monica},
        title = "{The spatial evolution of star clusters in NGC 628 with JWST}",
      journal = {\mnras},
         year = 2026,
        month = jan,
       volume = {545},
       number = {3},
          eid = {staf2025},
        pages = {staf2025},
          doi = {10.1093/mnras/staf2025},
archivePrefix = {arXiv},
       eprint = {2511.11115},
 primaryClass = {astro-ph.GA},
       adsurl = {https://ui.adsabs.harvard.edu/abs/2026MNRAS.545f2025B}
}

@ARTICLE{ck25,
       author = {{Ck}, Avinash and {Mayya}, Divakara and {Bressan}, Alessandro and {Alzate Trujillo}, Jairo Andres},
        title = "{Spatially-resolved TRGB JWST Color─Magnitude Diagram as a Tool to Measure Fossil Stellar Metallicity Gradients in Disk Galaxies: NGC 628}",
      journal = {\apj},
         year = 2025,
        month = may,
       volume = {984},
       number = {1},
          eid = {21},
        pages = {21},
          doi = {10.3847/1538-4357/adc105},
archivePrefix = {arXiv},
       eprint = {2503.10438},
 primaryClass = {astro-ph.GA},
       adsurl = {https://ui.adsabs.harvard.edu/abs/2025ApJ...984...21C}
}

@ARTICLE{vieira25,
       author = {{Faustino Vieira}, Helena and {Duarte-Cabral}, Ana and {Smith}, Matthew W.~L. and {Colombo}, Dario and {Davis}, Timothy A. and {Bazzi}, Zein},
        title = "{Cloud populations versus galactic environment in NGC 4689, NGC 628, NGC 1566, and NGC 4321}",
      journal = {\mnras},
         year = 2025,
        month = apr,
       volume = {538},
       number = {4},
        pages = {2744-2767},
          doi = {10.1093/mnras/staf411},
archivePrefix = {arXiv},
       eprint = {2503.08212},
 primaryClass = {astro-ph.GA},
       adsurl = {https://ui.adsabs.harvard.edu/abs/2025MNRAS.538.2744F}
}

@ARTICLE{nishimoto25,
       author = {{Nishimoto}, Shimpei and {Onishi}, Toshikazu and {Nishimura}, Atsushi and {Fujita}, Shinji and {Kawanishi}, Yasutomo and {Nakatani}, Shuyo and {Tokuda}, Kazuki and {Shimajiri}, Yoshito and {Kaneko}, Hiroyuki and {Miyamoto}, Yusuke and {Inoue}, Tsuyoshi and {Ito}, Atsushi M.},
        title = "{Infrared bubble recognition in the Milky Way and beyond using deep learning}",
      journal = {\pasj},
         year = 2025,
        month = apr,
       volume = {77},
       number = {2},
        pages = {403-424},
          doi = {10.1093/pasj/psaf008},
archivePrefix = {arXiv},
       eprint = {2504.03367},
 primaryClass = {astro-ph.GA},
       adsurl = {https://ui.adsabs.harvard.edu/abs/2025PASJ...77..403N}
}

@ARTICLE{pillsworth26,
       author = {{Pillsworth}, Rachel and {Pudritz}, Ralph E. and {Koch}, Eric W. and {O'Neill}, Theo J.},
        title = "{Filamentary Hierarchies and Superbubbles. II. Impact of Superbubbles and Galactic Dynamics on Filament Formation and Fragmentation}",
      journal = {\apj},
         year = 2026,
        month = apr,
       volume = {1001},
       number = {2},
          eid = {220},
        pages = {220},
          doi = {10.3847/1538-4357/ae578e},
archivePrefix = {arXiv},
       eprint = {2602.22326},
 primaryClass = {astro-ph.GA},
       adsurl = {https://ui.adsabs.harvard.edu/abs/2026ApJ..1001..220P}
}

@ARTICLE{pillsworth25,
       author = {{Pillsworth}, Rachel and {Roscoe}, Erica and {Pudritz}, Ralph E. and {Koch}, Eric W.},
        title = "{Filamentary Hierarchies and Superbubbles. I. Characterizing Filament Properties across a Simulated Spiral Galaxy}",
      journal = {\apj},
         year = 2025,
        month = jul,
       volume = {987},
       number = {1},
          eid = {20},
        pages = {20},
          doi = {10.3847/1538-4357/add68a},
archivePrefix = {arXiv},
       eprint = {2504.01099},
 primaryClass = {astro-ph.GA},
       adsurl = {https://ui.adsabs.harvard.edu/abs/2025ApJ...987...20P}
}

@ARTICLE{anand21,
       author = {{Anand}, Gagandeep S. and {Lee}, Janice C. and {Van Dyk}, Schuyler D. and {Leroy}, Adam K. and {Rosolowsky}, Erik and {Schinnerer}, Eva and {Larson}, Kirsten and {Kourkchi}, Ehsan and {Kreckel}, Kathryn and {Scheuermann}, Fabian and {Rizzi}, Luca and {Thilker}, David and {Tully}, R. Brent and {Bigiel}, Frank and {Blanc}, Guillermo A. and {Boquien}, M{\'e}d{\'e}ric and {Chandar}, Rupali and {Dale}, Daniel and {Emsellem}, Eric and {Deger}, Sinan and {Glover}, Simon C.~O. and {Grasha}, Kathryn and {Groves}, Brent and {S. Klessen}, Ralf and {Kruijssen}, J.~M. Diederik and {Querejeta}, Miguel and {S{\'a}nchez-Bl{\'a}zquez}, Patricia and {Schruba}, Andreas and {Turner}, Jordan and {Ubeda}, Leonardo and {Williams}, Thomas G. and {Whitmore}, Brad},
        title = "{Distances to PHANGS galaxies: New tip of the red giant branch measurements and adopted distances}",
      journal = {\mnras},
         year = 2021,
        month = mar,
       volume = {501},
       number = {3},
        pages = {3621-3639},
          doi = {10.1093/mnras/staa3668},
archivePrefix = {arXiv},
       eprint = {2012.00757},
 primaryClass = {astro-ph.GA},
       adsurl = {https://ui.adsabs.harvard.edu/abs/2021MNRAS.501.3621A}
}

@ARTICLE{pedrini25,
       author = {{Pedrini}, Alex and {Adamo}, Angela and {Bik}, Arjan and {Calzetti}, Daniela and {Linden}, Sean T. and {Gregg}, Benjamin and {Bajaj}, Varun and {Ryon}, Jenna E. and {Buckner}, Anne S.~M. and {Bortolini}, Giacomo and {Cignoni}, Michele and {Correnti}, Matteo and {Duarte-Cabral}, Ana and {Elmegreen}, Bruce G. and {Faustino Vieira}, Helena and {Gallagher}, John S. and {Grasha}, Kathryn and {Johnson}, Kelsey E. and {Krumholz}, Mark R. and {Lapeer}, Drew and {Lai}, Thomas S.-Y. and {Messa}, Matteo and {{\"O}stlin}, G{\"o}ran and {Roos}, Linn and {Smith}, Linda J. and {Tosi}, Monica},
        title = "{The Near Infrared Spectral Energy Distribution of Young Star Clusters in the FEAST Galaxies: Missing Ingredients at 1─5 {\ensuremath{\mu}}m}",
      journal = {\apj},
         year = 2025,
        month = oct,
       volume = {992},
       number = {1},
          eid = {96},
        pages = {96},
          doi = {10.3847/1538-4357/ae0182},
archivePrefix = {arXiv},
       eprint = {2509.01670},
 primaryClass = {astro-ph.GA},
       adsurl = {https://ui.adsabs.harvard.edu/abs/2025ApJ...992...96P}
}

@ARTICLE{elmegreen06,
       author = {{Elmegreen}, Bruce G. and {Elmegreen}, Debra Meloy and {Chandar}, Rupali and {Whitmore}, Brad and {Regan}, Michael},
        title = "{Hierarchical Star Formation in the Spiral Galaxy NGC 628}",
      journal = {\apj},
         year = 2006,
        month = jun,
       volume = {644},
       number = {2},
        pages = {879-889},
          doi = {10.1086/503797},
archivePrefix = {arXiv},
       eprint = {astro-ph/0605523},
 primaryClass = {astro-ph},
       adsurl = {https://ui.adsabs.harvard.edu/abs/2006ApJ...644..879E}
}

@ARTICLE{grasha15,
       author = {{Grasha}, K. and {Calzetti}, D. and {Adamo}, A. and {Kim}, H. and {Elmegreen}, B.~G. and {Gouliermis}, D.~A. and {Aloisi}, A. and {Bright}, S.~N. and {Christian}, C. and {Cignoni}, M. and {Dale}, D.~A. and {Dobbs}, C. and {Elmegreen}, D.~M. and {Fumagalli}, M. and {Gallagher}, III, J.~S. and {Grebel}, E.~K. and {Johnson}, K.~E. and {Lee}, J.~C. and {Messa}, M. and {Smith}, L.~J. and {Ryon}, J.~E. and {Thilker}, D. and {Ubeda}, L. and {Wofford}, A.},
        title = "{The Spatial Distribution of the Young Stellar Clusters in the Star-forming Galaxy NGC 628}",
      journal = {\apj},
         year = 2015,
        month = dec,
       volume = {815},
       number = {2},
          eid = {93},
        pages = {93},
          doi = {10.1088/0004-637X/815/2/93},
archivePrefix = {arXiv},
       eprint = {1511.02233},
 primaryClass = {astro-ph.GA},
       adsurl = {https://ui.adsabs.harvard.edu/abs/2015ApJ...815...93G}
}

@ARTICLE{grasha17b,
       author = {{Grasha}, K. and {Elmegreen}, B.~G. and {Calzetti}, D. and {Adamo}, A. and {Aloisi}, A. and {Bright}, S.~N. and {Cook}, D.~O. and {Dale}, D.~A. and {Fumagalli}, M. and {Gallagher}, III, J.~S. and {Gouliermis}, D.~A. and {Grebel}, E.~K. and {Kahre}, L. and {Kim}, H. and {Krumholz}, M.~R. and {Lee}, J.~C. and {Messa}, M. and {Ryon}, J.~E. and {Ubeda}, L.},
        title = "{Hierarchical Star Formation in Turbulent Media: Evidence from Young Star Clusters}",
      journal = {\apj},
         year = 2017,
        month = jun,
       volume = {842},
       number = {1},
          eid = {25},
        pages = {25},
          doi = {10.3847/1538-4357/aa740b},
archivePrefix = {arXiv},
       eprint = {1705.06281},
 primaryClass = {astro-ph.GA},
       adsurl = {https://ui.adsabs.harvard.edu/abs/2017ApJ...842...25G}
}

@ARTICLE{grasha17a,
       author = {{Grasha}, K. and {Calzetti}, D. and {Adamo}, A. and {Kim}, H. and {Elmegreen}, B.~G. and {Gouliermis}, D.~A. and {Dale}, D.~A. and {Fumagalli}, M. and {Grebel}, E.~K. and {Johnson}, K.~E. and {Kahre}, L. and {Kennicutt}, R.~C. and {Messa}, M. and {Pellerin}, A. and {Ryon}, J.~E. and {Smith}, L.~J. and {Shabani}, F. and {Thilker}, D. and {Ubeda}, L.},
        title = "{The Hierarchical Distribution of the Young Stellar Clusters in Six Local Star-forming Galaxies}",
      journal = {\apj},
         year = 2017,
        month = may,
       volume = {840},
       number = {2},
          eid = {113},
        pages = {113},
          doi = {10.3847/1538-4357/aa6f15},
archivePrefix = {arXiv},
       eprint = {1704.06321},
 primaryClass = {astro-ph.GA},
       adsurl = {https://ui.adsabs.harvard.edu/abs/2017ApJ...840..113G}
}

@ARTICLE{lapeer26,
       author = {{Lapeer}, Drew and {Calzetti}, Daniela and {Grasha}, Kathryn and {Adamo}, Angela and {Elmegreen}, Bruce G. and {Bik}, Arjan and {Bortolini}, Giacomo and {Buckner}, Anne and {Cignoni}, Michele and {Correnti}, Matteo and {Elmegreen}, Debra Meloy and {Faustino Vieira}, H. and {Hamilton}, Max and {Johnson}, Kelsey and {Lai}, Thomas S.-Y. and {Linden}, Sean T. and {Maji}, Subhransu and {Messa}, Matteo and {{\"O}stlin}, G{\"o}ran and {Pedrini}, Alex and {Sabbi}, E. and {Smith}, Linda J.},
        title = "{FEAST: Probing Hierarchical Star Formation with the Spatial Distributions of Young Star Clusters}",
      journal = {\apj},
         year = 2026,
        month = mar,
       volume = {999},
       number = {2},
          eid = {189},
        pages = {189},
          doi = {10.3847/1538-4357/ae3a92},
archivePrefix = {arXiv},
       eprint = {2601.11434},
 primaryClass = {astro-ph.GA},
       adsurl = {https://ui.adsabs.harvard.edu/abs/2026ApJ...999..189L}
}

@ARTICLE{pedrini26,
       author = {{Pedrini}, Alex and {Adamo}, Angela and {Calzetti}, Daniela and {Bik}, Arjan and {Haworth}, Thomas J. and {Elmegreen}, Bruce G. and {Krumholz}, Mark R. and {Linden}, Sean T. and {Gregg}, Benjamin and {Faustino Vieira}, Helena and {Bajaj}, Varun and {Ryon}, Jenna E. and {Ali}, Ahmad A. and {Andersson}, Eric P. and {Bortolini}, Giacomo and {Cignoni}, Michele and {Duarte-Cabral}, Ana and {Grasha}, Kathryn and {Lah{\'e}n}, Natalia and {Lai}, Thomas S.-Y. and {Lapeer}, Drew and {Messa}, Matteo and {{\"O}stlin}, G{\"o}ran and {Sabbi}, Elena and {Smith}, Linda J. and {Tosi}, Monica},
        title = "{The emerging timescale of young star clusters regulated by cluster stellar mass}",
      journal = {Nature Astronomy},
         year = 2026,
        month = may,
          doi = {10.1038/s41550-026-02857-y},
archivePrefix = {arXiv},
       eprint = {2603.09867},
 primaryClass = {astro-ph.GA},
       adsurl = {https://ui.adsabs.harvard.edu/abs/2026NatAs.tmp..112P}
}

@ARTICLE{shabani18,
       author = {{Shabani}, F. and {Grebel}, E.~K. and {Pasquali}, A. and {D'Onghia}, E. and {Gallagher}, J.~S. and {Adamo}, A. and {Messa}, M. and {Elmegreen}, B.~G. and {Dobbs}, C. and {Gouliermis}, D.~A. and {Calzetti}, D. and {Grasha}, K. and {Elmegreen}, D.~M. and {Cignoni}, M. and {Dale}, D.~A. and {Aloisi}, A. and {Smith}, L.~J. and {Tosi}, M. and {Thilker}, D.~A. and {Lee}, J.~C. and {Sabbi}, E. and {Kim}, H. and {Pellerin}, A.},
        title = "{Search for star cluster age gradients across spiral arms of three LEGUS disc galaxies}",
      journal = {\mnras},
         year = 2018,
        month = aug,
       volume = {478},
       number = {3},
        pages = {3590-3604},
          doi = {10.1093/mnras/sty1277},
archivePrefix = {arXiv},
       eprint = {1805.05643},
 primaryClass = {astro-ph.GA},
       adsurl = {https://ui.adsabs.harvard.edu/abs/2018MNRAS.478.3590S}
}

@ARTICLE{williams21,
       author = {{Williams}, Thomas G. and {Schinnerer}, Eva and {Emsellem}, Eric and {Meidt}, Sharon and {Querejeta}, Miguel and {Belfiore}, Francesco and {Be{\v{s}}li{\'c}}, Ivana and {Bigiel}, Frank and {Chevance}, M{\'e}lanie and {Dale}, Daniel A. and {Glover}, Simon C.~O. and {Grasha}, Kathryn and {Klessen}, Ralf S. and {Kruijssen}, J.~M. Diederik and {Leroy}, Adam K. and {Pan}, Hsi-An and {Pety}, J{\'e}r{\^o}me and {Pessa}, Ismael and {Rosolowsky}, Erik and {Saito}, Toshiki and {Santoro}, Francesco and {Schruba}, Andreas and {Sormani}, Mattia C. and {Sun}, Jiayi and {Watkins}, Elizabeth J.},
        title = "{Applying the Tremaine-Weinberg Method to Nearby Galaxies: Stellar-mass-based Pattern Speeds and Comparisons with ISM Kinematics}",
      journal = {\aj},
         year = 2021,
        month = apr,
       volume = {161},
       number = {4},
          eid = {185},
        pages = {185},
          doi = {10.3847/1538-3881/abe243},
archivePrefix = {arXiv},
       eprint = {2102.01091},
 primaryClass = {astro-ph.GA},
       adsurl = {https://ui.adsabs.harvard.edu/abs/2021AJ....161..185W}
}

@ARTICLE{meidt21,
       author = {{Meidt}, Sharon E. and {Leroy}, Adam K. and {Querejeta}, Miguel and {Schinnerer}, Eva and {Sun}, Jiayi and {van der Wel}, Arjen and {Emsellem}, Eric and {Henshaw}, Jonathan and {Hughes}, Annie and {Kruijssen}, J.~M. Diederik and {Rosolowsky}, Erik and {Schruba}, Andreas and {Barnes}, Ashley and {Bigiel}, Frank and {Blanc}, Guillermo A. and {Chevance}, Melanie and {Cao}, Yixian and {Dale}, Daniel A. and {Faesi}, Christopher and {Glover}, Simon C.~O. and {Grasha}, Kathryn and {Groves}, Brent and {Herrera}, Cynthia and {Klessen}, Ralf S. and {Kreckel}, Kathryn and {Liu}, Daizhong and {Pan}, Hsi-An and {Pety}, Jerome and {Saito}, Toshiki and {Usero}, Antonio and {Watkins}, Elizabeth and {Williams}, Thomas G.},
        title = "{The Organization of Cloud-scale Gas Density Structure: High-resolution CO versus 3.6 {\ensuremath{\mu}}m Brightness Contrasts in Nearby Galaxies}",
      journal = {\apj},
         year = 2021,
        month = jun,
       volume = {913},
       number = {2},
          eid = {113},
        pages = {113},
          doi = {10.3847/1538-4357/abf35b},
archivePrefix = {arXiv},
       eprint = {2103.13247},
 primaryClass = {astro-ph.GA},
       adsurl = {https://ui.adsabs.harvard.edu/abs/2021ApJ...913..113M}
}

@ARTICLE{thilker00,
       author = {{Thilker}, David A. and {Braun}, Robert and {Walterbos}, Ren{\'e} A.~M.},
        title = "{HIIPHOT: Automated Photometry of H II Regions Applied to M51}",
      journal = {\aj},
         year = 2000,
        month = dec,
       volume = {120},
       number = {6},
        pages = {3070-3087},
          doi = {10.1086/316852},
archivePrefix = {arXiv},
       eprint = {astro-ph/0006026},
 primaryClass = {astro-ph},
       adsurl = {https://ui.adsabs.harvard.edu/abs/2000AJ....120.3070T}
}

@ARTICLE{scoville01,
       author = {{Scoville}, N.~Z. and {Polletta}, M. and {Ewald}, S. and {Stolovy}, S.~R. and {Thompson}, R. and {Rieke}, M.},
        title = "{High-Mass, OB Star Formation in M51: Hubble Space Telescope H{\ensuremath{\alpha}} and Pa{\ensuremath{\alpha}} Imaging}",
      journal = {\aj},
         year = 2001,
        month = dec,
       volume = {122},
       number = {6},
        pages = {3017-3045},
          doi = {10.1086/323445},
archivePrefix = {arXiv},
       eprint = {astro-ph/0108248},
 primaryClass = {astro-ph},
       adsurl = {https://ui.adsabs.harvard.edu/abs/2001AJ....122.3017S}
}

@ARTICLE{lada03,
       author = {{Lada}, Charles J. and {Lada}, Elizabeth A.},
        title = "{Embedded Clusters in Molecular Clouds}",
      journal = {\araa},
         year = 2003,
        month = jan,
       volume = {41},
        pages = {57-115},
          doi = {10.1146/annurev.astro.41.011802.094844},
archivePrefix = {arXiv},
       eprint = {astro-ph/0301540},
 primaryClass = {astro-ph},
       adsurl = {https://ui.adsabs.harvard.edu/abs/2003ARA&A..41...57L}
}

@ARTICLE{larsen04,
       author = {{Larsen}, S.~S.},
        title = "{The structure and environment of young stellar clusters in spiral galaxies}",
      journal = {\aap},
         year = 2004,
        month = mar,
       volume = {416},
        pages = {537-553},
          doi = {10.1051/0004-6361:20034533},
archivePrefix = {arXiv},
       eprint = {astro-ph/0312338},
 primaryClass = {astro-ph},
       adsurl = {https://ui.adsabs.harvard.edu/abs/2004A&A...416..537L}
}

@INPROCEEDINGS{elmegreen95,
       author = {{Elmegreen}, B.~G.},
        title = "{Density waves and star formation: is there triggering?}",
    booktitle = {The Formation of the Milky Way},
         year = 1995,
       editor = {{Alfaro}, E.~J. and {Delgado}, A.~J.},
        month = jan,
        pages = {28-38},
       adsurl = {https://ui.adsabs.harvard.edu/abs/1995fmw..conf...28E}
}

@ARTICLE{speights25,
       author = {{Speights}, Jason C. and {Aust}, Virginia and {Lu}, Qinyan},
        title = "{Spiral Density Waves in the Multiple-armed Galaxy NGC 628}",
      journal = {\apj},
         year = 2025,
        month = mar,
       volume = {981},
       number = {2},
          eid = {115},
        pages = {115},
          doi = {10.3847/1538-4357/adb1b3},
       adsurl = {https://ui.adsabs.harvard.edu/abs/2025ApJ...981..115S}
}

@ARTICLE{turner19,
       author = {{Turner}, J.~A. and {Dale}, D.~A. and {Adamo}, A. and {Calzetti}, D. and {Grasha}, K. and {Grebel}, E.~K. and {Johnson}, K.~E. and {Lee}, J.~C. and {Smith}, L.~J. and {Yoon}, I.},
        title = "{An ALMA/HST Study of Millimeter Dust Emission and Star Clusters}",
      journal = {\apj},
         year = 2019,
        month = oct,
       volume = {884},
       number = {2},
          eid = {112},
        pages = {112},
          doi = {10.3847/1538-4357/ab3faa},
archivePrefix = {arXiv},
       eprint = {1909.04629},
 primaryClass = {astro-ph.GA},
       adsurl = {https://ui.adsabs.harvard.edu/abs/2019ApJ...884..112T}
}

@ARTICLE{grand12,
       author = {{Grand}, Robert J.~J. and {Kawata}, Daisuke and {Cropper}, Mark},
        title = "{Dynamics of stars around spiral arms in an N-body/SPH simulated barred spiral galaxy}",
      journal = {\mnras},
         year = 2012,
        month = oct,
       volume = {426},
       number = {1},
        pages = {167-180},
          doi = {10.1111/j.1365-2966.2012.21733.x},
archivePrefix = {arXiv},
       eprint = {1202.6387},
 primaryClass = {astro-ph.GA},
       adsurl = {https://ui.adsabs.harvard.edu/abs/2012MNRAS.426..167G}
}

@ARTICLE{chevance22,
       author = {{Chevance}, M{\'e}lanie and {Kruijssen}, J.~M. Diederik and {Krumholz}, Mark R. and {Groves}, Brent and {Keller}, Benjamin W. and {Hughes}, Annie and {Glover}, Simon C.~O. and {Henshaw}, Jonathan D. and {Herrera}, Cinthya N. and {Kim}, Jaeyeon and {Leroy}, Adam K. and {Pety}, J{\'e}r{\^o}me and {Razza}, Alessandro and {Rosolowsky}, Erik and {Schinnerer}, Eva and {Schruba}, Andreas and {Barnes}, Ashley T. and {Bigiel}, Frank and {Blanc}, Guillermo A. and {Dale}, Daniel A. and {Emsellem}, Eric and {Faesi}, Christopher M. and {Grasha}, Kathryn and {Klessen}, Ralf S. and {Kreckel}, Kathryn and {Liu}, Daizhong and {Longmore}, Steven N. and {Meidt}, Sharon E. and {Querejeta}, Miguel and {Saito}, Toshiki and {Sun}, Jiayi and {Usero}, Antonio},
        title = "{Pre-supernova feedback mechanisms drive the destruction of molecular clouds in nearby star-forming disc galaxies}",
      journal = {\mnras},
         year = 2022,
        month = jan,
       volume = {509},
       number = {1},
        pages = {272-288},
          doi = {10.1093/mnras/stab2938},
archivePrefix = {arXiv},
       eprint = {2010.13788},
 primaryClass = {astro-ph.GA},
       adsurl = {https://ui.adsabs.harvard.edu/abs/2022MNRAS.509..272C}
}

@ARTICLE{romanelli25,
       author = {{Romanelli}, Andrea and {Chevance}, M{\'e}lanie and {Kruijssen}, J.~M. Diederik and {Ramambason}, Lise and {Querejeta}, Miguel and {Boquien}, Mederic and {Dale}, Daniel A. and {den Brok}, Jakob and {Glover}, Simon C.~O. and {Grasha}, Kathryn and {Hughes}, Annie and {Kim}, Jaeyeon and {Longmore}, Steven and {Meidt}, Sharon E. and {Mendez-Delgado}, Jos{\'e} Eduardo and {Neumann}, Lukas and {Pety}, J{\'e}r{\^o}me and {Schinnerer}, Eva and {Smith}, Rowan and {Sun}, Jiayi and {Williams}, Thomas G.},
        title = "{The impact of spiral arms on the star formation life cycle}",
      journal = {\aap},
         year = 2025,
        month = jun,
       volume = {698},
          eid = {A296},
        pages = {A296},
          doi = {10.1051/0004-6361/202553895},
archivePrefix = {arXiv},
       eprint = {2505.10908},
 primaryClass = {astro-ph.GA},
       adsurl = {https://ui.adsabs.harvard.edu/abs/2025A&A...698A.296R}
}

@ARTICLE{kim25,
       author = {{Kim}, Jaeyeon and {Chevance}, M{\'e}lanie and {Ramambason}, Lise and {Kreckel}, Kathryn and {Klessen}, Ralf S. and {Dale}, Daniel A. and {Leroy}, Adam K. and {Sandstrom}, Karin and {Chown}, Ryan and {Williams}, Thomas G. and {Sarbadhicary}, Sumit K. and {Belfiore}, Francesco and {Bigiel}, Frank and {Congiu}, Enrico and {Egorov}, Oleg V. and {Emsellem}, Eric and {Glover}, Simon C.~O. and {Grasha}, Kathryn and {Hughes}, Annie and {Kruijssen}, J.~M. Diederik and {Lee}, Janice C. and {Pathak}, Debosmita and {Pessa}, Ismael and {Rosolowsky}, Erik and {Sun}, Jiayi and {Sutter}, Jessica and {Thilker}, David A.},
        title = "{Timescales of Polycyclic Aromatic Hydrocarbon and Dust Continuum Emission from Gas Clouds Compared to Molecular Gas Cloud Lifetimes in PHANGS-JWST Galaxies}",
      journal = {\apj},
         year = 2025,
        month = aug,
       volume = {988},
       number = {2},
          eid = {215},
        pages = {215},
          doi = {10.3847/1538-4357/ade443},
archivePrefix = {arXiv},
       eprint = {2506.10063},
 primaryClass = {astro-ph.GA},
       adsurl = {https://ui.adsabs.harvard.edu/abs/2025ApJ...988..215K}
}

@ARTICLE{Ramambason26,
       author = {{Ramambason}, Lise and {Chevance}, M{\'e}lanie and {Kim}, Jaeyeon and {Belfiore}, Francesco and {Kruijssen}, J.~M. Diederik and {Romanelli}, Andrea and {Amiri}, Amirnezam and {Boquien}, M{\'e}d{\'e}ric and {Chown}, Ryan and {Dale}, Daniel A. and {Dlamini}, Simthembile and {Egorov}, Oleg V. and {Gerasimov}, Ivan and {Glover}, Simon C.~O. and {Grasha}, Kathryn and {Hassani}, Hamid and {Kim}, Hwihyun and {Kreckel}, Kathryn and {Koziol}, Hannah and {Leroy}, Adam K. and {M{\'e}ndez-Delgado}, Jos{\'e} Eduardo and {Neumann}, Justus and {Neumann}, Lukas and {Pan}, Hsi-An and {Pathak}, Debosmita and {Sandstrom}, Karin and {Sarbadhicary}, Sumit K. and {Schinnerer}, Eva and {Sun}, Jiayi and {Sutter}, Jessica and {Thilker}, David A. and {Ubeda}, Leonardo and {Weinbeck}, Tony D. and {Whitmore}, Bradley C. and {Williams}, Thomas G.},
        title = "{Duration and properties of the embedded phase of star formation in 37 nearby galaxies from PHANGS-JWST}",
      journal = {\aap},
         year = 2026,
        month = feb,
       volume = {706},
          eid = {A186},
        pages = {A186},
          doi = {10.1051/0004-6361/202556206},
archivePrefix = {arXiv},
       eprint = {2507.01508},
 primaryClass = {astro-ph.GA},
       adsurl = {https://ui.adsabs.harvard.edu/abs/2026A&A...706A.186R}
}

@ARTICLE{rodriguez25,
       author = {{Rodr{\'\i}guez}, M. Jimena and {Lee}, Janice C. and {Indebetouw}, Remy and {Whitmore}, B.~C. and {Maschmann}, Daniel and {Williams}, Thomas G. and {Chandar}, Rupali and {Barnes}, A.~T. and {Gnedin}, Oleg Y. and {Sandstrom}, Karin M. and {Rosolowsky}, Erik and {Leroy}, Adam K. and {Thilker}, David A. and {Kim}, Hwihyun and {Sun}, Jiayi and {Klessen}, Ralf S. and {Groves}, Brent and {Wofford}, Aida and {Boquien}, M{\'e}d{\'e}ric and {Dale}, Daniel A. and {{\'U}beda}, Leonardo and {Larson}, Kirsten L. and {Grasha}, Kathryn and {Johnson}, Kelsey E. and {Levy}, Rebecca C. and {Bigiel}, Frank and {Hassani}, Hamid and {Sarbadhicary}, Sumit K.},
        title = "{Tracing the Earliest Stages of Star and Cluster Formation in 19 Nearby Galaxies with PHANGS-JWST and HST: Compact 3.3 {\ensuremath{\mu}}m Polycyclic Aromatic Hydrocarbon Emitters and Their Relation to the Optical Census of Star Clusters}",
      journal = {\apj},
         year = 2025,
        month = apr,
       volume = {983},
       number = {2},
          eid = {137},
        pages = {137},
          doi = {10.3847/1538-4357/adbb69},
archivePrefix = {arXiv},
       eprint = {2412.07862},
 primaryClass = {astro-ph.GA},
       adsurl = {https://ui.adsabs.harvard.edu/abs/2025ApJ...983..137R}
}

@ARTICLE{elmegreen00,
       author = {{Elmegreen}, Bruce G.},
        title = "{Star Formation in a Crossing Time}",
      journal = {\apj},
         year = 2000,
        month = feb,
       volume = {530},
       number = {1},
        pages = {277-281},
          doi = {10.1086/308361},
archivePrefix = {arXiv},
       eprint = {astro-ph/9911172},
 primaryClass = {astro-ph},
       adsurl = {https://ui.adsabs.harvard.edu/abs/2000ApJ...530..277E}
}

@ARTICLE{menon21,
       author = {{Menon}, Shyam H. and {Grasha}, Kathryn and {Elmegreen}, Bruce G. and {Federrath}, Christoph and {Krumholz}, Mark R. and {Calzetti}, Daniela and {S{\'a}nchez}, N{\'e}stor and {Linden}, Sean T. and {Adamo}, Angela and {Messa}, Matteo and {Cook}, David O. and {Dale}, Daniel A. and {Grebel}, Eva K. and {Fumagalli}, Michele and {Sabbi}, Elena and {Johnson}, Kelsey E. and {Smith}, Linda J. and {Kennicutt}, Robert C.},
        title = "{The dependence of the hierarchical distribution of star clusters on galactic environment}",
      journal = {\mnras},
         year = 2021,
        month = nov,
       volume = {507},
       number = {4},
        pages = {5542-5566},
          doi = {10.1093/mnras/stab2413},
archivePrefix = {arXiv},
       eprint = {2108.04387},
 primaryClass = {astro-ph.GA},
       adsurl = {https://ui.adsabs.harvard.edu/abs/2021MNRAS.507.5542M}
}

@ARTICLE{delmegreen80b,
       author = {{Elmegreen}, D.~M.},
        title = "{Properties of spurs in spiral galaxies}",
      journal = {\apj},
         year = 1980,
        month = dec,
       volume = {242},
        pages = {528-532},
          doi = {10.1086/158486},
       adsurl = {https://ui.adsabs.harvard.edu/abs/1980ApJ...242..528E}
}

@ARTICLE{elmegreen79,
       author = {{Elmegreen}, B.~G.},
        title = "{Gravitational collapse in dust lanes and the appearance of spiral structure in galaxies.}",
      journal = {\apj},
         year = 1979,
        month = jul,
       volume = {231},
        pages = {372-383},
          doi = {10.1086/157200},
       adsurl = {https://ui.adsabs.harvard.edu/abs/1979ApJ...231..372E}
}

@ARTICLE{kim02a,
       author = {{Kim}, Woong-Tae and {Ostriker}, Eve C.},
        title = "{Formation and Fragmentation of Gaseous Spurs in Spiral Galaxies}",
      journal = {\apj},
         year = 2002,
        month = may,
       volume = {570},
       number = {1},
        pages = {132-151},
          doi = {10.1086/339352},
archivePrefix = {arXiv},
       eprint = {astro-ph/0111398},
 primaryClass = {astro-ph},
       adsurl = {https://ui.adsabs.harvard.edu/abs/2002ApJ...570..132K}
}

@ARTICLE{shetty06,
       author = {{Shetty}, Rahul and {Ostriker}, Eve C.},
        title = "{Global Modeling of Spur Formation in Spiral Galaxies}",
      journal = {\apj},
         year = 2006,
        month = aug,
       volume = {647},
       number = {2},
        pages = {997-1017},
          doi = {10.1086/505594},
archivePrefix = {arXiv},
       eprint = {astro-ph/0605142},
 primaryClass = {astro-ph},
       adsurl = {https://ui.adsabs.harvard.edu/abs/2006ApJ...647..997S}
}

@ARTICLE{kim02b,
       author = {{Kim}, Woong-Tae and {Ostriker}, Eve C. and {Stone}, James M.},
        title = "{Three-dimensional Simulations of Parker, Magneto-Jeans, and Swing Instabilities in Shearing Galactic Gas Disks}",
      journal = {\apj},
         year = 2002,
        month = dec,
       volume = {581},
       number = {2},
        pages = {1080-1100},
          doi = {10.1086/344367},
archivePrefix = {arXiv},
       eprint = {astro-ph/0208414},
 primaryClass = {astro-ph},
       adsurl = {https://ui.adsabs.harvard.edu/abs/2002ApJ...581.1080K}
}

@ARTICLE{kim06a,
       author = {{Kim}, Woong-Tae and {Ostriker}, Eve C.},
        title = "{Formation of Spiral-Arm Spurs and Bound Clouds in Vertically Stratified Galactic Gas Disks}",
      journal = {\apj},
         year = 2006,
        month = jul,
       volume = {646},
       number = {1},
        pages = {213-231},
          doi = {10.1086/504677},
archivePrefix = {arXiv},
       eprint = {astro-ph/0603751},
 primaryClass = {astro-ph},
       adsurl = {https://ui.adsabs.harvard.edu/abs/2006ApJ...646..213K}
}

@ARTICLE{balbus88,
       author = {{Balbus}, Steven A.},
        title = "{Local Interstellar Gasdynamical Stability and Substructure in Spiral Arms}",
      journal = {\apj},
         year = 1988,
        month = jan,
       volume = {324},
        pages = {60},
          doi = {10.1086/165880},
       adsurl = {https://ui.adsabs.harvard.edu/abs/1988ApJ...324...60B}
}

@INPROCEEDINGS{lynds70,
       author = {{Lynds}, B.~T.},
        title = "{The Distribution of Dark Nebulae in Late-Type Spirals}",
    booktitle = {The Spiral Structure of our Galaxy},
         year = 1970,
       editor = {{Becker}, Wilhelm and {Kontopoulos}, Georgios Ioannou},
       series = {IAU Symposium},
       volume = {38},
        month = jan,
        pages = {26},
       adsurl = {https://ui.adsabs.harvard.edu/abs/1970IAUS...38...26L}
}

@ARTICLE{corder08,
       author = {{Corder}, Stuartt and {Sheth}, Kartik and {Scoville}, Nicholas Z. and {Koda}, Jin and {Vogel}, Stuart N. and {Ostriker}, Eve},
        title = "{Detection of Dense Molecular Gas in Interarm Spurs in M51}",
      journal = {\apj},
         year = 2008,
        month = dec,
       volume = {689},
       number = {1},
        pages = {148-152},
          doi = {10.1086/592557},
archivePrefix = {arXiv},
       eprint = {0811.2903},
 primaryClass = {astro-ph},
       adsurl = {https://ui.adsabs.harvard.edu/abs/2008ApJ...689..148C}
}

@INPROCEEDINGS{weaver70b,
       author = {{Weaver}, H.},
        title = "{Spiral Structure of the Galaxy Derived from the Hat Creek Survey of Neutral Hydrogen}",
    booktitle = {The Spiral Structure of our Galaxy},
         year = 1970,
       editor = {{Becker}, Wilhelm and {Kontopoulos}, Georgios Ioannou},
       series = {IAU Symposium},
       volume = {38},
        month = jan,
        pages = {126},
       adsurl = {https://ui.adsabs.harvard.edu/abs/1970IAUS...38..126W}
}

@ARTICLE{lavigne06,
       author = {{La Vigne}, Misty A. and {Vogel}, Stuart N. and {Ostriker}, Eve C.},
        title = "{A Hubble Space Telescope Archival Survey of Feathers in Spiral Galaxies}",
      journal = {\apj},
         year = 2006,
        month = oct,
       volume = {650},
       number = {2},
        pages = {818-834},
          doi = {10.1086/506589},
archivePrefix = {arXiv},
       eprint = {astro-ph/0606761},
 primaryClass = {astro-ph},
       adsurl = {https://ui.adsabs.harvard.edu/abs/2006ApJ...650..818L}
}

@INPROCEEDINGS{weaver70a,
       author = {{Weaver}, H.~F.},
        title = "{Some Characteristics of Interstellar Gas in the Galaxy}",
    booktitle = {Interstellar Gas Dynamics},
         year = 1970,
       editor = {{Habing}, Harm Jan},
       series = {IAU Symposium},
       volume = {39},
        month = jan,
        pages = {22},
       adsurl = {https://ui.adsabs.harvard.edu/abs/1970IAUS...39...22W}
}

@ARTICLE{elmegreen83,
       author = {{Elmegreen}, B.~G. and {Elmegreen}, D.~M.},
        title = "{Regular strings of HII regions and superclouds in spiral galaxies : clues to the origin of cloudy structure.}",
      journal = {\mnras},
         year = 1983,
        month = apr,
       volume = {203},
        pages = {31-45},
          doi = {10.1093/mnras/203.1.31},
       adsurl = {https://ui.adsabs.harvard.edu/abs/1983MNRAS.203...31E}
}

@ARTICLE{elmegreen87a,
       author = {{Elmegreen}, Bruce G. and {Elmegreen}, Debra Meloy},
        title = "{H i Superclouds in the Inner Galaxy}",
      journal = {\apj},
         year = 1987,
        month = sep,
       volume = {320},
        pages = {182},
          doi = {10.1086/165534},
       adsurl = {https://ui.adsabs.harvard.edu/abs/1987ApJ...320..182E}
}

@ARTICLE{allen73,
       author = {{Allen}, R.~J. and {Goss}, W.~M. and {van Woerden}, H.},
        title = "{The giant spiral galaxy M101: I. A high resolution map of the neutral hydrogen.}",
      journal = {\aap},
         year = 1973,
        month = jan,
       volume = {29},
        pages = {447-451},
       adsurl = {https://ui.adsabs.harvard.edu/abs/1973A&A....29..447A}
}

@ARTICLE{mcgee64,
       author = {{McGee}, R.~X. and {Milton}, Janice A.},
        title = "{A sky survey of neutral hydrogen at {\ensuremath{\lambda}} 21 cm. III. Gas at higher radial velocities}",
      journal = {Australian Journal of Physics},
         year = 1964,
        month = mar,
       volume = {17},
        pages = {128},
          doi = {10.1071/PH640128},
       adsurl = {https://ui.adsabs.harvard.edu/abs/1964AuJPh..17..128M}
}

@ARTICLE{henderson82,
       author = {{Henderson}, A.~P. and {Jackson}, P.~D. and {Kerr}, F.~J.},
        title = "{The distribution of neutral atomic hydrogen in our galaxy beyond the solar circle}",
      journal = {\apj},
         year = 1982,
        month = dec,
       volume = {263},
        pages = {116-122},
          doi = {10.1086/160486},
       adsurl = {https://ui.adsabs.harvard.edu/abs/1982ApJ...263..116H}
}

@ARTICLE{rand90,
       author = {{Rand}, Richard J. and {Kulkarni}, Shrinivas R.},
        title = "{M51: Molecular Spiral Arms, Giant Molecular Associations, and Superclouds}",
      journal = {\apjl},
         year = 1990,
        month = feb,
       volume = {349},
        pages = {L43},
          doi = {10.1086/185647},
       adsurl = {https://ui.adsabs.harvard.edu/abs/1990ApJ...349L..43R}
}

@ARTICLE{grabelsky87,
       author = {{Grabelsky}, D.~A. and {Cohen}, R.~S. and {Bronfman}, L. and {Thaddeus}, P. and {May}, J.},
        title = "{Molecular Clouds in the Carina Arm: Large-Scale Properties of Molecular Gas and Comparison with H i}",
      journal = {\apj},
         year = 1987,
        month = apr,
       volume = {315},
        pages = {122},
          doi = {10.1086/165118},
       adsurl = {https://ui.adsabs.harvard.edu/abs/1987ApJ...315..122G}
}

@ARTICLE{park23,
       author = {{Park}, Geumsook and {Koo}, Bon-Chul and {Kim}, Kee-Tae and {Elmegreen}, Bruce},
        title = "{Neutral Atomic and Molecular Clouds and Star Formation in the Outer Carina Arm}",
      journal = {\apj},
         year = 2023,
        month = sep,
       volume = {955},
       number = {1},
          eid = {59},
        pages = {59},
          doi = {10.3847/1538-4357/acebda},
archivePrefix = {arXiv},
       eprint = {2308.01577},
 primaryClass = {astro-ph.GA},
       adsurl = {https://ui.adsabs.harvard.edu/abs/2023ApJ...955...59P}
}

@INPROCEEDINGS{belmegreen80,
       author = {{Elmegreen}, B.~G. and {Elmegreen}, D.~M.},
        title = "{Optical and theoretical studies of giant clouds in spiral galaxies}",
    booktitle = {Interstellar Molecules},
         year = 1980,
       editor = {{Andrew}, B.~H.},
       series = {IAU Symposium},
       volume = {87},
        month = jan,
        pages = {191-194},
       adsurl = {https://ui.adsabs.harvard.edu/abs/1980IAUS...87..191E}
}

@ARTICLE{delmegreen80,
       author = {{Elmegreen}, D.~M.},
        title = "{An optical analysis of dust complexes in spiral galaxies}",
      journal = {\apjs},
         year = 1980,
        month = may,
       volume = {43},
        pages = {37-56},
          doi = {10.1086/190666},
       adsurl = {https://ui.adsabs.harvard.edu/abs/1980ApJS...43...37E}
}

@ARTICLE{dobbs06,
       author = {{Dobbs}, C.~L. and {Bonnell}, I.~A.},
        title = "{Spurs and feathering in spiral galaxies}",
      journal = {\mnras},
         year = 2006,
        month = apr,
       volume = {367},
       number = {3},
        pages = {873-878},
          doi = {10.1111/j.1365-2966.2006.10146.x},
archivePrefix = {arXiv},
       eprint = {astro-ph/0602100},
 primaryClass = {astro-ph},
       adsurl = {https://ui.adsabs.harvard.edu/abs/2006MNRAS.367..873D}
}

@ARTICLE{lee12,
       author = {{Lee}, Wing-Kit and {Shu}, Frank H.},
        title = "{Feathering Instability of Spiral Arms. I. Formulation of the Problem}",
      journal = {\apj},
         year = 2012,
        month = sep,
       volume = {756},
       number = {1},
          eid = {45},
        pages = {45},
          doi = {10.1088/0004-637X/756/1/45},
archivePrefix = {arXiv},
       eprint = {1207.0875},
 primaryClass = {astro-ph.GA},
       adsurl = {https://ui.adsabs.harvard.edu/abs/2012ApJ...756...45L}
}

@ARTICLE{li13,
       author = {{Li}, Guang-Xing and {Wyrowski}, Friedrich and {Menten}, Karl and {Belloche}, Arnaud},
        title = "{A 500 pc filamentary gas wisp in the disk of the Milky Way}",
      journal = {\aap},
         year = 2013,
        month = nov,
       volume = {559},
          eid = {A34},
        pages = {A34},
          doi = {10.1051/0004-6361/201322411},
archivePrefix = {arXiv},
       eprint = {1310.3267},
 primaryClass = {astro-ph.GA},
       adsurl = {https://ui.adsabs.harvard.edu/abs/2013A&A...559A..34L}
}

@ARTICLE{honig15,
       author = {{Honig}, Z.~N. and {Reid}, M.~J.},
        title = "{Characteristics of Spiral Arms in Late-type Galaxies}",
      journal = {\apj},
         year = 2015,
        month = feb,
       volume = {800},
       number = {1},
          eid = {53},
        pages = {53},
          doi = {10.1088/0004-637X/800/1/53},
archivePrefix = {arXiv},
       eprint = {1412.1012},
 primaryClass = {astro-ph.GA},
       adsurl = {https://ui.adsabs.harvard.edu/abs/2015ApJ...800...53H}
}

@ARTICLE{du17,
       author = {{Du}, Xinyu and {Xu}, Ye and {Yang}, Ji and {Sun}, Yan},
        title = "{The Molecular Structures of the Local Arm and Perseus Arm in the Galactic Region of l = [139.{\textdegree}75, 149.{\textdegree}75], b = [-5.{\textdegree}25, 5.{\textdegree}25]}",
      journal = {\apjs},
         year = 2017,
        month = apr,
       volume = {229},
       number = {2},
          eid = {24},
        pages = {24},
          doi = {10.3847/1538-4365/aa5d9d},
archivePrefix = {arXiv},
       eprint = {1702.00544},
 primaryClass = {astro-ph.GA},
       adsurl = {https://ui.adsabs.harvard.edu/abs/2017ApJS..229...24D}
}

@ARTICLE{jackson10,
       author = {{Jackson}, James M. and {Finn}, Susanna C. and {Chambers}, Edward T. and {Rathborne}, Jill M. and {Simon}, Robert},
        title = "{The ``Nessie'' Nebula: Cluster Formation in a Filamentary Infrared Dark Cloud}",
      journal = {\apjl},
         year = 2010,
        month = aug,
       volume = {719},
       number = {2},
        pages = {L185-L189},
          doi = {10.1088/2041-8205/719/2/L185},
archivePrefix = {arXiv},
       eprint = {1007.5492},
 primaryClass = {astro-ph.GA},
       adsurl = {https://ui.adsabs.harvard.edu/abs/2010ApJ...719L.185J}
}

@ARTICLE{goodman14,
       author = {{Goodman}, Alyssa A. and {Alves}, Jo{\~a}o and {Beaumont}, Christopher N. and {Benjamin}, Robert A. and {Borkin}, Michelle A. and {Burkert}, Andreas and {Dame}, Thomas M. and {Jackson}, James and {Kauffmann}, Jens and {Robitaille}, Thomas and {Smith}, Rowan J.},
        title = "{The Bones of the Milky Way}",
      journal = {\apj},
         year = 2014,
        month = dec,
       volume = {797},
       number = {1},
          eid = {53},
        pages = {53},
          doi = {10.1088/0004-637X/797/1/53},
archivePrefix = {arXiv},
       eprint = {1408.0001},
 primaryClass = {astro-ph.GA},
       adsurl = {https://ui.adsabs.harvard.edu/abs/2014ApJ...797...53G}
}

@ARTICLE{xu16,
       author = {{Xu}, Y. and {Reid}, M. and {Dame}, T. and {Menten}, K. and {Sakai}, N. and {Li}, J. and {Brunthaler}, A. and {Moscadelli}, L. and {Zhang}, B. and {Zheng}, X.},
        title = "{The local spiral structure of the Milky Way}",
      journal = {Science Advances},
         year = 2016,
        month = sep,
       volume = {2},
       number = {9},
        pages = {e1600878-e1600878},
          doi = {10.1126/sciadv.1600878},
archivePrefix = {arXiv},
       eprint = {1610.00242},
 primaryClass = {astro-ph.GA},
       adsurl = {https://ui.adsabs.harvard.edu/abs/2016SciA....2E0878X}
}

@ARTICLE{battersby14,
       author = {{Battersby}, Cara and {Ginsburg}, Adam and {Bally}, John and {Longmore}, Steve and {Dunham}, Miranda and {Darling}, Jeremy},
        title = "{The Onset of Massive Star Formation: The Evolution of Temperature and Density Structure in an Infrared Dark Cloud}",
      journal = {\apj},
         year = 2014,
        month = jun,
       volume = {787},
       number = {2},
          eid = {113},
        pages = {113},
          doi = {10.1088/0004-637X/787/2/113},
archivePrefix = {arXiv},
       eprint = {1405.3286},
 primaryClass = {astro-ph.GA},
       adsurl = {https://ui.adsabs.harvard.edu/abs/2014ApJ...787..113B}
}

@ARTICLE{arzoumanian11,
       author = {{Arzoumanian}, D. and {Andr{\'e}}, Ph. and {Didelon}, P. and {K{\"o}nyves}, V. and {Schneider}, N. and {Men'shchikov}, A. and {Sousbie}, T. and {Zavagno}, A. and {Bontemps}, S. and {di Francesco}, J. and {Griffin}, M. and {Hennemann}, M. and {Hill}, T. and {Kirk}, J. and {Martin}, P. and {Minier}, V. and {Molinari}, S. and {Motte}, F. and {Peretto}, N. and {Pezzuto}, S. and {Spinoglio}, L. and {Ward-Thompson}, D. and {White}, G. and {Wilson}, C.~D.},
        title = "{Characterizing interstellar filaments with Herschel in IC 5146}",
      journal = {\aap},
         year = 2011,
        month = may,
       volume = {529},
          eid = {L6},
        pages = {L6},
          doi = {10.1051/0004-6361/201116596},
archivePrefix = {arXiv},
       eprint = {1103.0201},
 primaryClass = {astro-ph.GA},
       adsurl = {https://ui.adsabs.harvard.edu/abs/2011A&A...529L...6A}
}

@ARTICLE{he26,
       author = {{He}, Qian-Ru and {Kim}, Won-Ju and {Fuller}, Gary A. and {Traficante}, Alessio and {Clarke}, Seamus D. and {Gao}, Yu and {Chen}, Xue-Peng and {Fang}, Min and {Wang}, Ke and {Chen}, En and {Baug}, Tapas and {Wang}, Xiao-Long and {Wang}, Chen and {Wang}, Yong-Xiong},
        title = "{The Structure of an 80 Parsec Long Massive Filament}",
      journal = {\aj},
         year = 2026,
        month = apr,
       volume = {171},
       number = {4},
          eid = {234},
        pages = {234},
          doi = {10.3847/1538-3881/ae3dd6},
archivePrefix = {arXiv},
       eprint = {2601.13616},
 primaryClass = {astro-ph.GA},
       adsurl = {https://ui.adsabs.harvard.edu/abs/2026AJ....171..234H}
}

@ARTICLE{zucker18,
       author = {{Zucker}, Catherine and {Battersby}, Cara and {Goodman}, Alyssa},
        title = "{Physical Properties of Large-scale Galactic Filaments}",
      journal = {\apj},
         year = 2018,
        month = sep,
       volume = {864},
       number = {2},
          eid = {153},
        pages = {153},
          doi = {10.3847/1538-4357/aacc66},
archivePrefix = {arXiv},
       eprint = {1712.09655},
 primaryClass = {astro-ph.GA},
       adsurl = {https://ui.adsabs.harvard.edu/abs/2018ApJ...864..153Z}
}

@ARTICLE{zucker15,
       author = {{Zucker}, Catherine and {Battersby}, Cara and {Goodman}, Alyssa},
        title = "{The Skeleton of the Milky Way}",
      journal = {\apj},
         year = 2015,
        month = dec,
       volume = {815},
       number = {1},
          eid = {23},
        pages = {23},
          doi = {10.1088/0004-637X/815/1/23},
archivePrefix = {arXiv},
       eprint = {1506.08807},
 primaryClass = {astro-ph.GA},
       adsurl = {https://ui.adsabs.harvard.edu/abs/2015ApJ...815...23Z}
}

@ARTICLE{alves20,
       author = {{Alves}, Jo{\~a}o and {Zucker}, Catherine and {Goodman}, Alyssa A. and {Speagle}, Joshua S. and {Meingast}, Stefan and {Robitaille}, Thomas and {Finkbeiner}, Douglas P. and {Schlafly}, Edward F. and {Green}, Gregory M.},
        title = "{A Galactic-scale gas wave in the solar neighbourhood}",
      journal = {\nat},
         year = 2020,
        month = feb,
       volume = {578},
       number = {7794},
        pages = {237-239},
          doi = {10.1038/s41586-019-1874-z},
archivePrefix = {arXiv},
       eprint = {2001.08748},
 primaryClass = {astro-ph.GA},
       adsurl = {https://ui.adsabs.harvard.edu/abs/2020Natur.578..237A}
}

@ARTICLE{kim20,
       author = {{Kim}, Woong-Tae and {Kim}, Chang-Goo and {Ostriker}, Eve C.},
        title = "{Local Simulations of Spiral Galaxies with the TIGRESS Framework. I. Star Formation and Arm Spurs/Feathers}",
      journal = {\apj},
         year = 2020,
        month = jul,
       volume = {898},
       number = {1},
          eid = {35},
        pages = {35},
          doi = {10.3847/1538-4357/ab9b87},
archivePrefix = {arXiv},
       eprint = {2006.05614},
 primaryClass = {astro-ph.GA},
       adsurl = {https://ui.adsabs.harvard.edu/abs/2020ApJ...898...35K}
}

@ARTICLE{koda09,
       author = {{Koda}, Jin and {Scoville}, Nick and {Sawada}, Tsuyoshi and {La Vigne}, Misty A. and {Vogel}, Stuart N. and {Potts}, Ashley E. and {Carpenter}, John M. and {Corder}, Stuartt A. and {Wright}, Melvyn C.~H. and {White}, Stephen M. and {Zauderer}, B. Ashley and {Patience}, Jenny and {Sargent}, Anneila I. and {Bock}, Douglas C.~J. and {Hawkins}, David and {Hodges}, Mark and {Kemball}, Athol and {Lamb}, James W. and {Plambeck}, Richard L. and {Pound}, Marc W. and {Scott}, Stephen L. and {Teuben}, Peter and {Woody}, David P.},
        title = "{Dynamically Driven Evolution of the Interstellar Medium in M51}",
      journal = {\apjl},
         year = 2009,
        month = aug,
       volume = {700},
       number = {2},
        pages = {L132-L136},
          doi = {10.1088/0004-637X/700/2/L132},
archivePrefix = {arXiv},
       eprint = {0907.1656},
 primaryClass = {astro-ph.CO},
       adsurl = {https://ui.adsabs.harvard.edu/abs/2009ApJ...700L.132K}
}

@ARTICLE{koda23,
       author = {{Koda}, Jin and {Hirota}, Akihiko and {Egusa}, Fumi and {Sakamoto}, Kazushi and {Sawada}, Tsuyoshi and {Heyer}, Mark and {Baba}, Junichi and {Boissier}, Samuel and {Calzetti}, Daniela and {Donovan Meyer}, Jennifer and {Elmegreen}, Bruce G. and {Gil de Paz}, Armando and {Harada}, Nanase and {Ho}, Luis C. and {Kobayashi}, Masato I.~N. and {Kuno}, Nario and {Lee}, Amanda M. and {Madore}, Barry F. and {Maeda}, Fumiya and {Mart{\'\i}n}, Sergio and {Muraoka}, Kazuyuki and {Nakanishi}, Kouichiro and {Onodera}, Sachiko and {Pineda}, Jorge L. and {Scoville}, Nick and {Watanabe}, Yoshimasa},
        title = "{Diverse Molecular Structures across the Whole Star-forming Disk of M83: High-fidelity Imaging at 40 pc Resolution}",
      journal = {\apj},
         year = 2023,
        month = jun,
       volume = {949},
       number = {2},
          eid = {108},
        pages = {108},
          doi = {10.3847/1538-4357/acc65e},
archivePrefix = {arXiv},
       eprint = {2303.12108},
 primaryClass = {astro-ph.GA},
       adsurl = {https://ui.adsabs.harvard.edu/abs/2023ApJ...949..108K}
}

@ARTICLE{wada11,
       author = {{Wada}, Keiichi and {Baba}, Junichi and {Saitoh}, Takayuki R.},
        title = "{Interplay between Stellar Spirals and the Interstellar Medium in Galactic Disks}",
      journal = {\apj},
         year = 2011,
        month = jul,
       volume = {735},
       number = {1},
          eid = {1},
        pages = {1},
          doi = {10.1088/0004-637X/735/1/1},
archivePrefix = {arXiv},
       eprint = {1104.1287},
 primaryClass = {astro-ph.GA},
       adsurl = {https://ui.adsabs.harvard.edu/abs/2011ApJ...735....1W}
}

@ARTICLE{lada88,
       author = {{Lada}, Charles J. and {Margulis}, Michael and {Sofue}, Yoshiaki and {Nakai}, Naomasa and {Handa}, Toshihiro},
        title = "{Observations of Molecular and Atomic Clouds in M31}",
      journal = {\apj},
         year = 1988,
        month = may,
       volume = {328},
        pages = {143},
          doi = {10.1086/166275},
       adsurl = {https://ui.adsabs.harvard.edu/abs/1988ApJ...328..143L}
}

@ARTICLE{schneider79,
       author = {{Schneider}, S. and {Elmegreen}, B.~G.},
        title = "{A catalog of dark globular filaments.}",
      journal = {\apjs},
         year = 1979,
        month = sep,
       volume = {41},
        pages = {87-95},
          doi = {10.1086/190609},
       adsurl = {https://ui.adsabs.harvard.edu/abs/1979ApJS...41...87S}
}

@INPROCEEDINGS{elmegreen87b,
       author = {{Elmegreen}, Bruce G.},
        title = "{Large scale star formation - Density waves, superassociations and propagation}",
    booktitle = {Star Forming Regions},
         year = 1987,
       editor = {{Peimbert}, Manuel and {Jugaku}, Jun},
       series = {IAU Symposium},
       volume = {115},
        month = jan,
        pages = {457-479},
       adsurl = {https://ui.adsabs.harvard.edu/abs/1987IAUS..115..457E}
}

@ARTICLE{willner04,
       author = {{Willner}, S.~P. and {Ashby}, M.~L.~N. and {Barmby}, P. and {Fazio}, G.~G. and {Pahre}, M. and {Smith}, H.~A. and {Kennicutt}, Jr., Robert C. and {Calzetti}, Daniela and {Dale}, Daniel A. and {Draine}, B.~T. and {Regan}, Michael W. and {Malhotra}, S. and {Thornley}, Michele D. and {Appleton}, P.~N. and {Frayer}, D. and {Helou}, G. and {Stolovy}, S. and {Storrie-Lombardi}, L.},
        title = "{Infrared Array Camera (IRAC) Observations of M81}",
      journal = {\apjs},
         year = 2004,
        month = sep,
       volume = {154},
       number = {1},
        pages = {222-228},
          doi = {10.1086/422913},
archivePrefix = {arXiv},
       eprint = {astro-ph/0405626},
 primaryClass = {astro-ph},
       adsurl = {https://ui.adsabs.harvard.edu/abs/2004ApJS..154..222W}
}

@ARTICLE{schinnerer13,
       author = {{Schinnerer}, Eva and {Meidt}, Sharon E. and {Pety}, J{\'e}r{\^o}me and {Hughes}, Annie and {Colombo}, Dario and {Garc{\'\i}a-Burillo}, Santiago and {Schuster}, Karl F. and {Dumas}, Ga{\"e}lle and {Dobbs}, Clare L. and {Leroy}, Adam K. and {Kramer}, Carsten and {Thompson}, Todd A. and {Regan}, Michael W.},
        title = "{The PdBI Arcsecond Whirlpool Survey (PAWS). I. A Cloud-scale/Multi-wavelength View of the Interstellar Medium in a Grand-design Spiral Galaxy}",
      journal = {\apj},
         year = 2013,
        month = dec,
       volume = {779},
       number = {1},
          eid = {42},
        pages = {42},
          doi = {10.1088/0004-637X/779/1/42},
archivePrefix = {arXiv},
       eprint = {1304.1801},
 primaryClass = {astro-ph.CO},
       adsurl = {https://ui.adsabs.harvard.edu/abs/2013ApJ...779...42S}
}

@ARTICLE{schinnerer17,
       author = {{Schinnerer}, Eva and {Meidt}, Sharon E. and {Colombo}, Dario and {Chandar}, Rupali and {Dobbs}, Clare L. and {Garc{\'\i}a-Burillo}, Santiago and {Hughes}, Annie and {Leroy}, Adam K. and {Pety}, J{\'e}r{\^o}me and {Querejeta}, Miguel and {Kramer}, Carsten and {Schuster}, Karl F.},
        title = "{The PdBI Arcsecond Whirlpool Survey (PAWS): The Role of Spiral Arms in Cloud and Star Formation}",
      journal = {\apj},
         year = 2017,
        month = feb,
       volume = {836},
       number = {1},
          eid = {62},
        pages = {62},
          doi = {10.3847/1538-4357/836/1/62},
archivePrefix = {arXiv},
       eprint = {1701.02184},
 primaryClass = {astro-ph.GA},
       adsurl = {https://ui.adsabs.harvard.edu/abs/2017ApJ...836...62S}
}

@ARTICLE{dobbs10,
       author = {{Dobbs}, C.~L. and {Pringle}, J.~E.},
        title = "{Age distributions of star clusters in spiral and barred galaxies as a test for theories of spiral structure}",
      journal = {\mnras},
         year = 2010,
        month = nov,
       volume = {409},
       number = {1},
        pages = {396-404},
          doi = {10.1111/j.1365-2966.2010.17323.x},
archivePrefix = {arXiv},
       eprint = {1007.1399},
 primaryClass = {astro-ph.GA},
       adsurl = {https://ui.adsabs.harvard.edu/abs/2010MNRAS.409..396D}
}

@ARTICLE{chandar11,
       author = {{Chandar}, Rupali and {Whitmore}, Bradley C. and {Calzetti}, Daniela and {Di Nino}, Daiana and {Kennicutt}, Robert C. and {Regan}, Michael and {Schinnerer}, Eva},
        title = "{New Constraints on Mass-dependent Disruption of Star Clusters in M51}",
      journal = {\apj},
         year = 2011,
        month = feb,
       volume = {727},
       number = {2},
          eid = {88},
        pages = {88},
          doi = {10.1088/0004-637X/727/2/88},
archivePrefix = {arXiv},
       eprint = {1101.2869},
 primaryClass = {astro-ph.GA},
       adsurl = {https://ui.adsabs.harvard.edu/abs/2011ApJ...727...88C}
}

@ARTICLE{hou21,
       author = {{Hou}, L.~G.},
        title = "{The spiral structure in the Solar neighbourhood}",
      journal = {Frontiers in Astronomy and Space Sciences},
         year = 2021,
        month = aug,
       volume = {8},
          eid = {103},
        pages = {103},
          doi = {10.3389/fspas.2021.671670},
archivePrefix = {arXiv},
       eprint = {2110.04446},
 primaryClass = {astro-ph.GA},
       adsurl = {https://ui.adsabs.harvard.edu/abs/2021FrASS...8..103H}
}

@ARTICLE{veena21,
       author = {{Veena}, V.~S. and {Schilke}, P. and {S{\'a}nchez-Monge}, {\'A}. and {Sormani}, M.~C. and {Klessen}, R.~S. and {Schuller}, F. and {Colombo}, D. and {Csengeri}, T. and {Mattern}, M. and {Urquhart}, J.~S.},
        title = "{A Kiloparsec-scale Molecular Wave in the Inner Galaxy: Feather of the Milky Way?}",
      journal = {\apjl},
         year = 2021,
        month = nov,
       volume = {921},
       number = {2},
          eid = {L42},
        pages = {L42},
          doi = {10.3847/2041-8213/ac341f},
archivePrefix = {arXiv},
       eprint = {2110.13938},
 primaryClass = {astro-ph.GA},
       adsurl = {https://ui.adsabs.harvard.edu/abs/2021ApJ...921L..42V}
}

@ARTICLE{ssun24,
       author = {{Sun}, Shenglan and {Wang}, Ke and {Liu}, Xunchuan and {Xu}, Fengwei},
        title = "{The Formation of Milky Way ``Bones'': Ubiquitous HI Narrow Self-absorption Associated with CO Emission}",
      journal = {\apjl},
         year = 2024,
        month = sep,
       volume = {973},
       number = {1},
          eid = {L27},
        pages = {L27},
          doi = {10.3847/2041-8213/ad77ce},
archivePrefix = {arXiv},
       eprint = {2409.01895},
 primaryClass = {astro-ph.GA},
       adsurl = {https://ui.adsabs.harvard.edu/abs/2024ApJ...973L..27S}
}

@ARTICLE{feher24,
       author = {{Feh{\'e}r}, Orsolya and {Ragan}, S.~E. and {Priestley}, F.~D. and {Clark}, P.~C. and {Moore}, T.~J.~T.},
        title = "{N$_{2}$H$^{+}$(1-0) as a tracer of dense gas in and between spiral arms}",
      journal = {\mnras},
         year = 2024,
        month = may,
       volume = {530},
       number = {2},
        pages = {1311-1327},
          doi = {10.1093/mnras/stae918},
archivePrefix = {arXiv},
       eprint = {2403.19269},
 primaryClass = {astro-ph.GA},
       adsurl = {https://ui.adsabs.harvard.edu/abs/2024MNRAS.530.1311F}
}

@ARTICLE{ge23,
       author = {{Ge}, Y. and {Wang}, K. and {Duarte-Cabral}, A. and {Pettitt}, A.~R. and {Dobbs}, C.~L. and {S{\'a}nchez-Monge}, A. and {Neralwar}, K.~R. and {Urquhart}, J.~S. and {Colombo}, D. and {Dur{\'a}n-Camacho}, E. and {Beuther}, H. and {Bronfman}, L. and {Rigby}, A.~J. and {Eden}, D. and {Neupane}, S. and {Barnes}, P. and {Henning}, T. and {Yang}, A.~Y.},
        title = "{Large-scale velocity-coherent filaments in the SEDIGISM survey: Association with spiral arms and the fraction of dense gas}",
      journal = {\aap},
         year = 2023,
        month = jul,
       volume = {675},
          eid = {A119},
        pages = {A119},
          doi = {10.1051/0004-6361/202245784},
archivePrefix = {arXiv},
       eprint = {2305.07353},
 primaryClass = {astro-ph.GA},
       adsurl = {https://ui.adsabs.harvard.edu/abs/2023A&A...675A.119G}
}

@ARTICLE{chen23,
       author = {{Chen}, Xuepeng and {Sun}, Li and {Feng}, Jiancheng and {Zhang}, Shiyu and {Guo}, Weihua and {Xu}, Xiaoyun and {Su}, Yang and {Sun}, Yan and {Zhang}, Shaobo and {Zhou}, Xin and {Chen}, Zhiwei and {Yan}, Qing-Zeng and {Zhang}, Miaomiao and {Fang}, Min and {Yang}, Ji},
        title = "{The Cassiopeia Filament: A Blown Spur of the Local Arm}",
      journal = {\aj},
         year = 2023,
        month = jan,
       volume = {165},
       number = {1},
          eid = {16},
        pages = {16},
          doi = {10.3847/1538-3881/ac9ea2},
archivePrefix = {arXiv},
       eprint = {2211.00810},
 primaryClass = {astro-ph.GA},
       adsurl = {https://ui.adsabs.harvard.edu/abs/2023AJ....165...16C}
}

@ARTICLE{li22b,
       author = {{Li}, Guang-Xing and {Chen}, Bing-Qiu},
        title = "{Discovery of a coherent, wave-like velocity pattern for the Radcliffe wave}",
      journal = {\mnras},
         year = 2022,
        month = nov,
       volume = {517},
       number = {1},
        pages = {L102-L107},
          doi = {10.1093/mnrasl/slac050},
archivePrefix = {arXiv},
       eprint = {2205.03218},
 primaryClass = {astro-ph.GA},
       adsurl = {https://ui.adsabs.harvard.edu/abs/2022MNRAS.517L.102L}
}

@ARTICLE{elmegreen97,
       author = {{Elmegreen}, Bruce G. and {Efremov}, Yuri N.},
        title = "{A Universal Formation Mechanism for Open and Globular Clusters in Turbulent Gas}",
      journal = {\apj},
         year = 1997,
        month = may,
       volume = {480},
       number = {1},
        pages = {235-245},
          doi = {10.1086/303966},
       adsurl = {https://ui.adsabs.harvard.edu/abs/1997ApJ...480..235E}
}

@ARTICLE{ryon17,
       author = {{Ryon}, J.~E. and {Gallagher}, J.~S. and {Smith}, L.~J. and {Adamo}, A. and {Calzetti}, D. and {Bright}, S.~N. and {Cignoni}, M. and {Cook}, D.~O. and {Dale}, D.~A. and {Elmegreen}, B.~E. and {Fumagalli}, M. and {Gouliermis}, D.~A. and {Grasha}, K. and {Grebel}, E.~K. and {Kim}, H. and {Messa}, M. and {Thilker}, D. and {Ubeda}, L.},
        title = "{Effective Radii of Young, Massive Star Clusters in Two LEGUS Galaxies}",
      journal = {\apj},
         year = 2017,
        month = jun,
       volume = {841},
       number = {2},
          eid = {92},
        pages = {92},
          doi = {10.3847/1538-4357/aa719e},
archivePrefix = {arXiv},
       eprint = {1705.02692},
 primaryClass = {astro-ph.GA},
       adsurl = {https://ui.adsabs.harvard.edu/abs/2017ApJ...841...92R}
}

@ARTICLE{chernin95,
       author = {{Chernin}, Arthur D. and {Efremov}, Yury N. and {Voinovich}, Peter A.},
        title = "{Superassociations: violent star formation induced by shock-shock collisions}",
      journal = {\mnras},
         year = 1995,
        month = jul,
       volume = {275},
       number = {2},
        pages = {313-326},
          doi = {10.1093/mnras/275.2.313},
       adsurl = {https://ui.adsabs.harvard.edu/abs/1995MNRAS.275..313C}
}

@ARTICLE{zhao24,
       author = {{Zhao}, Bo and {Pudritz}, Ralph E. and {Pillsworth}, Rachel and {Robinson}, Hector and {Wadsley}, James},
        title = "{Filamentary Hierarchies and Superbubbles: Galactic Multiscale Magnetohydrodynamic Simulations of Giant Molecular Cloud to Star Cluster Formation}",
      journal = {\apj},
         year = 2024,
        month = oct,
       volume = {974},
       number = {2},
          eid = {240},
        pages = {240},
          doi = {10.3847/1538-4357/ad67e2},
archivePrefix = {arXiv},
       eprint = {2405.18474},
 primaryClass = {astro-ph.GA},
       adsurl = {https://ui.adsabs.harvard.edu/abs/2024ApJ...974..240Z}
}

@ARTICLE{israel78,
       author = {{Israel}, F.~P.},
        title = "{H II regions and CO clouds: the blister model.}",
      journal = {\aap},
         year = 1978,
        month = dec,
       volume = {70},
        pages = {769-775},
       adsurl = {https://ui.adsabs.harvard.edu/abs/1978A&A....70..769I}
}

@ARTICLE{foyle10,
       author = {{Foyle}, K. and {Rix}, H.-W. and {Walter}, F. and {Leroy}, A.~K.},
        title = "{Arm and Interarm Star Formation in Spiral Galaxies}",
      journal = {\apj},
         year = 2010,
        month = dec,
       volume = {725},
       number = {1},
        pages = {534-541},
          doi = {10.1088/0004-637X/725/1/534},
archivePrefix = {arXiv},
       eprint = {1010.0678},
 primaryClass = {astro-ph.CO},
       adsurl = {https://ui.adsabs.harvard.edu/abs/2010ApJ...725..534F}
}

@ARTICLE{moore12,
       author = {{Moore}, T.~J.~T. and {Urquhart}, J.~S. and {Morgan}, L.~K. and {Thompson}, M.~A.},
        title = "{The effect of spiral arms on star formation in the Galaxy}",
      journal = {\mnras},
         year = 2012,
        month = oct,
       volume = {426},
       number = {1},
        pages = {701-707},
          doi = {10.1111/j.1365-2966.2012.21740.x},
archivePrefix = {arXiv},
       eprint = {1204.1578},
 primaryClass = {astro-ph.GA},
       adsurl = {https://ui.adsabs.harvard.edu/abs/2012MNRAS.426..701M}
}

@ARTICLE{bagetakos11,
       author = {{Bagetakos}, I. and {Brinks}, E. and {Walter}, F. and {de Blok}, W.~J.~G. and {Usero}, A. and {Leroy}, A.~K. and {Rich}, J.~W. and {Kennicutt}, Jr., R.~C.},
        title = "{The Fine-scale Structure of the Neutral Interstellar Medium in Nearby Galaxies}",
      journal = {\aj},
         year = 2011,
        month = jan,
       volume = {141},
       number = {1},
          eid = {23},
        pages = {23},
          doi = {10.1088/0004-6256/141/1/23},
archivePrefix = {arXiv},
       eprint = {1008.1845},
 primaryClass = {astro-ph.CO},
       adsurl = {https://ui.adsabs.harvard.edu/abs/2011AJ....141...23B}
}

@ARTICLE{lahen19,
       author = {{Lah{\'e}n}, Natalia and {Naab}, Thorsten and {Johansson}, Peter H. and {Elmegreen}, Bruce and {Hu}, Chia-Yu and {Walch}, Stefanie},
        title = "{The Formation of Low-metallicity Globular Clusters in Dwarf Galaxy Mergers}",
      journal = {\apjl},
         year = 2019,
        month = jul,
       volume = {879},
       number = {2},
          eid = {L18},
        pages = {L18},
          doi = {10.3847/2041-8213/ab2a13},
archivePrefix = {arXiv},
       eprint = {1905.09840},
 primaryClass = {astro-ph.GA},
       adsurl = {https://ui.adsabs.harvard.edu/abs/2019ApJ...879L..18L}
}

@ARTICLE{cook23,
       author = {{Cook}, D.~O. and {Lee}, J.~C. and {Adamo}, A. and {Calzetti}, D. and {Chandar}, R. and {Whitmore}, B.~C. and {Aloisi}, A. and {Cignoni}, M. and {Dale}, D.~A. and {Elmegreen}, B.~G. and {Fumagalli}, M. and {Grasha}, K. and {Johnson}, K.~E. and {Kennicutt}, R.~C. and {Kim}, H. and {Linden}, S.~T. and {Messa}, M. and {{\"O}stlin}, G. and {Ryon}, J.~E. and {Sacchi}, E. and {Thilker}, D.~A. and {Tosi}, M. and {Wofford}, A.},
        title = "{Fraction of stars in clusters for the LEGUS dwarf galaxies}",
      journal = {\mnras},
         year = 2023,
        month = mar,
       volume = {519},
       number = {3},
        pages = {3749-3775},
          doi = {10.1093/mnras/stac3748},
archivePrefix = {arXiv},
       eprint = {2212.07519},
 primaryClass = {astro-ph.GA},
       adsurl = {https://ui.adsabs.harvard.edu/abs/2023MNRAS.519.3749C}
}

@ARTICLE{chandar17,
       author = {{Chandar}, Rupali and {Fall}, S. Michael and {Whitmore}, Bradley C. and {Mulia}, Alexander J.},
        title = "{The Fraction of Stars That Form in Clusters in Different Galaxies}",
      journal = {\apj},
         year = 2017,
        month = nov,
       volume = {849},
       number = {2},
          eid = {128},
        pages = {128},
          doi = {10.3847/1538-4357/aa92ce},
archivePrefix = {arXiv},
       eprint = {1806.11537},
 primaryClass = {astro-ph.GA},
       adsurl = {https://ui.adsabs.harvard.edu/abs/2017ApJ...849..128C}
}

@ARTICLE{elmegreen11,
       author = {{Elmegreen}, Bruce G.},
        title = "{Gravitational Instabilities in Two-component Galaxy Disks with Gas Dissipation}",
      journal = {\apj},
         year = 2011,
        month = aug,
       volume = {737},
       number = {1},
          eid = {10},
        pages = {10},
          doi = {10.1088/0004-637X/737/1/10},
archivePrefix = {arXiv},
       eprint = {1106.1580},
 primaryClass = {astro-ph.GA},
       adsurl = {https://ui.adsabs.harvard.edu/abs/2011ApJ...737...10E}
}

@ARTICLE{kim06b,
       author = {{Kim}, Chang-Goo and {Kim}, Woong-Tae and {Ostriker}, Eve C.},
        title = "{Interstellar Turbulence Driving by Galactic Spiral Shocks}",
      journal = {\apjl},
         year = 2006,
        month = sep,
       volume = {649},
       number = {1},
        pages = {L13-L16},
          doi = {10.1086/508160},
archivePrefix = {arXiv},
       eprint = {astro-ph/0608161},
 primaryClass = {astro-ph},
       adsurl = {https://ui.adsabs.harvard.edu/abs/2006ApJ...649L..13K}
}

@ARTICLE{lau25,
       author = {{Lau}, Cheryl S.~C. and {Bonnell}, Ian A.},
        title = "{Semi-confined supernova feedback in H II region bubbles}",
      journal = {\mnras},
         year = 2025,
        month = jun,
       volume = {540},
       number = {1},
        pages = {1124-1143},
          doi = {10.1093/mnras/staf756},
archivePrefix = {arXiv},
       eprint = {2410.21255},
 primaryClass = {astro-ph.GA},
       adsurl = {https://ui.adsabs.harvard.edu/abs/2025MNRAS.540.1124L}
}

@ARTICLE{billett02,
       author = {{Billett}, Olivia H. and {Hunter}, Deidre A. and {Elmegreen}, Bruce G.},
        title = "{Compact Star Clusters in Nearby Dwarf Irregular Galaxies}",
      journal = {\aj},
         year = 2002,
        month = mar,
       volume = {123},
       number = {3},
        pages = {1454-1475},
          doi = {10.1086/339181},
archivePrefix = {arXiv},
       eprint = {astro-ph/0112260},
 primaryClass = {astro-ph},
       adsurl = {https://ui.adsabs.harvard.edu/abs/2002AJ....123.1454B}
}

@ARTICLE{reina17,
       author = {{Reina-Campos}, Marta and {Kruijssen}, J.~M. Diederik},
        title = "{A unified model for the maximum mass scales of molecular clouds, stellar clusters and high-redshift clumps}",
      journal = {\mnras},
         year = 2017,
        month = aug,
       volume = {469},
       number = {2},
        pages = {1282-1298},
          doi = {10.1093/mnras/stx790},
archivePrefix = {arXiv},
       eprint = {1704.00732},
 primaryClass = {astro-ph.GA},
       adsurl = {https://ui.adsabs.harvard.edu/abs/2017MNRAS.469.1282R}
}

@ARTICLE{adamo15,
       author = {{Adamo}, A. and {Kruijssen}, J.~M.~D. and {Bastian}, N. and {Silva-Villa}, E. and {Ryon}, J.},
        title = "{Probing the role of the galactic environment in the formation of stellar clusters, using M83 as a test bench}",
      journal = {\mnras},
         year = 2015,
        month = sep,
       volume = {452},
       number = {1},
        pages = {246-260},
          doi = {10.1093/mnras/stv1203},
archivePrefix = {arXiv},
       eprint = {1505.07475},
 primaryClass = {astro-ph.GA},
       adsurl = {https://ui.adsabs.harvard.edu/abs/2015MNRAS.452..246A}
}

@ARTICLE{larsen02,
       author = {{Larsen}, S{\o}ren S.},
        title = "{The Luminosity Function of Star Clusters in Spiral Galaxies}",
      journal = {\aj},
         year = 2002,
        month = sep,
       volume = {124},
       number = {3},
        pages = {1393-1409},
          doi = {10.1086/342381},
archivePrefix = {arXiv},
       eprint = {astro-ph/0206295},
 primaryClass = {astro-ph},
       adsurl = {https://ui.adsabs.harvard.edu/abs/2002AJ....124.1393L}
}

@ARTICLE{whitmore00,
       author = {{Whitmore}, Bradley C.},
        title = "{The Formation of Star Clusters}",
      journal = {arXiv e-prints},
         year = 2000,
        month = dec,
          eid = {astro-ph/0012546},
        pages = {astro-ph/0012546},
          doi = {10.48550/arXiv.astro-ph/0012546},
archivePrefix = {arXiv},
       eprint = {astro-ph/0012546},
 primaryClass = {astro-ph},
       adsurl = {https://ui.adsabs.harvard.edu/abs/2000astro.ph.12546W}
}

@ARTICLE{johnson03,
       author = {{Johnson}, Kelsey E. and {Kobulnicky}, Henry A.},
        title = "{The Spectral Energy Distributions of Infant Super-Star Clusters in Henize 2-10 from 7 Millimeters to 6 Centimeters}",
      journal = {\apj},
         year = 2003,
        month = nov,
       volume = {597},
       number = {2},
        pages = {923-928},
          doi = {10.1086/378585},
archivePrefix = {arXiv},
       eprint = {astro-ph/0308303},
 primaryClass = {astro-ph},
       adsurl = {https://ui.adsabs.harvard.edu/abs/2003ApJ...597..923J}
}

@ARTICLE{johnson18,
       author = {{Johnson}, Kelsey E. and {Brogan}, Crystal L. and {Indebetouw}, Remy and {Testi}, Leonardo and {Wilner}, David J. and {Reines}, Amy E. and {Chen}, C.-H. Rosie and {Vanzi}, Leonardo},
        title = "{The Association of Molecular Gas and Natal Super Star Clusters in Henize 2-10}",
      journal = {\apj},
         year = 2018,
        month = feb,
       volume = {853},
       number = {2},
          eid = {125},
        pages = {125},
          doi = {10.3847/1538-4357/aa9ff8},
archivePrefix = {arXiv},
       eprint = {1712.02791},
 primaryClass = {astro-ph.GA},
       adsurl = {https://ui.adsabs.harvard.edu/abs/2018ApJ...853..125J}
}

@ARTICLE{costa21,
       author = {{Costa}, Allison H. and {Johnson}, Kelsey E. and {Indebetouw}, Remy and {Finn}, Molly K. and {Brogan}, Crystal L. and {Reines}, Amy},
        title = "{Toward a More Complex Understanding of Natal Super Star Clusters with Multiwavelength Observations}",
      journal = {\apj},
         year = 2021,
        month = sep,
       volume = {918},
       number = {2},
          eid = {76},
        pages = {76},
          doi = {10.3847/1538-4357/ac0e93},
archivePrefix = {arXiv},
       eprint = {2107.02695},
 primaryClass = {astro-ph.GA},
       adsurl = {https://ui.adsabs.harvard.edu/abs/2021ApJ...918...76C}
}

@ARTICLE{johnson15,
       author = {{Johnson}, K.~E. and {Leroy}, A.~K. and {Indebetouw}, R. and {Brogan}, C.~L. and {Whitmore}, B.~C. and {Hibbard}, J. and {Sheth}, K. and {Evans}, A.~S.},
        title = "{The Physical Conditions in a Pre-super Star Cluster Molecular Cloud in the Antennae Galaxies}",
      journal = {\apj},
         year = 2015,
        month = jun,
       volume = {806},
       number = {1},
          eid = {35},
        pages = {35},
          doi = {10.1088/0004-637X/806/1/35},
archivePrefix = {arXiv},
       eprint = {1503.06477},
 primaryClass = {astro-ph.GA},
       adsurl = {https://ui.adsabs.harvard.edu/abs/2015ApJ...806...35J}
}

@ARTICLE{finn19,
       author = {{Finn}, Molly K. and {Johnson}, Kelsey E. and {Brogan}, Crystal L. and {Wilson}, Christine D. and {Indebetouw}, Remy and {Harris}, William E. and {Kamenetzky}, Julia and {Bemis}, Ashley},
        title = "{New Insights into the Physical Conditions and Internal Structure of a Candidate Proto-globular Cluster}",
      journal = {\apj},
         year = 2019,
        month = apr,
       volume = {874},
       number = {2},
          eid = {120},
        pages = {120},
          doi = {10.3847/1538-4357/ab0d1e},
archivePrefix = {arXiv},
       eprint = {1903.08669},
 primaryClass = {astro-ph.GA},
       adsurl = {https://ui.adsabs.harvard.edu/abs/2019ApJ...874..120F}
}

@ARTICLE{krahm24,
       author = {{Krahm}, Grace and {Finn}, Molly K. and {Indebetouw}, Remy and {Johnson}, Kelsey E. and {Kamenetzky}, Julia and {Bemis}, Ashley},
        title = "{Physical Properties of Molecular Clouds in the Overlap Region of the Merging Antennae Galaxies}",
      journal = {\apj},
         year = 2024,
        month = apr,
       volume = {964},
       number = {2},
          eid = {166},
        pages = {166},
          doi = {10.3847/1538-4357/ad2451},
archivePrefix = {arXiv},
       eprint = {2401.04835},
 primaryClass = {astro-ph.GA},
       adsurl = {https://ui.adsabs.harvard.edu/abs/2024ApJ...964..166K}
}

@ARTICLE{finn24a,
       author = {{Finn}, Molly K. and {Johnson}, Kelsey E. and {Indebetouw}, Remy and {Costa}, Allison H. and {Adamo}, Angela and {Aloisi}, Alessandra and {Bittle}, Lauren and {Calzetti}, Daniela and {Dale}, Daniel A. and {Dobbs}, Clare L. and {Donovan Meyer}, Jennifer and {Elmegreen}, Bruce G. and {Elmegreen}, Debra M. and {Fumagalli}, Michele and {Gallagher}, J.~S. and {Grasha}, Kathryn and {Grebel}, Eva K. and {Kennicutt}, Robert C. and {Krumholz}, Mark R. and {Lee}, Janice C. and {Messa}, Matteo and {Nair}, Preethi and {Sabbi}, Elena and {Smith}, Linda J. and {Thilker}, David A. and {Whitmore}, Bradley C. and {Wofford}, Aida},
        title = "{ALMA-LEGUS. I. The Influence of Galaxy Morphology on Molecular Cloud Properties}",
      journal = {\apj},
         year = 2024,
        month = mar,
       volume = {964},
       number = {1},
          eid = {12},
        pages = {12},
          doi = {10.3847/1538-4357/ad1eec},
archivePrefix = {arXiv},
       eprint = {2401.01450},
 primaryClass = {astro-ph.GA},
       adsurl = {https://ui.adsabs.harvard.edu/abs/2024ApJ...964...12F}
}

@ARTICLE{finn24b,
       author = {{Finn}, Molly K. and {Johnson}, Kelsey E. and {Indebetouw}, Remy and {Costa}, Allison H. and {Adamo}, Angela and {Aloisi}, Alessandra and {Bittle}, Lauren and {Calzetti}, Daniela and {Dale}, Daniel A. and {Dobbs}, Clare L. and {Donovan Meyer}, Jennifer and {Elmegreen}, Bruce G. and {Elmegreen}, Debra M. and {Fumagalli}, Michele and {Gallagher}, J.~S. and {Grasha}, Kathryn and {Grebel}, Eva K. and {Kennicutt}, Robert C. and {Krumholz}, Mark R. and {Lee}, Janice C. and {Messa}, Matteo and {Nair}, Preethi and {Sabbi}, Elena and {Smith}, Linda J. and {Thilker}, David A. and {Whitmore}, Bradley C. and {Wofford}, Aida},
        title = "{ALMA-LEGUS. II. The Influence of Subgalactic Environments on Molecular Cloud Properties}",
      journal = {\apj},
         year = 2024,
        month = mar,
       volume = {964},
       number = {1},
          eid = {13},
        pages = {13},
          doi = {10.3847/1538-4357/ad198a},
archivePrefix = {arXiv},
       eprint = {2401.01451},
 primaryClass = {astro-ph.GA},
       adsurl = {https://ui.adsabs.harvard.edu/abs/2024ApJ...964...13F}
}

@ARTICLE{calzetti15,
       author = {{Calzetti}, D. and {Johnson}, K.~E. and {Adamo}, A. and {Gallagher}, III, J.~S. and {Andrews}, J.~E. and {Smith}, L.~J. and {Clayton}, G.~C. and {Lee}, J.~C. and {Sabbi}, E. and {Ubeda}, L. and {Kim}, H. and {Ryon}, J.~E. and {Thilker}, D. and {Bright}, S.~N. and {Zackrisson}, E. and {Kennicutt}, R.~C. and {de Mink}, S.~E. and {Whitmore}, B.~C. and {Aloisi}, A. and {Chandar}, R. and {Cignoni}, M. and {Cook}, D. and {Dale}, D.~A. and {Elmegreen}, B.~G. and {Elmegreen}, D.~M. and {Evans}, A.~S. and {Fumagalli}, M. and {Gouliermis}, D.~A. and {Grasha}, K. and {Grebel}, E.~K. and {Krumholz}, M.~R. and {Walterbos}, R. and {Wofford}, A. and {Brown}, T.~M. and {Christian}, C. and {Dobbs}, C. and {Herrero}, A. and {Kahre}, L. and {Messa}, M. and {Nair}, P. and {Nota}, A. and {{\"O}stlin}, G. and {Pellerin}, A. and {Sacchi}, E. and {Schaerer}, D. and {Tosi}, M.},
        title = "{The Brightest Young Star Clusters in NGC 5253.}",
      journal = {\apj},
         year = 2015,
        month = oct,
       volume = {811},
       number = {2},
          eid = {75},
        pages = {75},
          doi = {10.1088/0004-637X/811/2/75},
archivePrefix = {arXiv},
       eprint = {1508.04476},
 primaryClass = {astro-ph.GA},
       adsurl = {https://ui.adsabs.harvard.edu/abs/2015ApJ...811...75C}
}

@ARTICLE{lopez12,
       author = {{L{\'o}pez-S{\'a}nchez}, {\'A}. R. and {Koribalski}, B.~S. and {van Eymeren}, J. and {Esteban}, C. and {Kirby}, E. and {Jerjen}, H. and {Lonsdale}, N.},
        title = "{The intriguing H I gas in NGC 5253: an infall of a diffuse, low-metallicity H I cloud?}",
      journal = {\mnras},
         year = 2012,
        month = jan,
       volume = {419},
       number = {2},
        pages = {1051-1069},
          doi = {10.1111/j.1365-2966.2011.19762.x},
archivePrefix = {arXiv},
       eprint = {1109.0806},
 primaryClass = {astro-ph.CO},
       adsurl = {https://ui.adsabs.harvard.edu/abs/2012MNRAS.419.1051L}
}

@ARTICLE{johnson12,
       author = {{Johnson}, Megan and {Hunter}, Deidre A. and {Oh}, Se-Heon and {Zhang}, Hong-Xin and {Elmegreen}, Bruce and {Brinks}, Elias and {Tollerud}, Erik and {Herrmann}, Kimberly},
        title = "{The Stellar and Gas Kinematics of the LITTLE THINGS Dwarf Irregular Galaxy NGC 1569}",
      journal = {\aj},
         year = 2012,
        month = nov,
       volume = {144},
       number = {5},
          eid = {152},
        pages = {152},
          doi = {10.1088/0004-6256/144/5/152},
archivePrefix = {arXiv},
       eprint = {1209.2453},
 primaryClass = {astro-ph.CO},
       adsurl = {https://ui.adsabs.harvard.edu/abs/2012AJ....144..152J}
}

@ARTICLE{elmegreen96,
       author = {{Elmegreen}, Bruce G. and {Falgarone}, Edith},
        title = "{A Fractal Origin for the Mass Spectrum of Interstellar Clouds}",
      journal = {\apj},
         year = 1996,
        month = nov,
       volume = {471},
        pages = {816},
          doi = {10.1086/178009},
       adsurl = {https://ui.adsabs.harvard.edu/abs/1996ApJ...471..816E}
}

@ARTICLE{fleck96,
       author = {{Fleck}, Jr., Robert C.},
        title = "{Scaling Relations for the Turbulent, Non--Self-gravitating, Neutral Component of the Interstellar Medium}",
      journal = {\apj},
         year = 1996,
        month = feb,
       volume = {458},
        pages = {739},
          doi = {10.1086/176853},
       adsurl = {https://ui.adsabs.harvard.edu/abs/1996ApJ...458..739F}
}

@ARTICLE{brown21,
       author = {{Brown}, Gillen and {Gnedin}, Oleg Y.},
        title = "{Radii of young star clusters in nearby galaxies}",
      journal = {\mnras},
         year = 2021,
        month = dec,
       volume = {508},
       number = {4},
        pages = {5935-5953},
          doi = {10.1093/mnras/stab2907},
archivePrefix = {arXiv},
       eprint = {2106.12420},
 primaryClass = {astro-ph.GA},
       adsurl = {https://ui.adsabs.harvard.edu/abs/2021MNRAS.508.5935B}
}

@ARTICLE{peeters04,
       author = {{Peeters}, E. and {Allamandola}, L.~J. and {Bauschlicher}, Jr., C.~W. and {Hudgins}, D.~M. and {Sandford}, S.~A. and {Tielens}, A.~G.~G.~M.},
        title = "{Deuterated Interstellar Polycyclic Aromatic Hydrocarbons}",
      journal = {\apj},
         year = 2004,
        month = mar,
       volume = {604},
       number = {1},
        pages = {252-257},
          doi = {10.1086/381866},
       adsurl = {https://ui.adsabs.harvard.edu/abs/2004ApJ...604..252P}
}

@ARTICLE{toomre69,
       author = {{Toomre}, Alar},
        title = "{Group Velocity of Spiral Waves in Galactic Disks}",
      journal = {\apj},
         year = 1969,
        month = dec,
       volume = {158},
        pages = {899},
          doi = {10.1086/150250},
       adsurl = {https://ui.adsabs.harvard.edu/abs/1969ApJ...158..899T}
}
\bibliographystyle{aasjournalv7}

\newpage
\begin{appendix}
\setcounter{figure}{0}
\renewcommand{\thefigure}{\thesection.\arabic{figure}}

\section{Additional Figures of eYSC superposed on the F770W image}
\begin{figure}[h]
\includegraphics[width=16.cm]{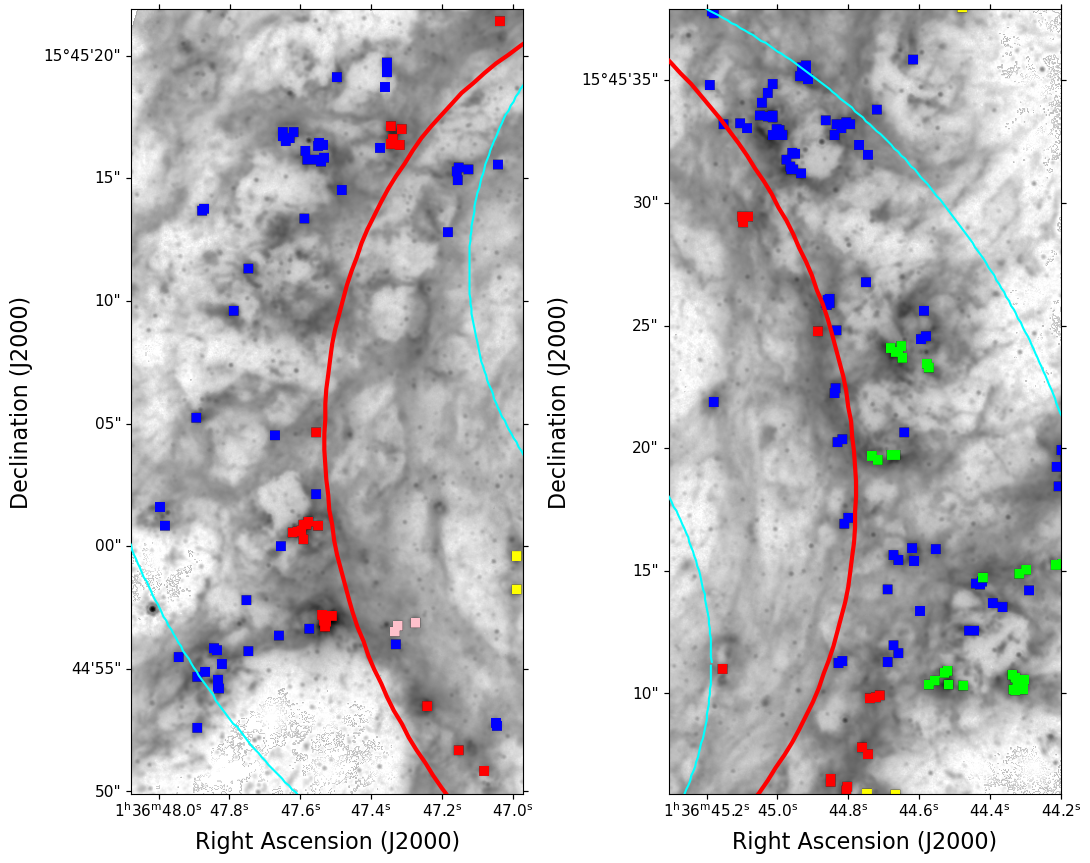} 
\caption{Enlargements of the eastern (left) and western (right) portions of the F770W image with clusters superposed as squares in the same color scheme as Fig. \ref{fig:8mubig}.  The red curve is the fit to the logarithmic spiral arm and the cyan curves are circles at the limiting radii for the fit. Bubble structure dominates the F770W image, often making it difficult to distinguish between $7.7\mu$m clouds that are in bubble walls and other clouds that are part of a long spiral filament.   The F770W image is displayed with a logarithmic intensity stretch.
}
\label{fig:8mu_2x1}
\end{figure}

\begin{figure}
\begin{center}
\includegraphics[width=16.cm]{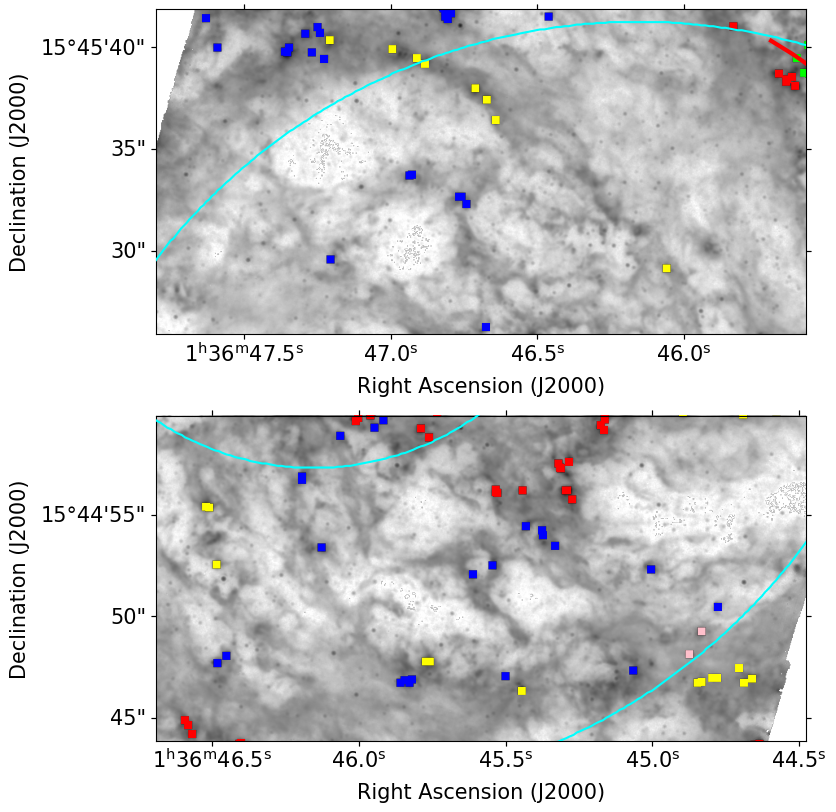} 
\caption{Enlargements of the northern (top) and southern (bottom) portions of the F770W image with clusters superposed as squares in the same color scheme as Fig. \ref{fig:8mubig}.   A small piece of the red curve for the spiral arm fit is in the top right; the cyan curves denote the limiting radii for the fit. These fields of view highlight the two most prominent interarm feathers, which both contain many clusters (yellow squares) and yet have few peripheral clusters inside bubble walls (blue).  The F770W image is displayed with a logarithmic intensity stretch.
}
\label{fig:8mu_2x1_feathers}
\end{center}
\end{figure}

\begin{figure}
\begin{center}
\includegraphics[width=16.cm]{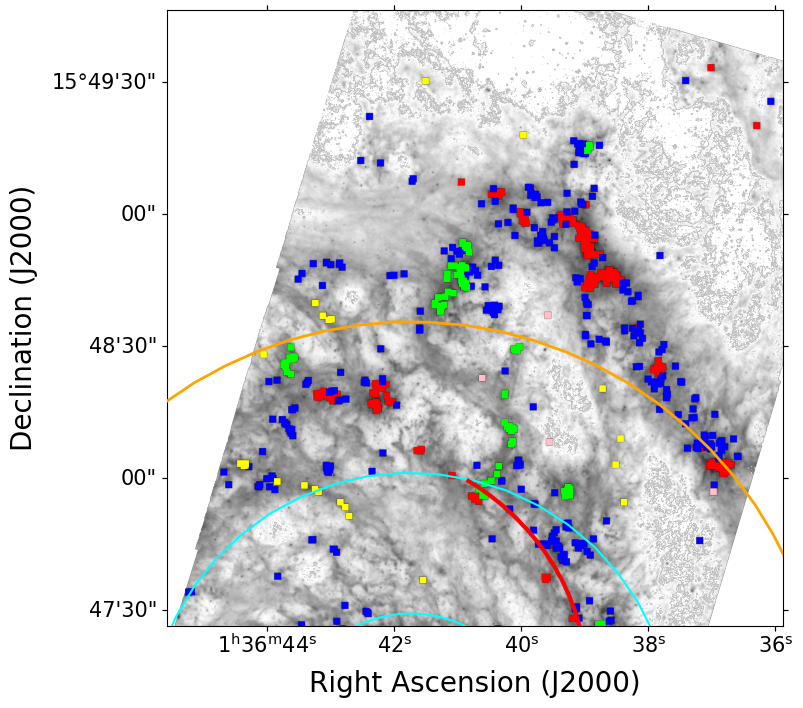}  
\caption{The northern part of the F770W image of NGC 628 with eYSC superposed using the color scheme related to local $7.7\mu$m morphology: red, green and yellow for $7.7\mu$m emission in spiral arm filaments, spiral arm spurs and interarm feathers, and blue for $7.7\mu$m emission in bubbles. The two lower cyan circles with radii of $28^{\prime\prime}$ and $60^{\prime\prime}$ are the limits used to fit the stellar density wave to a logarithmic spiral, and the orange circle at a radius of 4.5 kpc ($94^{\prime\prime}$) is an estimate for corotation from \cite{williams22}.  The red curve is the spiral fit.  The relative position of the eYSC in spiral arm filaments and bubbles appears to differ on the two sides of corotation with bubble eYSCs generally clockwise compared to spiral eYSCs inside corotation and more mixed outside. An exception is the prominent spur where the bubble eYSCs are clockwise relative to the spur eYSCs even though this is nominally outside corotation.   The F770W image is displayed with a logarithmic intensity stretch.
}
\label{fig:8mu_top}
\end{center}
\end{figure}

\begin{figure}
\begin{center}
\includegraphics[width=16.cm]{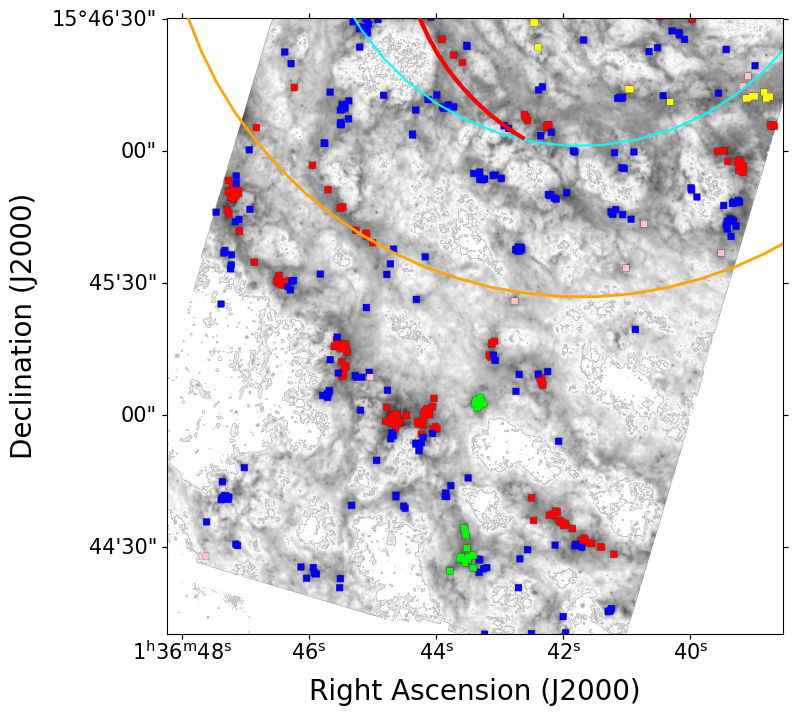}  
\caption{The southern part of the F770W image of NGC 628 with eYSC superposed using the same color scheme related to local $7.7\mu$m morphology: red, green and yellow for $7.7\mu$m emission in spiral arm filaments, spiral arm spurs and interarm feathers, and blue for $7.7\mu$m emission in bubbles. The orange circle is corotation from \cite{williams22} and the upper one is the outer limit used to fit the stellar density wave to a logarithmic spiral.  The red curve is the logarithmic spiral fit to the peak of the stellar density wave. The southern arm has a larger proportion of eYSCs in spiral arm filaments compared to bubbles than the northern arm.  The F770W image is displayed with a logarithmic intensity stretch.
}
\label{fig:8mu_bottom}
\end{center}
\end{figure}

\end{appendix}

\end{document}